\pdfoutput=1
\documentclass[11pt,twoside,a4paper,cmspaper,final,collab]{cms-tdr}
\def\svnVersion{4ee8c8a}\def\svnDate{2024/12/30}\def\cmsCernNoTag{CERN-EP-2024-231}\def\cmsCernDate{\today}\def\cmsMessage{Published in Physical Review D as \href{http://dx.doi.org/10.1103/PhysRevD.110.112016}{\doi{10.1103/PhysRevD.110.112016}.}}
\begin{document}\cmsNoteHeader{TOP-23-007}

\ifthenelse{\boolean{cms@external}}{%
    \providecommand{\cmsLeft}{upper\xspace}
    \providecommand{\cmsRight}{lower\xspace}
    \providecommand{\cmsUpperLeft}{upper\xspace}
    \providecommand{\cmsUpperRight}{middle\xspace}
}{%
    \providecommand{\cmsLeft}{left\xspace}
    \providecommand{\cmsRight}{right\xspace}
    \providecommand{\cmsUpperLeft}{upper left\xspace}
    \providecommand{\cmsUpperRight}{upper right\xspace}
}

\newcommand{\tqh}{\ensuremath{\PQt_{\mathrm{h}}}\xspace}
\newcommand{\Wh}{\ensuremath{\PW_{\mathrm{h}}}\xspace}
\newcommand{\tql}{\ensuremath{\PQt_{\Pell}}\xspace}
\newcommand{\Jbl}{\ensuremath{\PQb_{\Pell}}\xspace}

\newcommand{\Mtop}{\ensuremath{m_{\PQt}}\xspace}
\newcommand{\mur}{\ensuremath{\mu_{\mathrm{r}}}\xspace}
\newcommand{\muf}{\ensuremath{\mu_{\mathrm{f}}}\xspace}
\newcommand{\SNN}{\ensuremath{S_{\text{NN}}}\xspace}
\newcommand{\Shigh}{\ensuremath{S_{\text{high}}}\xspace}
\newcommand{\Slow}{\ensuremath{S_{\text{low}}}\xspace}
\newcommand{\Qone}{\ensuremath{D}\xspace}
\newcommand{\Qonen}{\ensuremath{\Qone_n}\xspace}
\newcommand{\Dtilde}{\ensuremath{\tilde{D}}\xspace}
\newcommand{\Dtilden}{\ensuremath{\Dtilde_n}\xspace}
\newcommand{\chitilde}{\ensuremath{\tilde{\chi}}\xspace}
\newcommand{\DEcrit}{\ensuremath{\Delta_{\text{E\,crit}}}\xspace}
\newcommand{\DEsep}{\ensuremath{\Delta_{\text{E\,sep}}}\xspace}
\newcommand{\DEmax}{\ensuremath{\Delta_{\text{E\,max}}}\xspace}
\newcommand{\DE}{\ensuremath{\Delta_{\text{E}}}\xspace}
\newcommand{\lpj}{\ensuremath{\Pe/\Pgm\text{+jets}}\xspace}
\newcommand{\Wjets}{\ensuremath{\PW\text{+jets}}\xspace}

\newcommand{\tpt}{\ensuremath{\pt(\PQt)}\xspace}
\newcommand{\tbarpt}{\ensuremath{\pt(\PAQt)}\xspace}
\newcommand{\ttm}{\ensuremath{m(\ttbar)}\xspace}
\newcommand{\coschi}{\ensuremath{\cos(\chi)}\xspace}
\newcommand{\coschit}{\ensuremath{\cos(\chitilde)}\xspace}
\newcommand{\thetat}{\ensuremath{\theta}\xspace}
\newcommand{\costh}{\ensuremath{\cos(\thetat)}\xspace}
\newcommand{\absct}{\ensuremath{\abs{\costh}}\xspace}
\newcommand{\ttmcth}{\mbox{\ttm \vs \absct}\xspace}
\newcommand{\tptcth}{\mbox{\tpt \vs \absct}\xspace}

\newcommand{\Qm}{\ensuremath{Q_m}\xspace}
\newcommand{\QmMC}{\ensuremath{\Qm^{\text{MC}}}\xspace}
\newcommand{\Tm}{\ensuremath{T_m}\xspace}

\newcommand{\MINNLO}{\textsc{minnlo}\xspace}
\newcommand{\HATHOR}{\textsc{hathor}\xspace}
\newcommand{\PWGHRW}{\POWHEG{}+\HERWIG}
\newcommand{\PWGPYT}{\POWHEG{}+\PYTHIA}
\newcommand{\PWGPYTETA}{\POWHEG{}+\PYTHIA{}+\etat}
\newcommand{\MGHPYT}{\MGvATNLO{}+\PYTHIA}
\newcommand{\MINPYT}{\MINNLO{}+\PYTHIA}

\newcommand{\gluglu}{\ensuremath{\Pg\Pg}\xspace}
\newcommand{\etat}{{\HepParticle{\PGh}{\PQt}{}}\xspace}

\newcommand{\Sigtot}{\ensuremath{\Sigma_{\text{tot}}}\xspace}
\newcommand{\SigtotMC}{\ensuremath{\Sigma_{\text{tot}}^{\text{MC}}}\xspace}
\newcommand{\Sigm}{\ensuremath{\Sigma_m}\xspace}
\newcommand{\LSigm}{\ensuremath{L\Sigm}\xspace}
\newcommand{\decprod}{\ensuremath{\mathrm{p}}\xspace}
\newcommand{\decprodb}{\ensuremath{\bar{\mathrm{p}}}\xspace}
\newcommand{\porpbar}{\ensuremath{\decprod(\decprodb)}\xspace}
\newcommand{\thetappb}{\ensuremath{\theta_{\porpbar}}\xspace}
\newcommand{\costhetap}{\ensuremath{\cos(\theta_{\decprod})}\xspace}
\newcommand{\costhetapb}{\ensuremath{\cos(\theta_{\decprodb})}\xspace}
\newcommand{\phippb}{\ensuremath{\phi_{\porpbar}}\xspace}
\newcommand{\phip}{\ensuremath{\phi_{\decprod}}\xspace}
\newcommand{\phipb}{\ensuremath{\phi_{\decprodb}}\xspace}
\newcommand{\signorm}{\ensuremath{\sigma_{\text{norm}}}\xspace}

\newcommand{\Cnn}{\ensuremath{C_{nn}}\xspace}
\newcommand{\Crr}{\ensuremath{C_{rr}}\xspace}
\newcommand{\Ckk}{\ensuremath{C_{kk}}\xspace}

\newcommand{\an}{\ensuremath{a_n}\xspace}
\newcommand{\Qmn}{\ensuremath{Q_{mn}}\xspace}
\newcommand{\setQmn}{\ensuremath{\{\Qmn\}}\xspace}
\newcommand{\nuk}{\ensuremath{\nu_k}\xspace}
\newcommand{\setnuk}{\ensuremath{\{\nuk\}}\xspace}

\newcommand{\oneb}{\ensuremath{1\PQb}\xspace}
\newcommand{\twob}{\ensuremath{2\PQb}\xspace}
\newcommand{\onebhigh}{\mbox{\oneb \Shigh}\xspace}
\newcommand{\oneblow}{\mbox{\oneb \Slow}\xspace}
\newcommand{\twobhigh}{\mbox{\twob \Shigh}\xspace}
\newcommand{\twoblow}{\mbox{\twob \Slow}\xspace}

\cmsNoteHeader{TOP-23-007}
\title{Measurements of polarization and spin correlation and observation of entanglement in top quark pairs using lepton+jets events from proton-proton collisions at \texorpdfstring{$\sqrt{s}=13\TeV$}{sqrt(s) = 13 TeV}}
\date{\today}

\abstract{Measurements of the polarization and spin correlation in top quark pairs (\ttbar) are presented using events with a single electron or muon and jets in the final state. The measurements are based on proton-proton collision data from the LHC at $\sqrt{s}=13\TeV$ collected by the CMS experiment, corresponding to an integrated luminosity of 138\fbinv. All coefficients of the polarization vectors and the spin correlation matrix are extracted simultaneously by performing a binned likelihood fit to the data. The measurement is performed inclusively and in bins of additional observables, such as the mass of the \ttbar system and the top quark scattering angle in the \ttbar rest frame. The measured polarization and spin correlation are in agreement with the standard model. From the measured spin correlation, conclusions on the \ttbar spin entanglement are drawn by applying the Peres--Horodecki criterion. The standard model predicts entangled spins for \ttbar states at the production threshold and at high masses of the \ttbar system. Entanglement is observed for the first time in events at high \ttbar mass, where a large fraction of the \ttbar decays are space-like separated, with an expected and observed significance of above 5 standard deviations.}

\hypersetup{
pdfauthor={CMS Collaboration},
pdftitle={Measurements of polarization and spin correlation and observation of entanglement in top quark pairs using lepton+jets events from proton-proton collisions at sqrt(s) = 13 TeV},
pdfsubject={CMS},
pdfkeywords={CMS, top quark pairs, polarization, spin correlation, entanglement}}

\maketitle

\section{Introduction}
\label{introduction}

The top quark is the most massive known fundamental particle with a lifetime of the order of $10^{-25}\unit{s}$. This is shorter than the quantum chromodynamics (QCD) hadronization time scale $1/\Lambda_{\text{QCD}}\approx10^{-24}\unit{s}$, and the spin decorrelation time scale $\Mtop/\Lambda_{\text{QCD}}^2\approx10^{-21}\unit{s}$, where \Mtop is the top quark mass~\cite{ParticleDataGroup:2024, Mahlon:2010gw}.  Consequently, the top quark usually decays before hadronization, thus preserving its spin information in the angular distribution of the decay products. This makes top quark and antiquark (\ttbar) pairs excellent candidates for studying polarization and spin correlation.

We present measurements of the polarization and spin correlation in \ttbar pairs using proton-proton collisions at a center-of-mass energy of 13\TeV at the CERN LHC. The measurements are performed using data collected by the CMS detector between 2016 and 2018, corresponding to an integrated luminosity of 138\fbinv. Once produced, a top quark decays via the weak interaction into a \PW boson and a \PQb quark. The \PW boson further decays into either two quarks, which subsequently hadronize into jets, or a charged lepton and a neutrino. In this analysis, we focus on the final state with two \PQb jets, two jets from one \PW boson, and an electron or muon paired with a neutrino from the other \PW boson. This decay channel is referred to as the \lpj channel. Events with tau leptons are treated as \ttbar background and not included in the \lpj category.

At the LHC \ttbar pairs are produced through gluon-gluon (\gluglu) fusion and quark-antiquark (\qqbar) annihilation. The top quarks and antiquarks are unpolarized at leading order (LO). However, their spins are expected to be strongly correlated~\cite{Baumgart:2012twist}. The complete spin correlation is encoded in a $3{\times}3$ matrix that depends on the \ttbar production mechanism, the invariant mass of the \ttbar system \ttm, and the scattering angle of the top quark. Evidence for \ttbar spin correlation was first reported by the \DZERO experiment at the Tevatron in Refs.~\cite{D0:2011spincorr, D0:2015spincorr}. The ATLAS and CMS experiments have performed a number of top quark polarization and spin correlation measurements using various observables and data sets~\cite{CMS:2013sc, CMS:2015sc_lpjets, CMS:2016sc, CMS:2019spincorrdilep, ATLAS:2012sc, ATLAS:2014sc, ATLAS:2014spincorr7tev, ATLAS:2019spincorrdilep,ATLAS:2023fsd, CMS-PAS-TOP-23-001}.

The top quark polarization and spin correlation measurement is interesting in its own right as a test of the standard model (SM)~\cite{Wu:2018xiz,Maltoni:2024tul}, but it also provides new opportunities for testing quantum mechanics (QM) at high energies using the decay products of unstable particles as probes. This is not possible in experiments with stable particles, such as electrons and photons. An important prediction of QM is quantum entanglement, which has been studied in connection with particle physics at high energies only in recent years~\cite{Afik:2020topentangled, Fabbrichesi:2022topqubit, Fabbrichesi:2021bell, Dong:2023xiw}. The ATLAS and CMS experiments reported the observation of entanglement in the \ttbar system at the production threshold ~\cite{ATLAS:2023fsd, CMS-PAS-TOP-23-001} using events where both \PW bosons from the \ttbar pair decay into leptons. In this paper we include measurements of entanglement at the production threshold and at high \ttm for \ttbar events in the \lpj channel. In \cite{Abel:1992kz} it was argued that since collider experiments do not measure spins of the produced particles directly, but rather infer the spin directions statistically from the distribution of their decay products, a local hidden variable theory can always be found that explains the observation. Following \cite{Han:2023fci} we assume that the \ttbar system is described by QM and we characterize the state as separable or entangled.

This paper is organized in the following way. First, we outline the measurement strategy in Section~\ref{strategy}. We briefly describe the CMS detector in Section~\ref{cmsdetector} and then discuss the signal and background modeling used for this analysis in Section~\ref{samples}. The event selection and the \ttbar reconstruction are described in Sections\,\ref{objects} and \ref{reconstruction}, respectively, followed by a discussion of the background estimation in Section~\ref{background} and the extraction of the polarization and spin correlation coefficients in Section~\ref{signalfit}. Systematic uncertainties are detailed in Section~\ref{uncertainties}. Finally, the results are presented in Section~\ref{results} and the paper is summarized in Section~\ref{summary}. Tabulated results are provided in the HEPData record for this analysis~\cite{hepdata}.

\section{Measurement strategy}
\label{strategy}

We perform a measurement of the \ttbar polarization and spin correlation in the helicity basis $\{n,r,k\}$ following Ref.~\cite{Bernreuther:2015yna}. This basis is defined by boosting the top quarks and their decay products from the laboratory frame into the \ttbar rest frame. Afterward, based on the unit vectors in the top quark direction $\hat{k}$ and the beam in the positive $z$-direction $\hat{b}$, the axes of the new coordinate system are given by
\begin{equation}
    \hat{n}=\frac{\hat{b}\times\hat{k}}{\sin(\thetat)},
    \quad
    \hat{r}=\frac{\hat{b}-\costh\hat{k}}{\sin(\thetat)},
    \quad
    \hat{k},
\end{equation}
where \thetat is the scattering angle of the top quark, \ie, $\costh=\hat{b}\cdot\hat{k}$. The Bose--Einstein symmetry of the \gluglu initial state~\cite{Bernreuther:2015yna} requires a redefinition of the $n$ and $r$ axes (which are odd under Bose--Einstein symmetry) to allow nonzero values of the polarization and spin correlation coefficients involving an odd number of these axes. This is done by multiplying the $n$ and $r$ directions by the sign of \costh, which is also odd under Bose--Einstein symmetry, such that all axes are even under Bose--Einstein symmetry:
\begin{equation}
    \{n,r,k\}\to\big\{\mathrm{sgn}\big(\!\costh\big)n,\,\mathrm{sgn}\big(\!\costh\big)r,\,k\big\}.
\end{equation}
Finally, the top quark and antiquark are boosted individually into their rest frames together with their corresponding decay products.

In this basis, the unit vector
\ifthenelse{\boolean{cms@external}}{\begin{multline}
    \mathbf{\Omega}(\mathbf{\bar{\Omega}})=\big(\sin(\thetappb)\cos(\phippb), \\
    \sin(\thetappb)\sin(\phippb),\cos(\thetappb)\big)
\end{multline}}{\begin{equation}
    \mathbf{\Omega}(\mathbf{\bar{\Omega}})=\big(\sin(\thetappb)\cos(\phippb),\sin(\thetappb)\sin(\phippb),\cos(\thetappb)\big)
\end{equation}}
describes the direction of a decay product \porpbar of the top (anti)quark, where \phippb is the azimuthal and \thetappb the polar angle of the decay product. The differential cross section as a function of the four variables \phippb and \thetappb has the form
\ifthenelse{\boolean{cms@external}}{\begin{multline}\label{method_eq1}
    \Sigtot(\phippb,\thetappb)=\frac{\rd^4\sigma}{\rd\phip\,\rd\costhetap\,\rd\phipb\,\rd\costhetapb} \\
    =\signorm\big(1+\kappa\mathbf{P}\cdot\mathbf{\Omega}+\bar{\kappa}\mathbf{\bar{P}}\cdot\mathbf{\bar{\Omega}}-\kappa\bar{\kappa}\mathbf{\Omega}\cdot(C\mathbf{\bar{\Omega}})\big),
\end{multline}}{\begin{equation}\label{method_eq1}\begin{aligned}
    \Sigtot(\phippb,\thetappb)&=\frac{\rd^4\sigma}{\rd\phip\,\rd\costhetap\,\rd\phipb\,\rd\costhetapb} \\
    &=\signorm\big(1+\kappa\mathbf{P}\cdot\mathbf{\Omega}+\bar{\kappa}\mathbf{\bar{P}}\cdot\mathbf{\bar{\Omega}}-\kappa\bar{\kappa}\mathbf{\Omega}\cdot(C\mathbf{\bar{\Omega}})\big),
\end{aligned}\end{equation}}
with two 3-dimensional polarization vectors $\mathbf{P}$ and $\mathbf{\bar{P}}$, and one $3{\times}3$ spin correlation matrix $C$. This means that \Sigtot depends linearly on 15 coefficients collectively referred to as $\Qm =\big\{P_n,P_r,P_k,\bar{P}_n,\dots,\Cnn, C_{nr},\dots,\Ckk\big\} $.
In this analysis, we perform a measurement of all 15 coefficients, which we subsequently refer to as the full matrix measurement. There is one additional coefficient \signorm that describes the overall normalization. The spin analyzing power $\kappa$ represents how much information from the top quark spin is transferred to its decay products. We use the down-type quarks and the charged leptons in the \PW boson decays.
The magnitude of $\kappa$ for these decay products have the maximum value of unity at LO. Including QCD corrections, the magnitude of $\kappa$ for down-type quarks is reduced to 0.966~\cite{Brandenburg:2002xr}. However, we perform the measurements using the LO values and leave the application of different $\kappa$ values for reinterpretations. For simplicity, we flip the sign of $\kappa$ for \PAQt decays following the convention of Ref.~\cite{Baumgart:2012twist}, instead of inverting the axes of the coordinate system as in Refs.~\cite{Bernreuther:2015yna, CMS-PAS-TOP-23-001}.

The top quark, being a spin-1/2 particle, can be described as a two-state quantum system known as a qubit. The minimal example of an entangled state consists of two qubits, \eg, a \ttbar pair, where the entanglement can be characterized by their spin correlation. The Peres--Horodecki criterion~\cite{Peres:1996sep, Horodecki:1997sep} can be used to determine if a quantum state is separable. If the state is not separable, it is considered entangled.
In general, a quantum state is described by a density matrix $\rho$, in this case a spin density matrix whose coefficients are probed by Eq.~\eqref{method_eq1}. A quantum state is said to be separable if $\rho$ can be factorized into individual states belonging to separate subspaces, \ie, $\rho=\sum_nq_n\rho_n^a\otimes\rho_n^b$, where $\rho_n^a$, $\rho_n^b$ are density matrices describing the quantum states of the subsystems $a$ and $b$ and the $q_n$ are the corresponding probabilities. If $\rho$ is a separable physical state, the state $\rho^{T_2}=\sum_nq_n\rho_n^a\otimes(\rho_n^b)^T$ resulting from taking the transpose for only subsystem $b$ should also be a physical state. It was demonstrated~\cite{Peres:1996sep, Horodecki:1997sep} that a sufficient condition for entanglement is that $\rho^{T_2}$ has at least one negative eigenvalue, meaning it is not a physical state. This translates to a sufficient condition for entanglement based on the diagonal elements of the spin correlation matrix~\cite{Afik:2020topentangled,Afik:2022kwm}:
\begin{equation}
    \DE=\Cnn+\abs{\Crr+\Ckk}>1.
    \label{method_eq3}
\end{equation}
Based on the measured values of the spin correlation matrix we apply this criterion to evaluate the entanglement of the \ttbar system in different regions of phase space.

As an alternative to the full matrix measurement, we measure angular variables directly sensitive to \DE. The trace of the spin correlation matrix $\Tr(C)$ can be probed using the opening angle $\chi$ between the two decay products in the helicity basis, $\coschi=\mathbf{\Omega}\cdot\mathbf{\bar{\Omega}}$. This observable is sensitive to the entanglement in the spin-singlet state~\cite{Afik:2020topentangled} expected from \gluglu fusion events at the \ttbar production threshold. The distribution of $\chi$ is given by
\ifthenelse{\boolean{cms@external}}{\begin{multline}\label{method_eq4}
    \frac{\rd\sigma}{\rd\coschi}=\signorm\big(1+\Qone\kappa\bar{\kappa}\coschi\big), \\
    \text{ where }
    \Qone=-\frac{1}{3}\Tr(C).
\end{multline}}{\begin{equation}\label{method_eq4}
    \frac{\rd\sigma}{\rd\coschi}=\signorm\big(1+\Qone\kappa\bar{\kappa}\coschi\big),
    \text{ where }
    \Qone=-\frac{1}{3}\Tr(C).
\end{equation}}
For \gluglu fusion events at low \ttm, both \Crr and \Ckk are positive~\cite{Baumgart:2012twist}, which simplifies the entanglement criterion to
\begin{equation}
    \DE=-3\Qone=\Tr(C)>1.
\end{equation}
The entanglement in a spin-triplet state, predicted in both \qqbar annihilation and \gluglu fusion events with high \ttm and low \absct, can be probed using a criterion based on~\cite{Aguilar-Saavedra:2022uye, Baumgart:2012twist}
\begin{equation}
    \Dtilde=\frac{1}{3}(\Cnn-\Crr-\Ckk).
\end{equation}
The signs of \Crr and \Ckk become negative at transverse momentum of the top quark $\tpt\sim\Mtop$~\cite{Baumgart:2012twist}, so the entanglement criterion based on \Dtilde in the high-\ttm region is:
\begin{equation}
    \DE=3\Dtilde>1.
\end{equation}
The extraction of \Dtilde is performed using $\chitilde=-\Omega_n\bar{\Omega}_n+\Omega_r\bar{\Omega}_r+\Omega_k\bar{\Omega}_k$, analogous to $\chi$ but with an inverted sign of the $n$-component of one of the decay products.

\section{The CMS detector}
\label{cmsdetector}

The central feature of the CMS apparatus is a superconducting solenoid of 6\unit{m} internal diameter, providing a magnetic field of 3.8\unit{T}. Within the solenoid volume are a silicon pixel and strip tracker, a lead tungstate crystal electromagnetic calorimeter (ECAL), and a brass and scintillator hadron calorimeter (HCAL), each composed of a barrel and two endcap sections. Forward calorimeters extend the pseudorapidity ($\eta$) coverage provided by the barrel and endcap detectors. Muons are measured in gas-ionization detectors embedded in the steel flux-return yoke outside the solenoid.

Events of interest are selected using a two-tiered trigger system. The first level (L1), composed of custom hardware processors, uses information from the calorimeters and muon detectors to select events at a rate of around 100\unit{kHz} within a fixed latency of about 4\mus~\cite{CMS:2020cmk}. The second level, known as the high-level trigger, consists of a farm of processors running a version of the full event reconstruction software optimized for fast processing, and reduces the event rate to around 1\unit{kHz} before data storage~\cite{CMS:2016ngn}. For this measurement events are selected using single electron and muon triggers for isolated leptons with minimum \pt requirements, depending on the year, of 24 and 27\GeV for muons and 27 and 32\GeV for electrons.

The primary vertex is taken to be the vertex corresponding to the hardest scattering in the event, evaluated using tracking information alone, as described in Section 9.4.1 of Ref.~\cite{CMS-TDR-15-02}. The particle-flow (PF) algorithm~\cite{PF} aims to reconstruct and identify each individual particle in an event, with an optimized combination of information from various elements of the CMS detector. The energy of photons is obtained from the ECAL measurement. The energy of electrons is determined from a combination of the electron momentum at the primary interaction vertex as determined by the tracker, the energy of the corresponding ECAL cluster, and the energy sum of all bremsstrahlung photons spatially compatible with originating from the electron track. The energy of muons is obtained from the curvature of the corresponding track. The energy of charged hadrons is determined from a combination of their momentum measured in the tracker and the matching ECAL and HCAL energy deposits, corrected for the response function of the calorimeters to hadronic showers. Finally, the energy of neutral hadrons is obtained from the corresponding corrected ECAL and HCAL energies.
A more detailed description of the CMS detector, together with a definition of the coordinate system used and the relevant kinematic variables, can be found in Ref.~\cite{CMS:2008xjf}.

\section{Signal and background modeling}
\label{samples}

The matrix element (ME) event generator \POWHEG \textsc{v2}~\cite{Nason:2004rx,Frixione:2007vw,Frixione:2007nw} is used to simulate \ttbar events with next-to-LO (NLO) QCD accuracy. It is subsequently combined with the parton shower (PS) simulation from \PYTHIA8.230~\cite{Sjostrand:2014zea}, using the underlying event tune CP5~\cite{CP5}. In addition, a sample of \ttbar production at next-to-NLO (NNLO) QCD is generated with \POWHEG \MINNLO~\cite{Mazzitelli_2021} in combination with the CP5 \PYTHIA tune. This is used to estimate uncertainties in the contribution from higher-order QCD. Similarly, for the estimation of electroweak corrections, we use the \HATHOR \textsc{2.1-b3} package~\cite{hathorart}. The measured coefficients are compared to the predictions obtained using \PWGHRW7.1~\cite{Bahr:2008pv} with tune CH3~\cite{CMS:2020dqt} and \MGvATNLO2.6.1~\cite{Alwall:2014hca} combining ME calculations at NLO QCD including up to two additional partons with the \PYTHIA PS using the FXFX algorithm~\cite{Frederix:2012ps}. In all \ttbar simulations, the decays of the top quarks including the spin correlation are evaluated at LO precision.

The \ttbar simulations are normalized to the inclusive cross section value of $832^{+40}_{-46}\unit{pb}$~\cite{Czakon:2011xx} which is calculated with NNLO precision including soft-gluon resummation at the level of next-to-next-to-leading-logarithm. The renormalization scale \mur and factorization scale \muf are taken to be equal to the average transverse mass of the top quark and antiquark, $\mT=0.5(\sqrt{\smash[b]{\Mtop^2+\tpt^2}}+\sqrt{\smash[b]{\Mtop^2+\tbarpt^2}})$, where \pt is the transverse momentum of the top quark evaluated in the \ttbar rest frame, and $\Mtop=172.5\GeV$~\cite{CMS:2015lbj} is used in all simulations, unless stated otherwise. The main background processes in this analysis are single top quark production, Drell--Yan (DY) and \PW boson production in association with jets, and events composed uniquely of jets produced through the strong interaction, referred to as QCD multijet events. Single top quark production via $t$ channel and top quark production in association with a \PW boson are generated using \PWGPYT, while the production via $s$ channel is generated at NLO QCD using \MGHPYT. The simulation of background from DY+jets and \Wjets production is performed at LO QCD using \MGHPYT with the MLM PS matching~\cite{Mangano:2006rw,MLM} of up to four partons. The QCD multijet processes are simulated at LO using \PYTHIA. The cross sections are taken from NNLO calculations for \Wjets and DY+jets events~\cite{Li:2012wna} and NLO calculations for single top quark events~\cite{Kant:2014oha,Kidonakis:2012rm}. The default parameterization of the parton distribution functions (PDFs) used in all simulations is the NNLO version of NNPDF 3.1~\cite{NNPDF31}.

The detector response is modeled using \GEANTfour~\cite{Agostinelli:2002hh}. Additional proton-proton interactions within the same or nearby bunch crossings (pileup) are overlaid on each simulated event. Simulated events are assigned event weights based on the number of pileup interactions to match the pileup distribution in data. The same reconstruction algorithms that are applied to the data are used for simulated events.

\section{Physics object reconstruction}
\label{objects}

The measurements presented in this analysis depend on the reconstruction and identification of electrons, muons, jets, and the missing transverse momentum \ptvecmiss associated with neutrinos.

Electrons~\cite{CMSelectrons} and muons~\cite{Sirunyan:2018} are selected if they are isolated and compatible with originating from the primary vertex. Moreover, they must have $\pt>30\GeV$ and $\abs{\eta}<2.4$. In the 2018 data set, the minimum \pt of electrons was raised to 34\GeV because of the increased trigger thresholds. Leptons are required to satisfy several quality criteria including isolation and compatibility with the primary vertex. The electron and muon reconstruction and selection efficiencies are measured in the data using the ``tag-and-probe'' technique~\cite{TNP}. Depending on \pt and $\eta$, the overall reconstruction and selection efficiency is 50--80\% for electrons and 75--85\% for muons.

Jets are clustered from PF candidates using the anti-\kt jet algorithm with a distance parameter of 0.4 implemented in the \FASTJET package~\cite{Cacciari:2008gp, Cacciari:2011ma}. Charged PF candidates originating from a pileup interaction vertex are excluded. The total energy of the jets is corrected for energy depositions from pileup. In addition, \pt- and $\eta$-dependent corrections are applied to account for detector response effects~\cite{JET}. If an isolated lepton with $\pt>15\GeV$ within $\DR=\sqrt{\smash[b]{(\Delta\phi)^2+(\Delta\eta)^2}} = 0.4$ around a jet exists, the jet is assumed to represent the isolated lepton and is discarded to prevent counting the lepton momentum twice. The jets are considered for analysis if they fulfill the kinematic requirements $\pt>30\GeV$ and $\abs{\eta}<2.4$.

For the identification of \PQb jets, the {DeepJet} algorithm~\cite{Bols_2020, BTV-16-002} is used. It is based on an artificial neural network (NN) that provides a discriminant to distinguish between \PQb and other flavored jets. Jets are categorized based on three thresholds of the discriminant and a jet belongs to the category with the highest threshold that is smaller than that jet's discriminant value. The tight, medium, and loose thresholds have, depending on the jet \pt and $\eta$, efficiencies of about 50--70, 70--82, and 85--92\%, respectively, and rejection probabilities of about 97, 85, and 55\% for \PQc jets and about 99.5, 98, and 90\% for non-heavy-flavor jets.

The missing transverse momentum vector \ptvecmiss is computed as the negative vector sum of the transverse momenta of all the PF candidates in an event, and its magnitude is denoted as \ptmiss~\cite{CMS:2019ctu}. The \ptvecmiss is modified to account for corrections to the energy scale of the reconstructed jets in the event.

The data was recorded in the years 2016--2018. For each year individual sets of simulations and correction factors are used according to the actual data-taking conditions and detector configurations. Because of significant changes in the detector configuration affecting the tracking efficiency, two separate sets of simulations and scale factors are used for 2016 data. Therefore, four different data-taking periods are analyzed.

\section{Event reconstruction}
\label{reconstruction}

The reconstruction of the \ttbar system is performed using an artificial NN. The goal is the correct identification of detector-level objects as decay products of the leptonically (\tql) and hadronically (\tqh) decaying top quarks in \lpj events. In the simulation, a quark or lepton at the generator level can be spatially matched to the corresponding detector-level object. Of the possible candidates, we select the highest \pt object within $\DR=0.2$. If a one-to-one assignment to a corresponding detector-level object is possible for all of the particles in the generator-level \ttbar system, the event is labeled ``reconstructable'', while all other \lpj events are called ``nonreconstructable''.

The input layer of the NN is a vector encoding kinematical information about the detector-level objects and \PQb tagging category for jets. The first four elements of the vector hold the four-momentum of the electron or muon ($p_x(\Pell)$, $p_y(\Pell)$, $p_z(\Pell)$, $E(\Pell)$), followed by ($p^{\mathrm{miss}}_x$, $p^{\mathrm{miss}}_y$). Finally, for up to eight jets the four-momentum and their \PQb tagging category ($p_x$, $p_y$, $p_z$, $E$, \mbox{\PQb cat.}) are stored. The NN is trained to assume the following order of the jets: the \PQb jet in the decay of \tql, the \PQb jet in the decay of \tqh, the down-type, and the up-type quark in the \PW boson decay. The remaining jets are added in descending order of their \pt. If there are fewer than eight jets, the rest of the input vector is filled with zeros. The \PQb tagging information also helps to identify \PQc jets from \PW boson decays, since 40--50\% of the \PQc jets are loosely \PQb tagged. The input layer is followed by seven fully connected layers, each with 220 nodes and the hyperbolic tangent as the activation function. The output layer consists of a single node whose value is transformed by a sigmoid function into the range $[0,1]$. In total, the network has about 300\,000 parameters.

The NN is trained using about 20M simulated \lpj events. Events with one selected electron or muon, no additional isolated electron or muon with $\pt>15\GeV$, and at least four jets are used. The NN is trained using a batch size of 128 events and the \textsc{adam} algorithm~\cite{adam} for the minimization of the logistic loss function. For each event in a batch, the network is provided with all possible permutations for the four jets from the \ttbar decay using up to eight jets per event, \ie, for 4 (5, 6, 7, 8) jets the training includes 24 (120, 360, 840, 1680) input vectors. Correct permutations are trained to have a response of one, while all other permutations should result in zero. The training sample includes ``nonreconstructable'' events, \ie, those with no correct permutation. During the training, the logistic loss function is monitored with a validation sample. The losses for the training and validation samples are compatible and no indication of overtraining is observed. During the inference, for each event, all possible permutations of assigning detector-level jets to the corresponding \ttbar decay products are successively provided as input to the NN and the permutation resulting in the highest NN score \SNN is used.

For the selected permutation, the neutrino four-momentum $p_{\PGn}$ is calculated using the \PW boson mass constraint $(p_{\PGn}+p_{\Pell})^2=m_{\PW}^2$, where \ptvecmiss is taken as the \pt of the neutrino. 
This constraint results in a quadratic equation for the longitudinal component of the neutrino momentum $p_z(\PGn)$. In the 39.1\% of events where this equation yields no real solution, the $x$ and $y$ components of \ptvecmiss are scaled separately to find a single solution under the condition of a minimum modification of \ptmiss, \ie, in the transverse plane we choose the point with the smallest distance from \ptvecmiss for which a solution exists. This modified \ptvecmiss together with the calculated solution for $p_z(\PGn)$ form the neutrino momentum. If there are two solutions of the quadratic equation, the invariant mass $m(\tql)^2=(p_{\PGn}+p_{\Pell}+p_{\Jbl})^2$ is calculated for both $p_{\PGn}$ and the solution with $m(\tql)$ closer to \Mtop is selected. This procedure identifies the correct solution in 69.9\% of the events. To enhance the fraction of correctly reconstructed events and reduce the background contribution, event selection requirements on the reconstructed particle masses $\abs{m(\tql)-172.5\GeV}<50\GeV$, $\abs{m(\tqh)-172.5\GeV}<50\GeV$, and $\abs{m(\Wh)-80.4\GeV}<30\GeV$ are imposed.

\begin{figure}[ht!]
\centering
\includegraphics[width=0.48\textwidth]{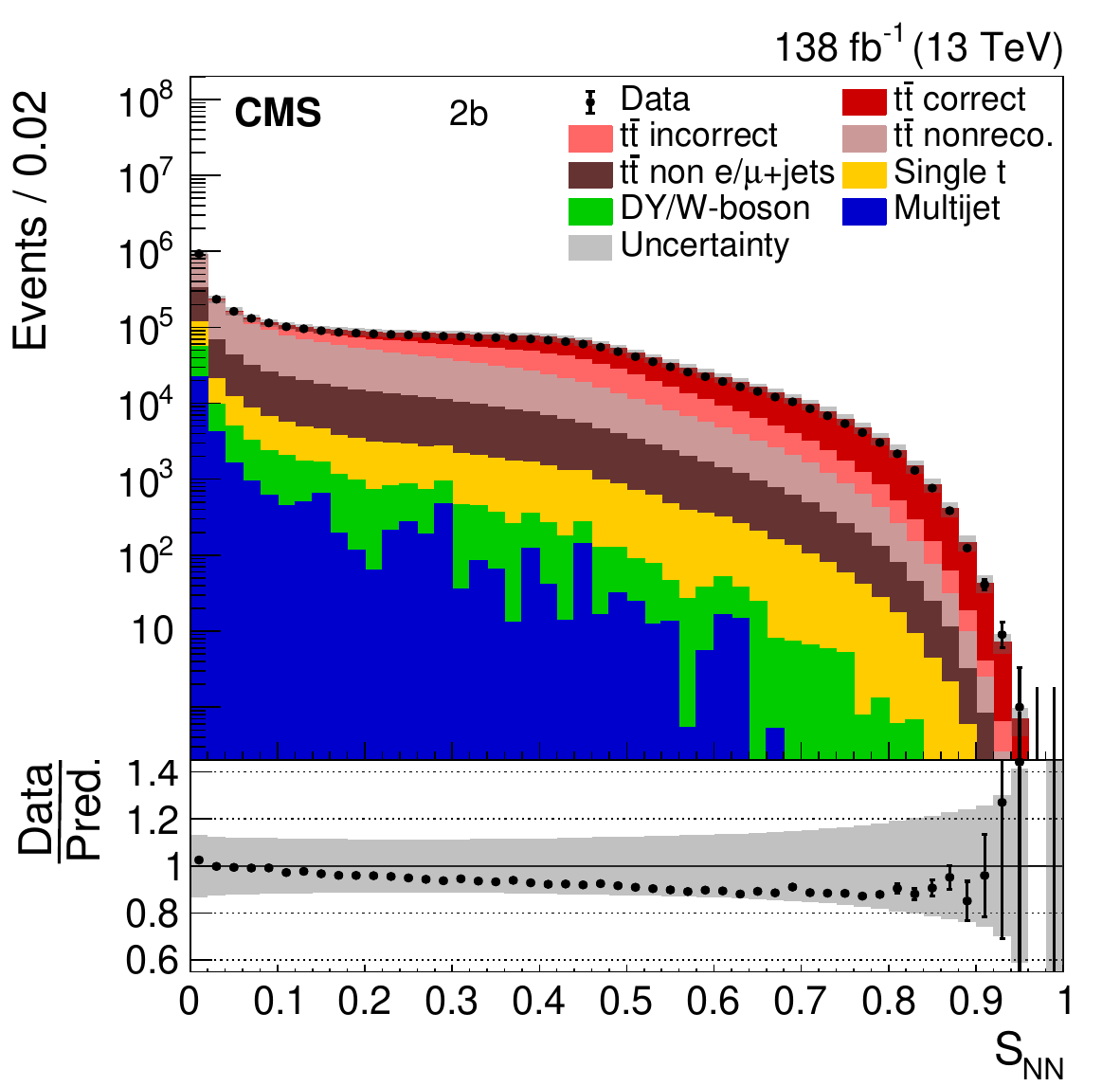}%
\hfill%
\includegraphics[width=0.48\textwidth]{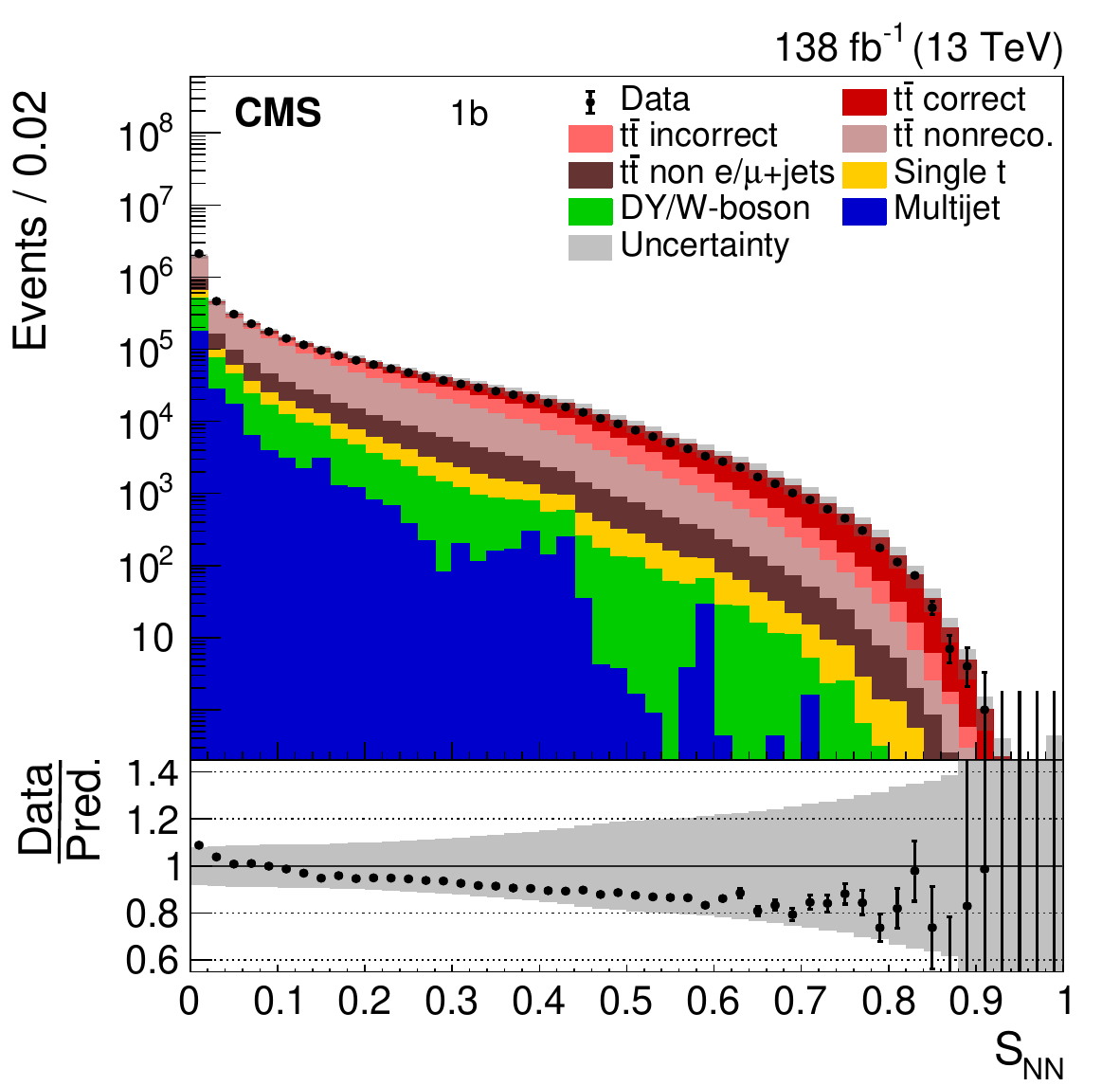}
\caption{Distribution of \SNN in the \twob (\cmsLeft) and \oneb (\cmsRight) categories. The data (points) are compared to the prediction (stacked histograms). The \ttbar contribution is split into the correctly and incorrectly reconstructed, ``nonreconstructable'', and non \lpj events. The gray uncertainty band indicates the combined statistical and systematic uncertainties in the prediction, while the vertical bars on the points show the statistical uncertainty of the data. The ratios of data to the predicted yields are provided in the lower panels.}
\label{reconstruction_fig_1}
\end{figure}

The distributions of \SNN are shown in Fig.~\ref{reconstruction_fig_1} for events in categories where either both (\twob) or exactly one (\oneb) of the jets identified as \PQb jets from the \ttbar decay are medium \PQb tagged. In these distributions, it can be seen that the data and the prediction are in agreement within the uncertainty bands. We reject all events with $\SNN<0.1$ due to the low fraction of correctly reconstructed events and the large contribution of background processes.

The \twob and \oneb categories are further split based on the value of \SNN. In the \oneb (\twob) category events belong to the \Shigh category if $\SNN>0.30$ (0.36), while the remaining events are placed in the \Slow category. These requirements define the signal categories for the analysis and were systematically optimized to minimize the uncertainties in the expected spin polarization and correlation coefficients.

\begin{figure}[ht!]
\centering
\includegraphics[width=0.48\textwidth]{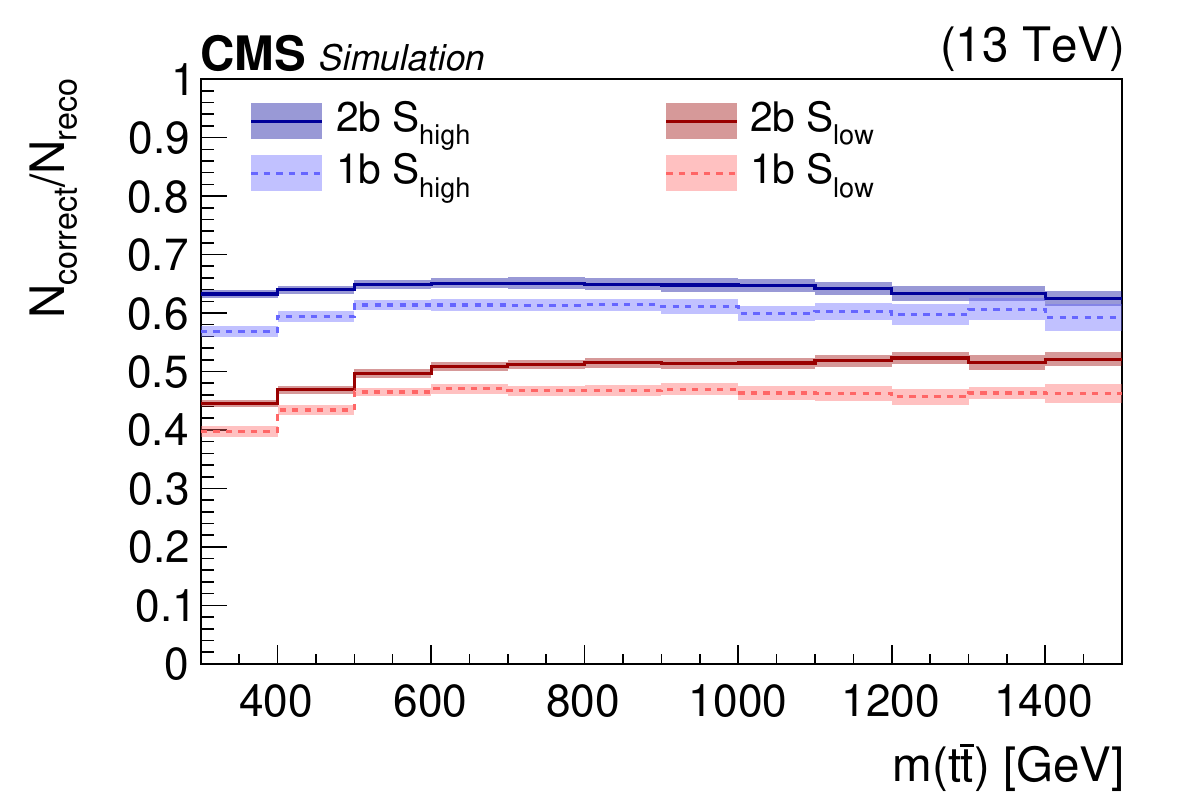}%
\hfill%
\includegraphics[width=0.48\textwidth]{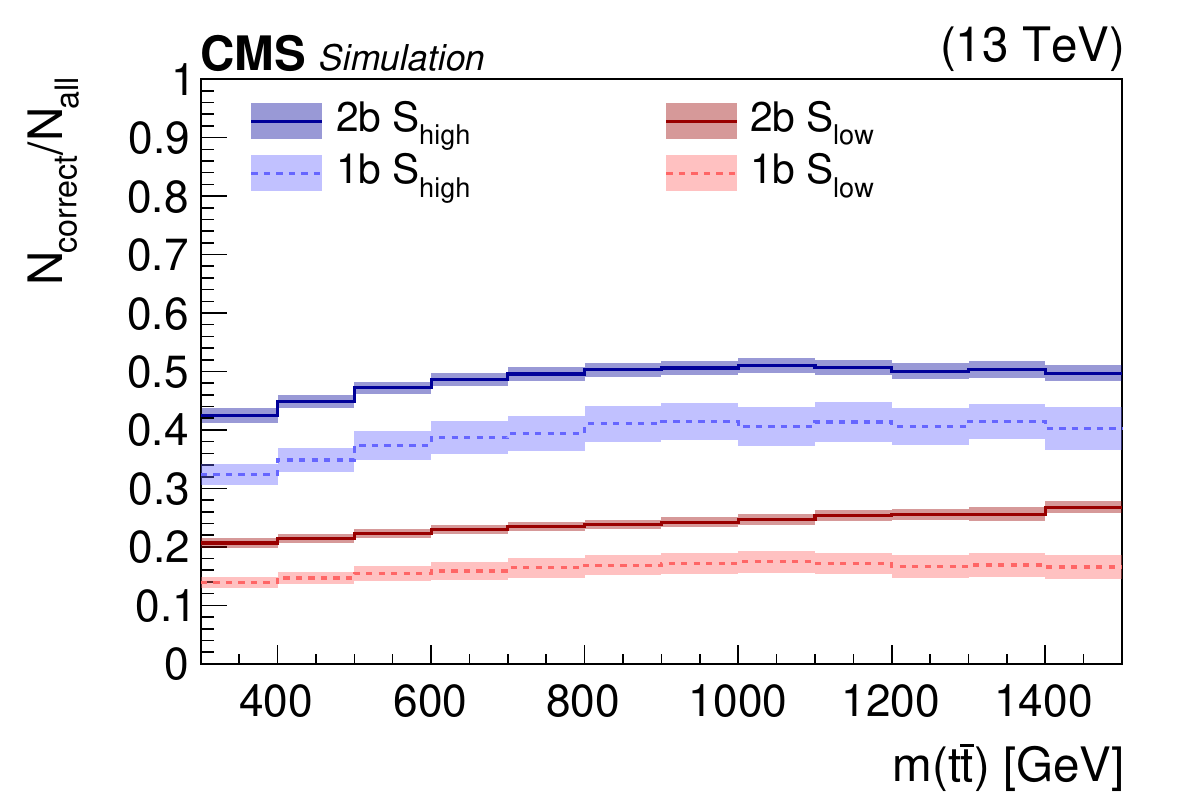}
\caption{Reconstruction efficiency of the NN (\cmsLeft) and the fraction of correctly reconstructed events (\cmsRight) as a function of \ttm estimated from the simulation. The values are shown separately for the \oneb and \twob categories with the \Slow and \Shigh selections. The event counts $N_{\text{correct}}$ and $N_{\text{reco}}$ refer to the number of correctly reconstructed and ``reconstructable'' \ttbar events, respectively. All reconstructed events regardless of the process are labeled $N_{\text{all}}$. The uncertainty bands include all systematic uncertainties as detailed in Section~\ref{uncertainties}.}
\label{reconstruction_fig_2}
\end{figure}

In the simulation, the fraction of reconstructable \lpj events is 73\% for \twobhigh, 47\% for \twoblow, 64\% for \onebhigh, and 38\% for \oneblow. The fractions of correctly reconstructed events with respect to all signal and background events in the various categories are 46\% for \twobhigh, 21\% for \twoblow, 37\% for \onebhigh, and 15\% for \oneblow. Figure~\ref{reconstruction_fig_2} shows these fractions as functions of \ttm together with the fraction of correctly reconstructed events with respect to all reconstructable events.

\section{Background estimation}
\label{background}

The main background contributions of non-\ttbar events are expected from QCD multijet, DY, \PW boson, and single top quark production.

\begin{figure}[ht!]
\centering
\includegraphics[width=0.46\textwidth]{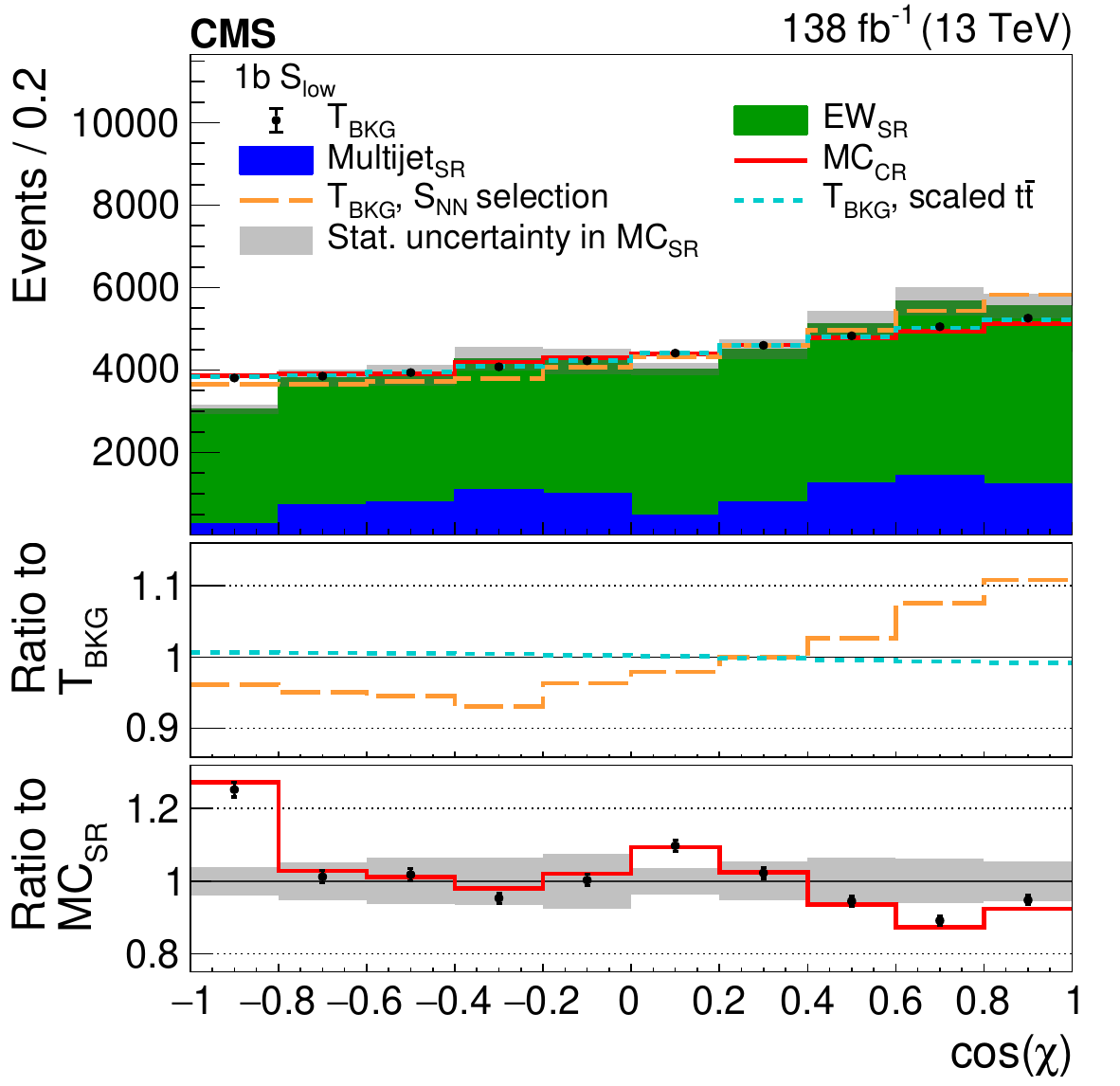}%
\hfill%
\includegraphics[width=0.46\textwidth]{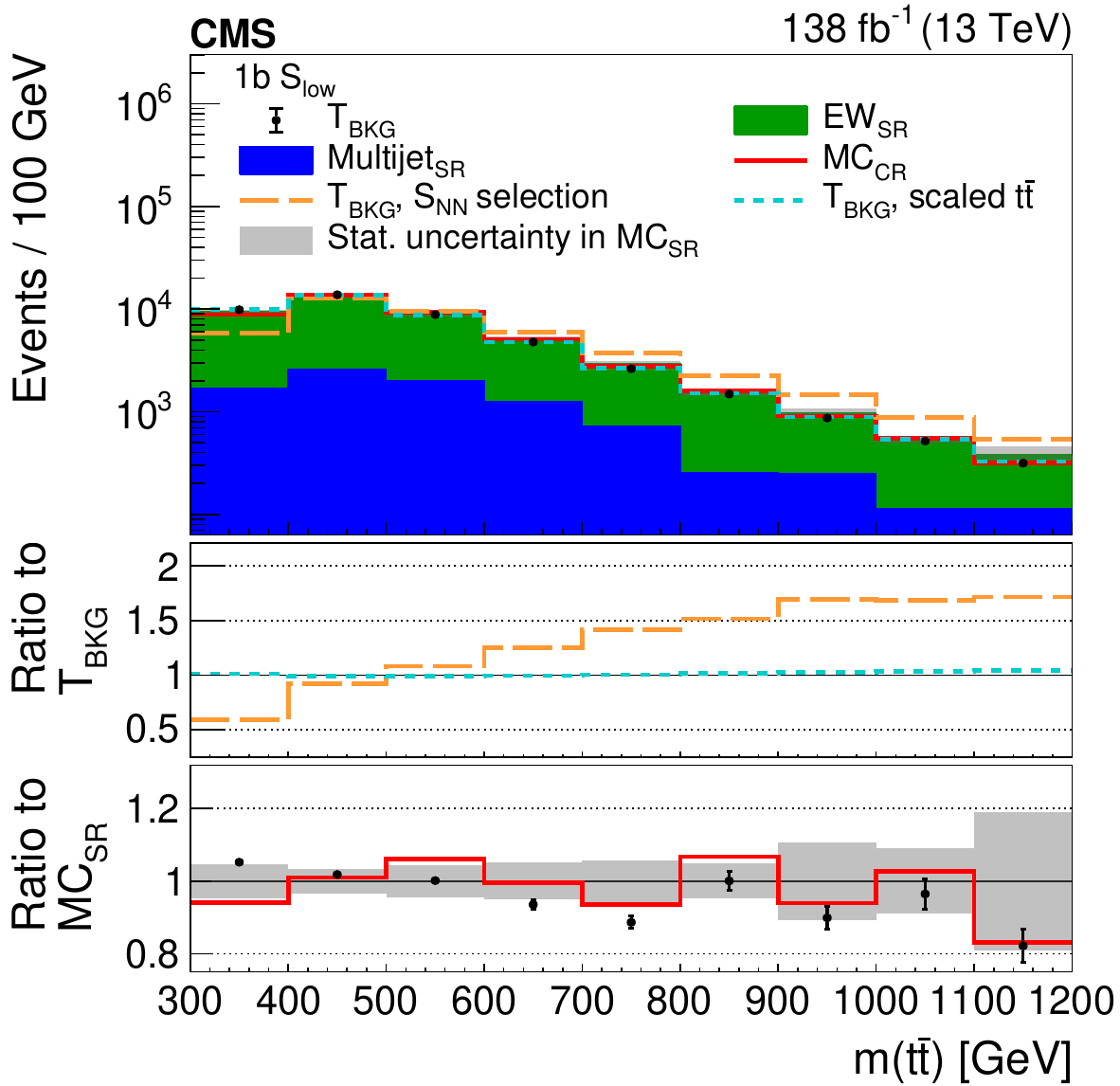}
\caption{Comparison of the \coschi (\cmsLeft) and \ttm (\cmsRight) distributions of the simulated background in the control region ($\text{MC}_{\text{CR}}$) shown as the red line, and in the \oneblow signal region ($\text{MC}_{\text{SR}}$) shown as the stacked histograms of the multijet and EW components. The estimated background template ($\text{T}_{\text{BKG}}$) shown as black markers corresponds to the data distribution in the CR after subtracting the predicted \ttbar and single top quark contributions. Variations of the $\text{T}_{\text{BKG}}$ are obtained applying the additional \SNN selection for the \oneblow category (orange line) and by taking into account the mismatch of the normalization in the CR when subtracting the \ttbar and single top quark contributions (blue line). All distributions are normalized to the event yields predicted by the $\text{MC}_{\text{SR}}$. The gray uncertainty band shows the statistical uncertainties in the $\text{MC}_{\text{SR}}$. The middle panels show the relative effects of the $\text{T}_{\text{BKG}}$ variations. The lower panels show the ratio of the $\text{MC}_{\text{CR}}$ and the $\text{T}_{\text{BKG}}$ to the $\text{MC}_{\text{SR}}$.}
\label{background_fig_1}
\end{figure}

The shapes of the QCD multijet (multijet), and DY and \PW boson (EW) background distributions are estimated using a combined template of these backgrounds that is obtained from a \PQb-jet depleted control region (CR). The simulation of these backgrounds suffers from large statistical uncertainties due to their high cross sections but low fraction of events that pass the selection. They contribute fractions of about 6.6\% (\oneblow), 2.4\% (\onebhigh), 1.1\% (\twoblow), and 0.2\% (\twobhigh), \ie, the contribution in the \twob categories is negligible. The shape of the background distribution is estimated by performing the NN reconstruction for events without any jet fulfilling the medium \PQb tagging requirement. No selections on \SNN are applied, but the mass ranges of \tql, \tqh, and the \Wh boson are required as introduced in Section~\ref{reconstruction}. These selection requirements define the CR. The expected contributions of \ttbar and single top quark events are subtracted from the data in the CR. The simulated kinematic distributions obtained in the CR are generally in agreement with the simulated distributions in the signal categories, as shown for the \oneb signal category in Fig.~\ref{background_fig_1}.

As systematic uncertainties in the background template shapes, we evaluate shape differences between the CR and the signal categories. The definition of the CR is inclusive in \SNN. This choice has the advantage of maximizing the number of events in the CR while minimizing the contribution from \ttbar events. We obtain alternative shapes from additional control regions, where the \Slow or \Shigh requirement of the corresponding signal category is also imposed. While it is expected that these distributions are more similar to the real background, they suffer from a small sample of events and large \ttbar contributions. Therefore, they are only used to evaluate the uncertainty in the background templates.

In the CR, there is an excess of about 20\% in data. This excess is within the uncertainty in the simulated event yield. However, to take into account a possible underestimation of \ttbar production in the CR, an additional systematic uncertainty in the shape is obtained by scaling the amount of subtracted \ttbar and single top quark events by the ratio of the total observed and simulated event yields in the CR, shown as the dashed light blue lines in Fig.~\ref{background_fig_1}.

The predicted ratio of the multijet and EW event yield (multijet+EW) in each signal category to the corresponding CR is applied to normalize the background templates. This normalization factor has a large statistical uncertainty due to the limited number of simulated events in the signal categories. In addition, the observed differences between the predicted and observed event yields in the CR are considered as a systematic uncertainty in the normalization of the background. As a result, the normalization uncertainties can be as large as 50\%, depending on the category.

The obtained background predictions with their shape and normalization uncertainties are included in the fits of spin polarization and correlation coefficients, as described in Section~\ref{signalfit}. The normalization uncertainties are treated as uncorrelated among all categories, because of their large statistical component from the simulation. The shape uncertainties are considered as uncorrelated among the categories to account for differences in their selection. In addition, the uncertainties are assumed uncorrelated among the data-taking periods, because of the differences in the \PQb tagging performances and selections. It has been verified that the results of the analysis are not sensitive to these assumptions.

The contribution of single top quark production is about 4.0\% (\oneblow), 2.2\% (\onebhigh), 2.4\% (\twoblow), and 1.4\% (\twobhigh). Templates according to their SM expectation are taken from the simulation. We evaluate the relevant uncertainties in these templates (as described in Section~\ref{uncertainties}): ME and PS scales, jet resolution and energy scales, and \PQb tagging and lepton efficiencies. The ME scale uncertainties are treated independently from the corresponding variations of the \ttbar simulation.

\begin{figure*}[p!]
\centering
\includegraphics[width=0.48\textwidth]{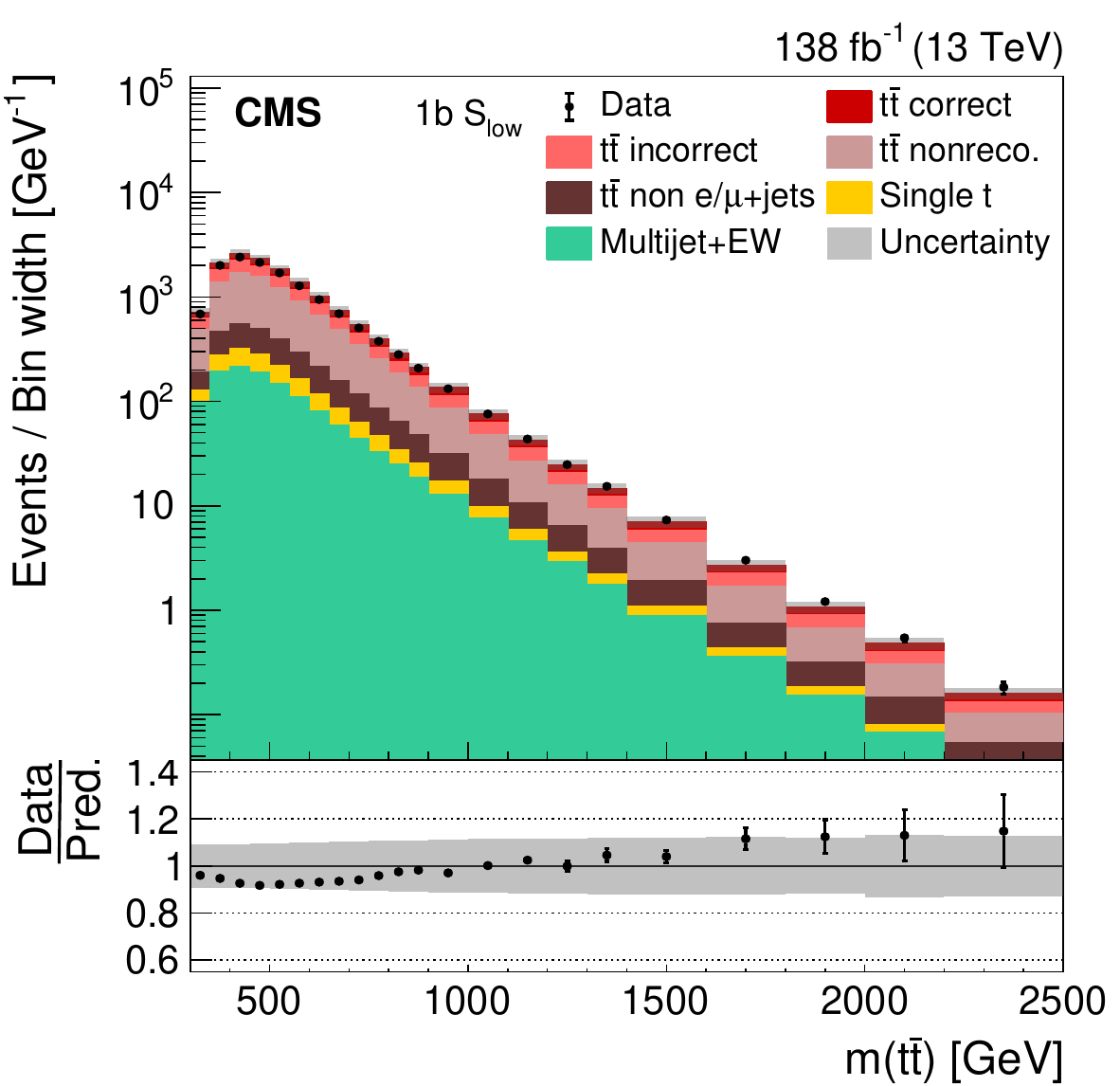}%
\hfill%
\includegraphics[width=0.48\textwidth]{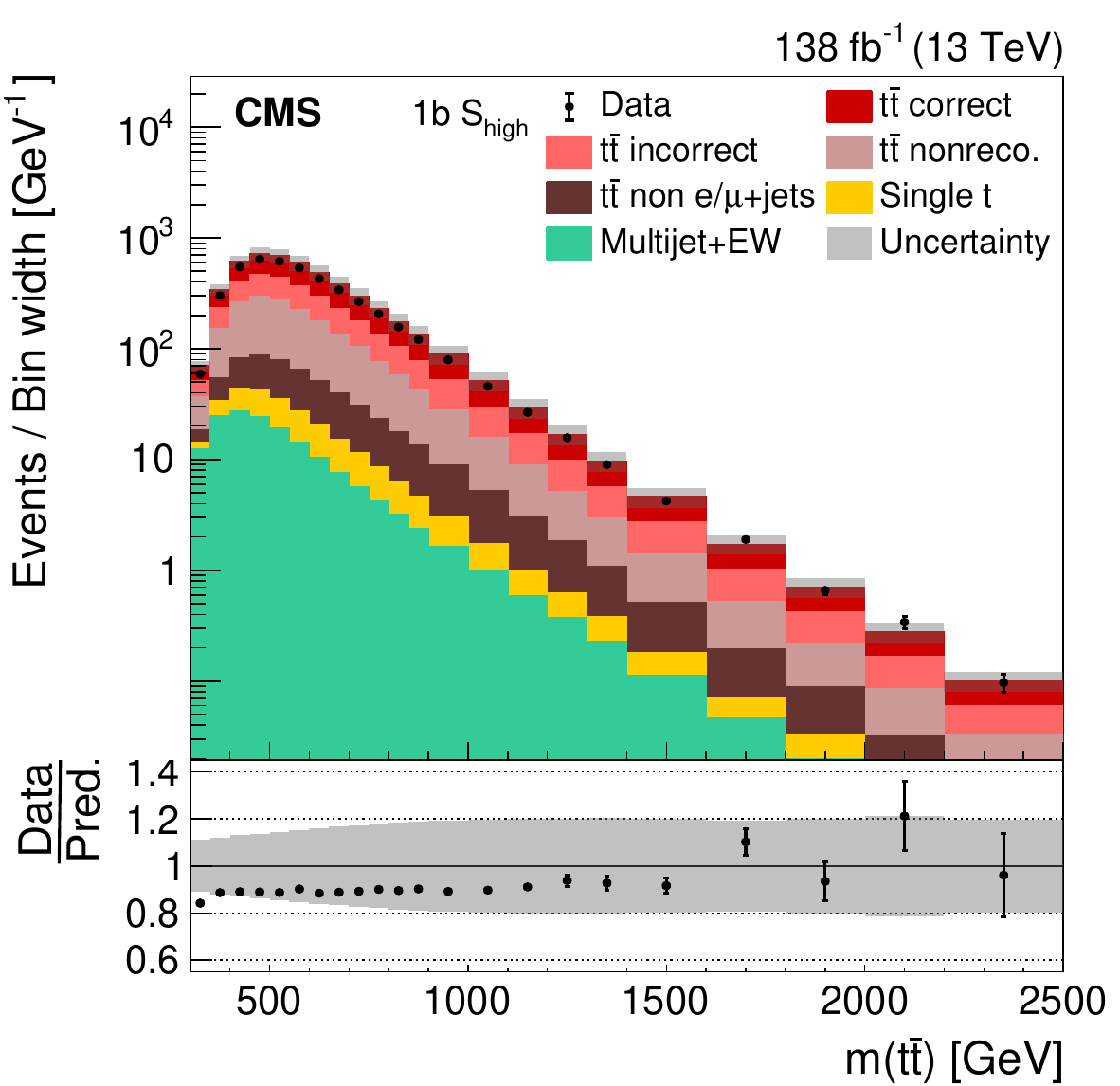} \\
\includegraphics[width=0.48\textwidth]{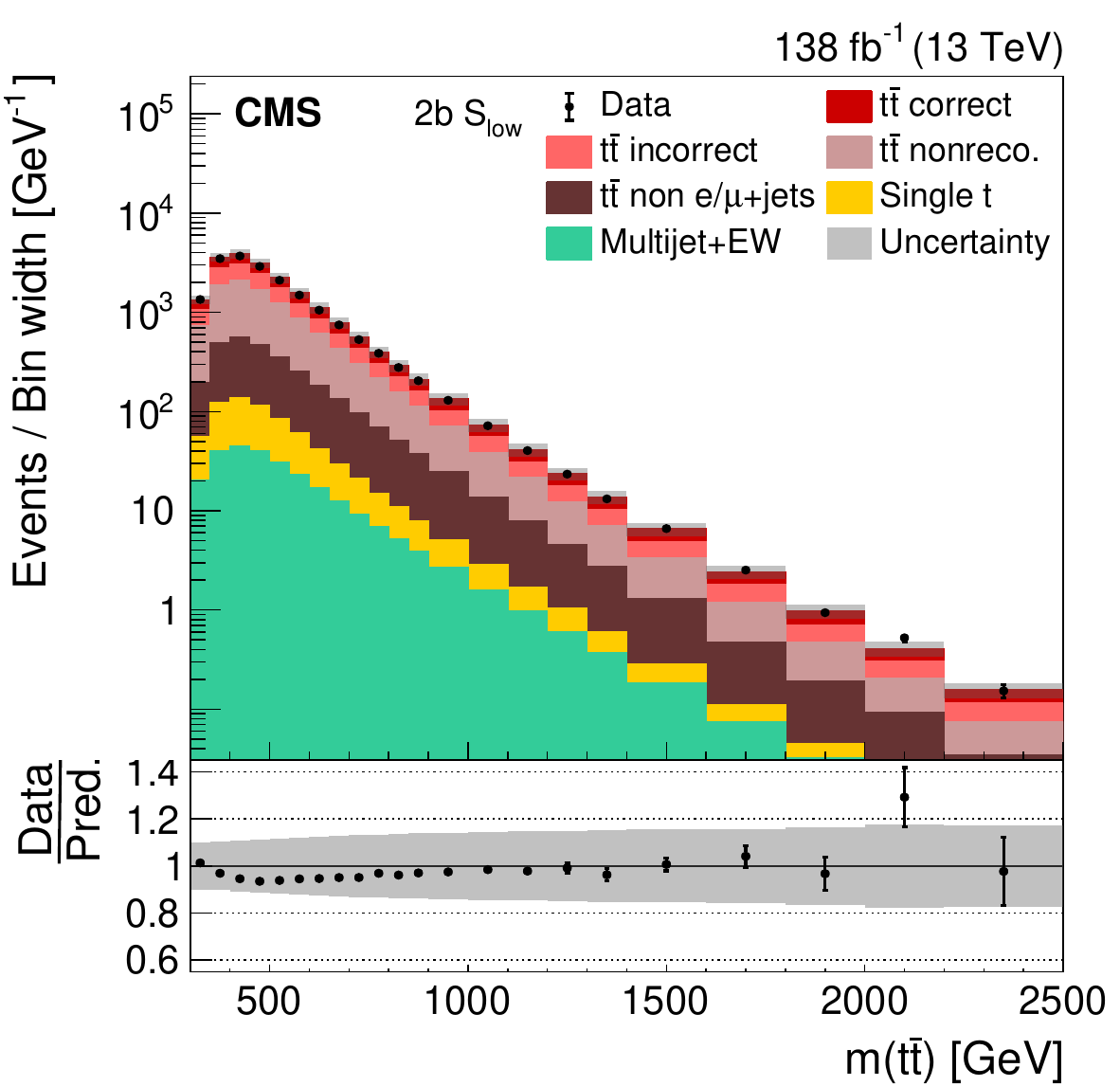}%
\hfill%
\includegraphics[width=0.48\textwidth]{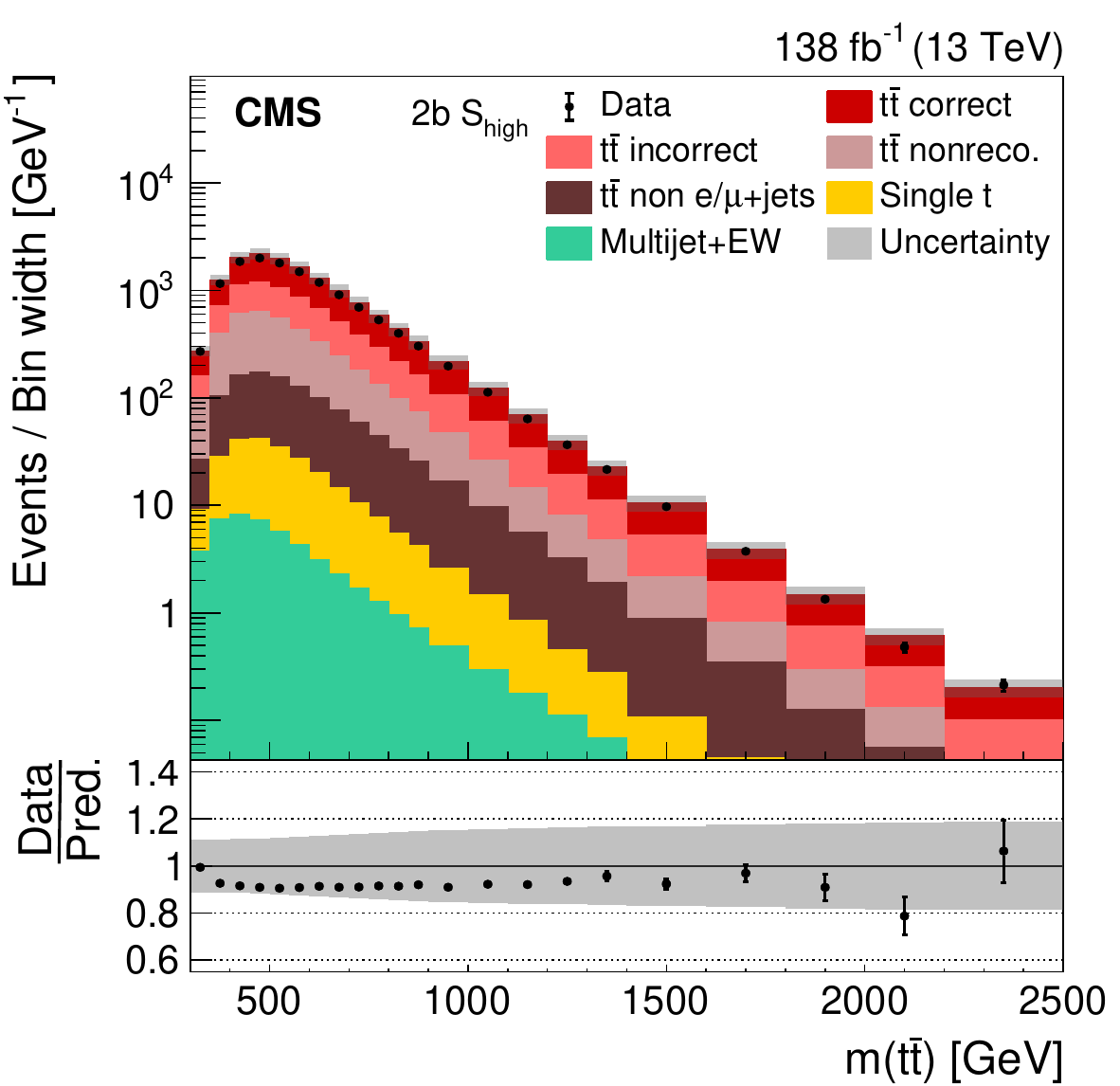}
\caption{Distribution of \ttm in all four categories. The data (points) are compared to the prediction (stacked histograms). The \ttbar and single top quark contributions are taken from the simulation, while the multijet+EW background is obtained from the CR. The \ttbar contribution is split into the correctly and incorrectly reconstructed, ``nonreconstructable'', and non \lpj events. The gray uncertainty band indicates the combined statistical and systematic uncertainties in the prediction. The vertical bars on the points show the statistical uncertainty of the data. Ratios to the predicted yields are provided in the lower panels.}
\label{background_fig_2}
\end{figure*}

\begin{figure*}[tp!]
\centering
\includegraphics[width=0.48\textwidth]{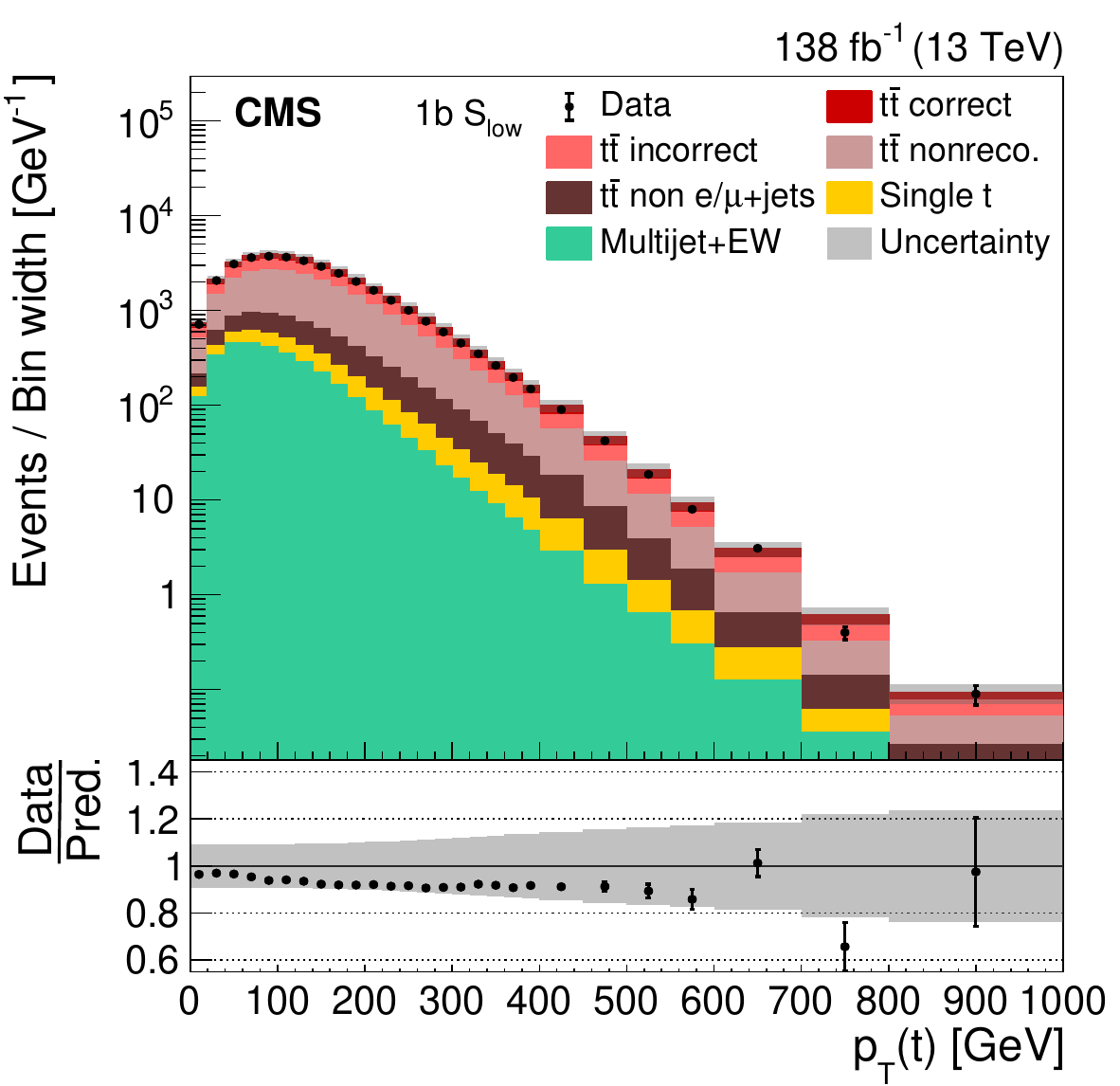}%
\hfill%
\includegraphics[width=0.48\textwidth]{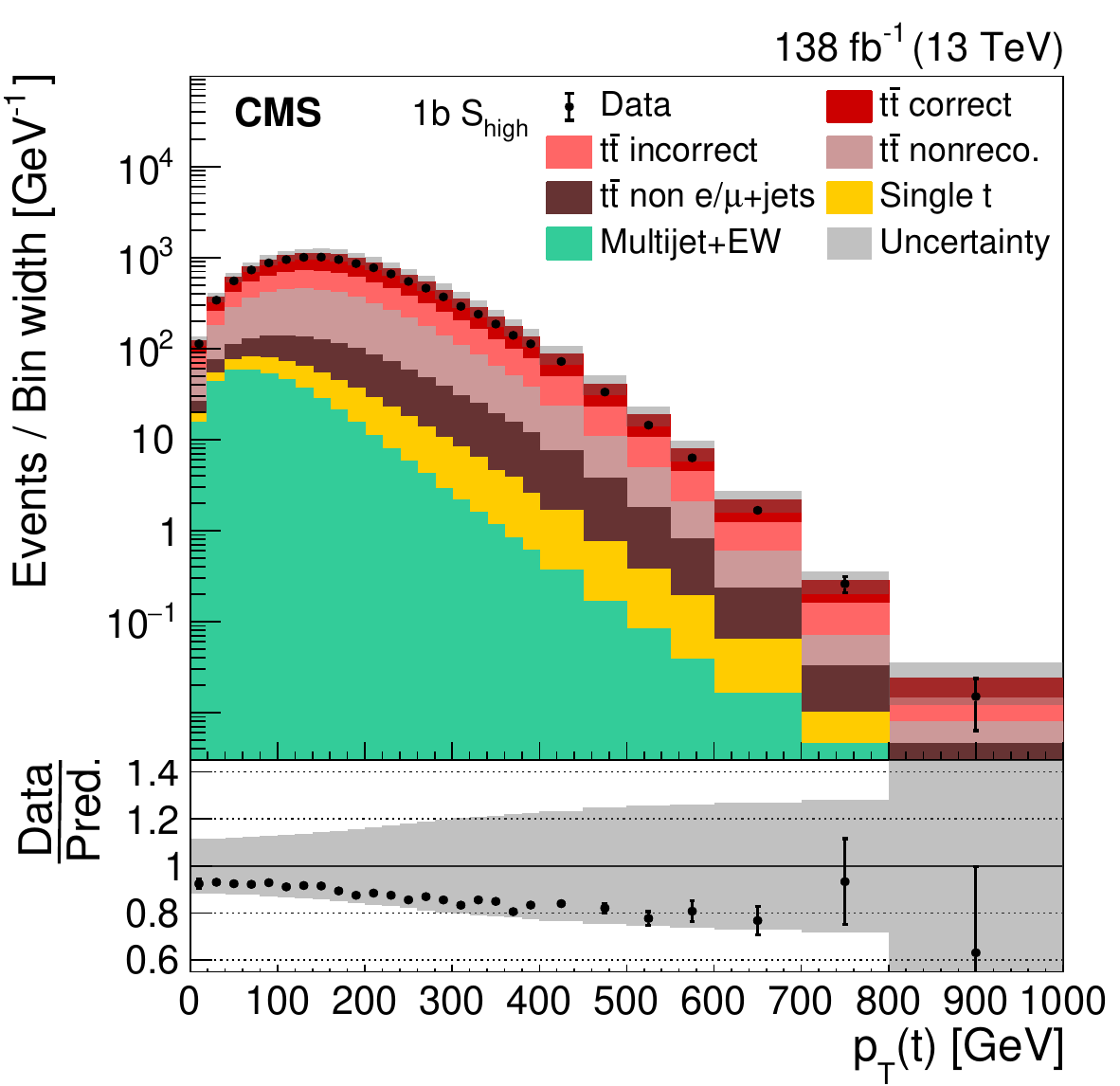} \\
\includegraphics[width=0.48\textwidth]{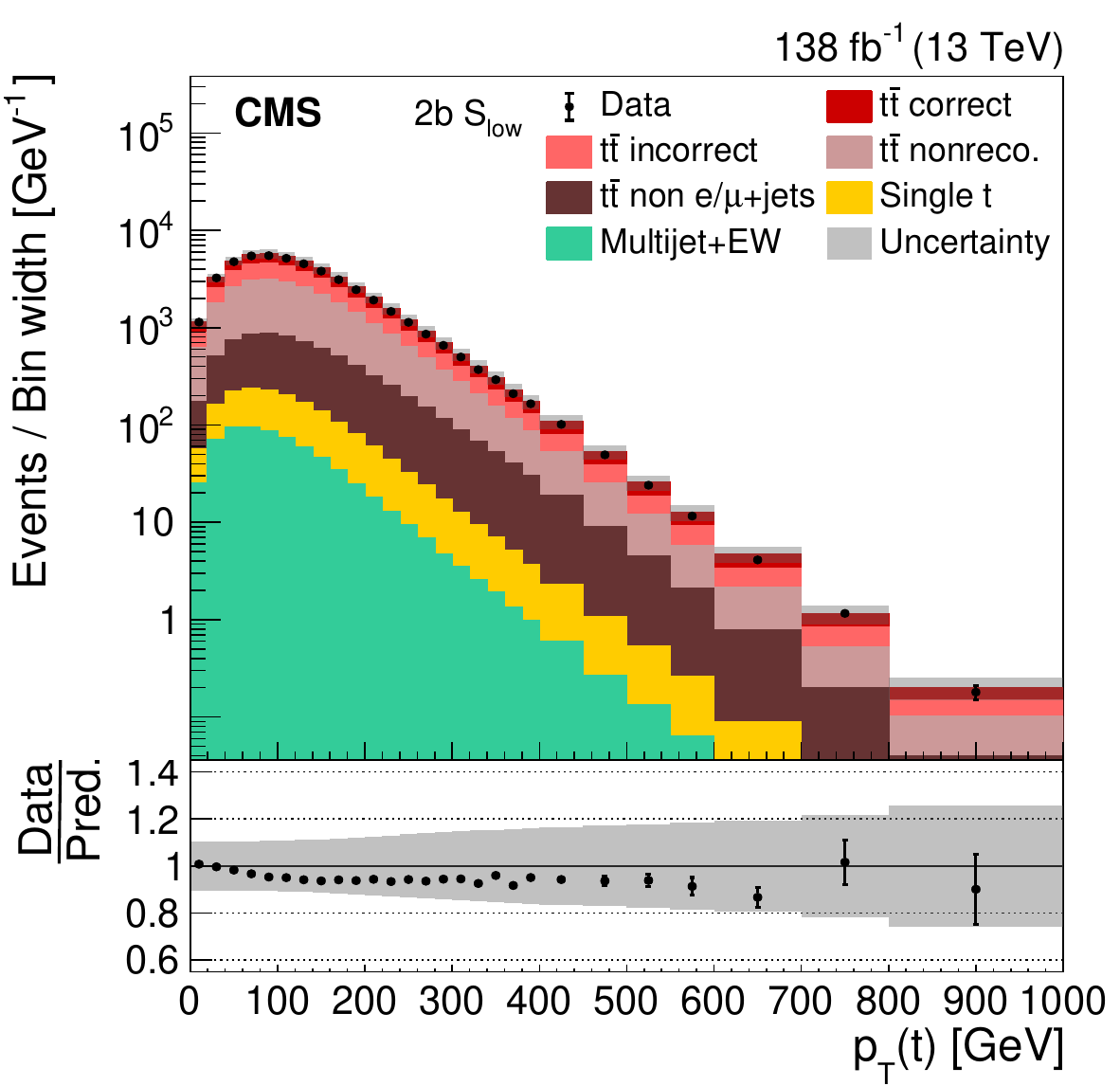}%
\hfill%
\includegraphics[width=0.48\textwidth]{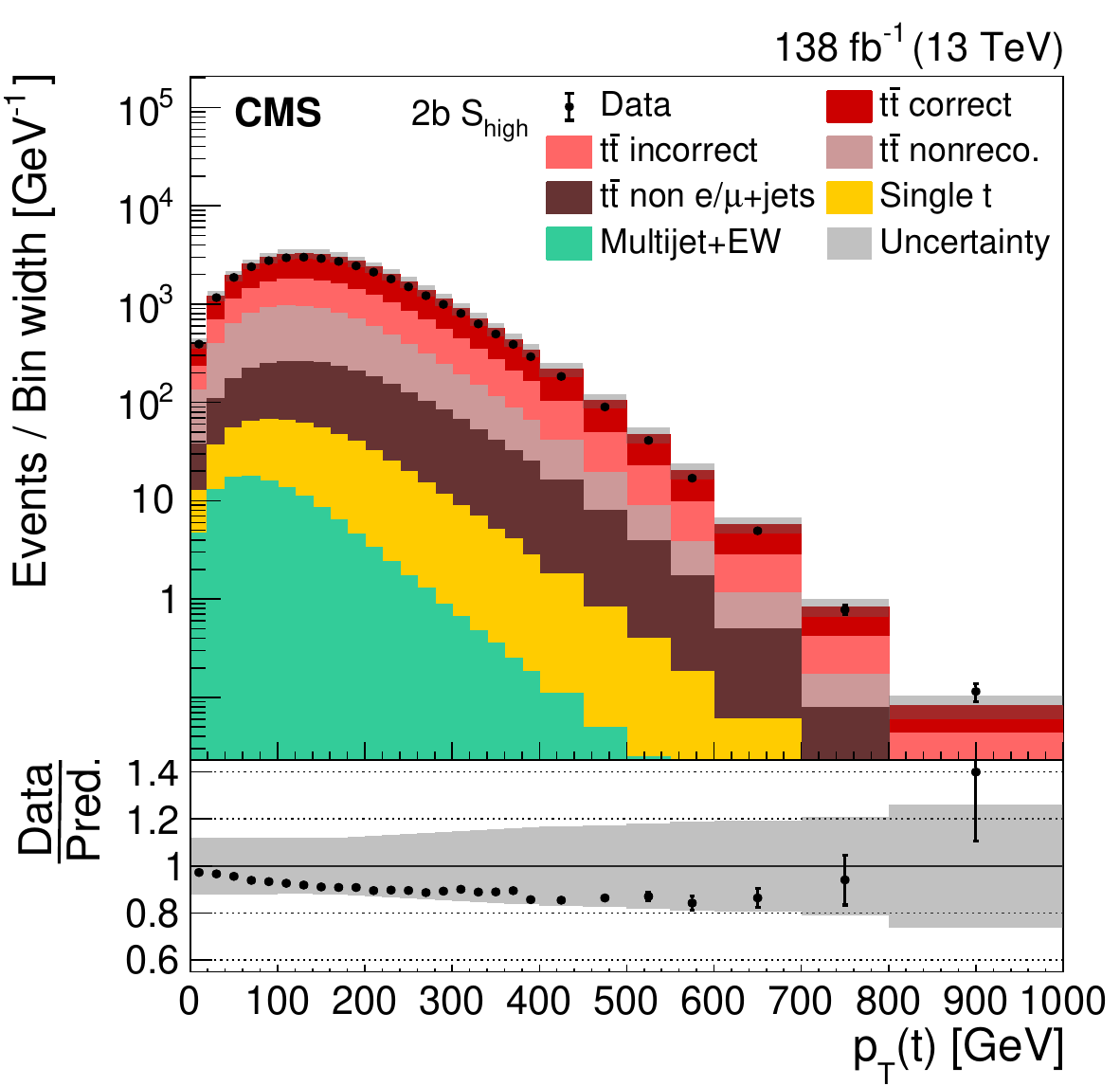}
\caption{Distribution of \tpt in all four categories. The data (points) are compared to the prediction (stacked histograms). The \ttbar and single top quark contributions are taken from the simulation, while the multijet+EW background is obtained from the CR. The \ttbar contribution is split into the correctly and incorrectly reconstructed, ``nonreconstructable'', and non \lpj events. The gray uncertainty band indicates the combined statistical and systematic uncertainties in the prediction. The vertical bars on the points show the statistical uncertainty of the data. Ratios to the predicted yields are provided in the lower panels.}
\label{background_fig_2b}
\end{figure*}

\begin{figure*}[p!]
\centering
\includegraphics[width=0.48\textwidth]{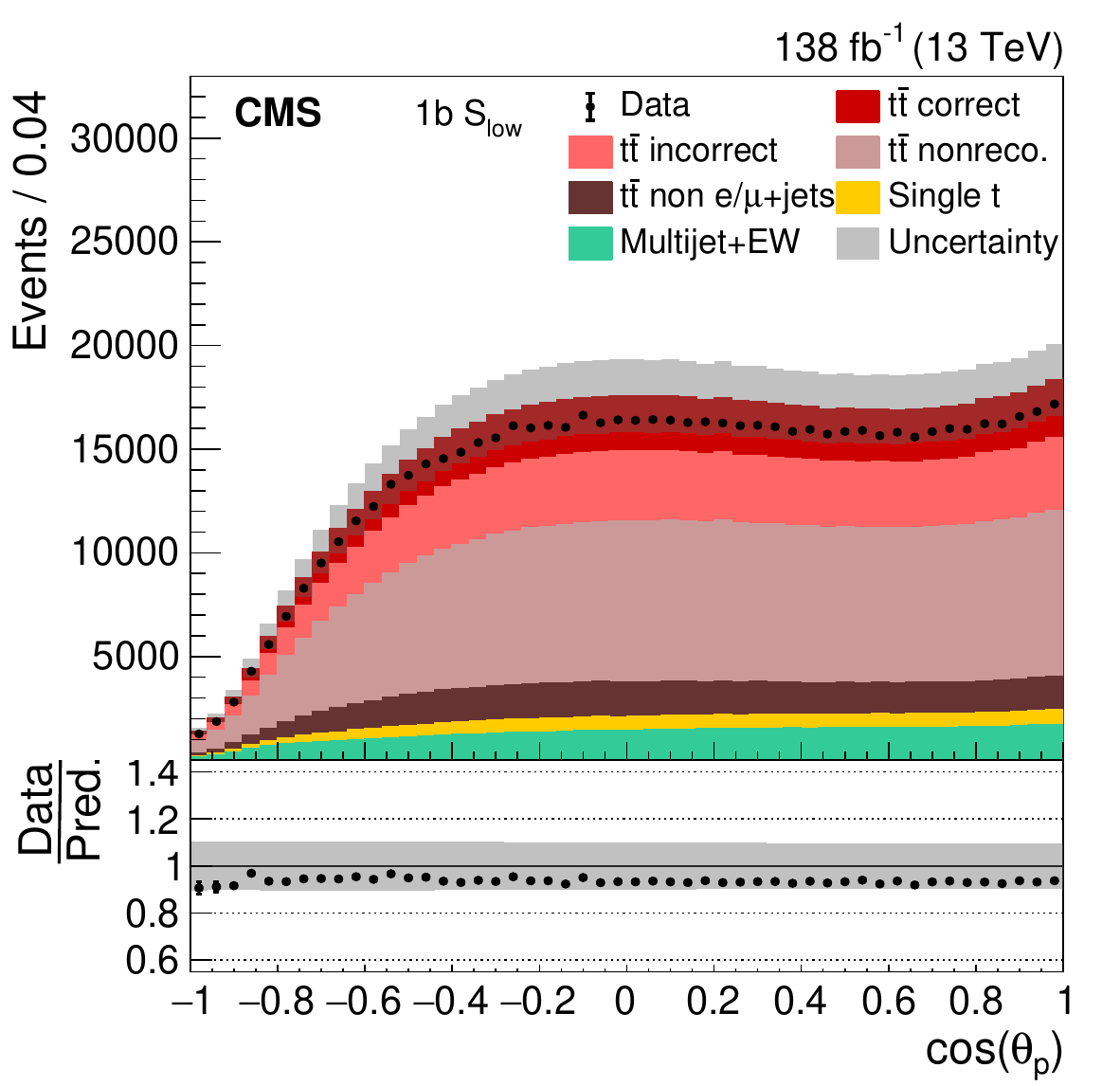}%
\hfill%
\includegraphics[width=0.48\textwidth]{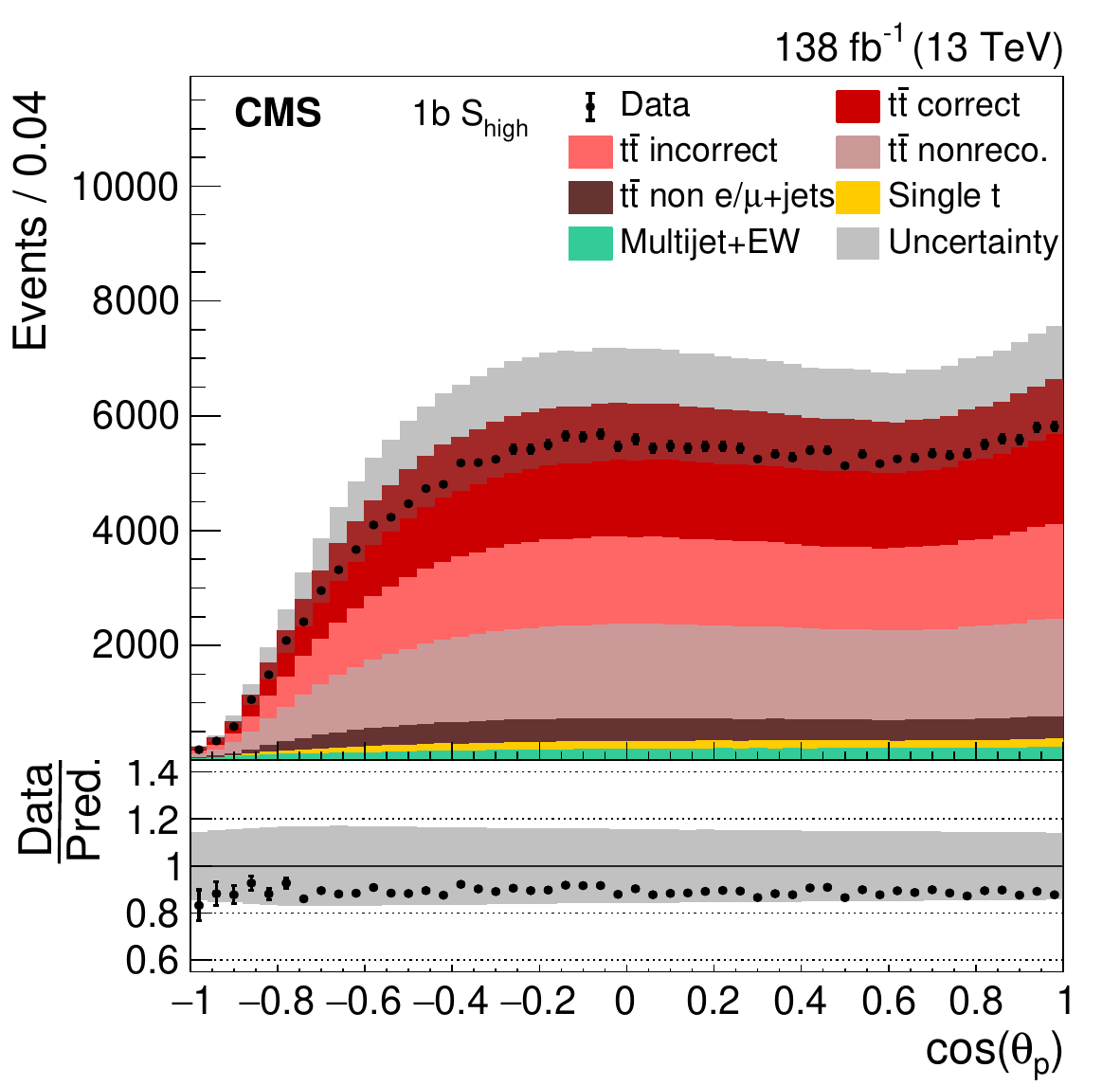} \\
\includegraphics[width=0.48\textwidth]{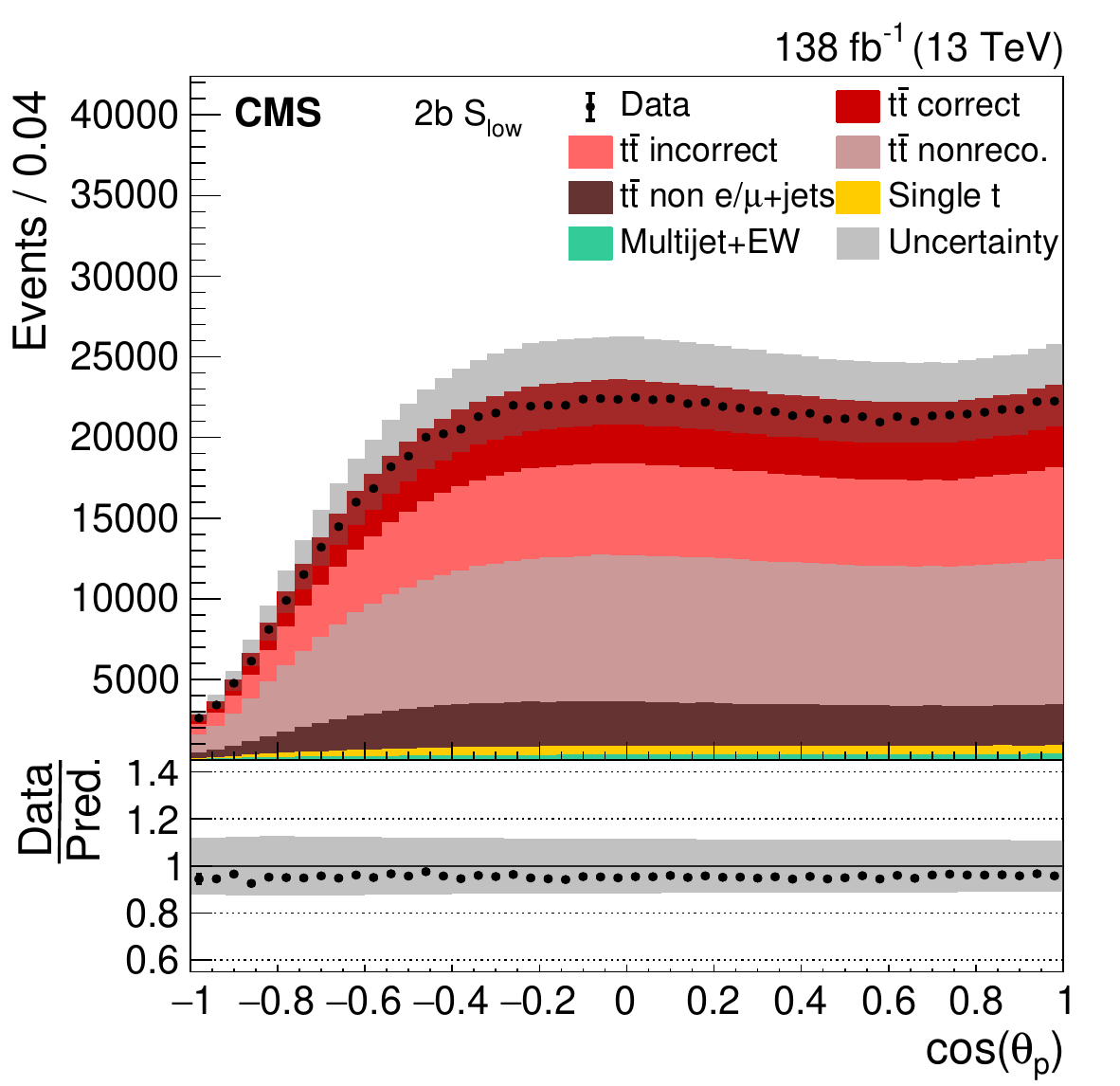}%
\hfill%
\includegraphics[width=0.48\textwidth]{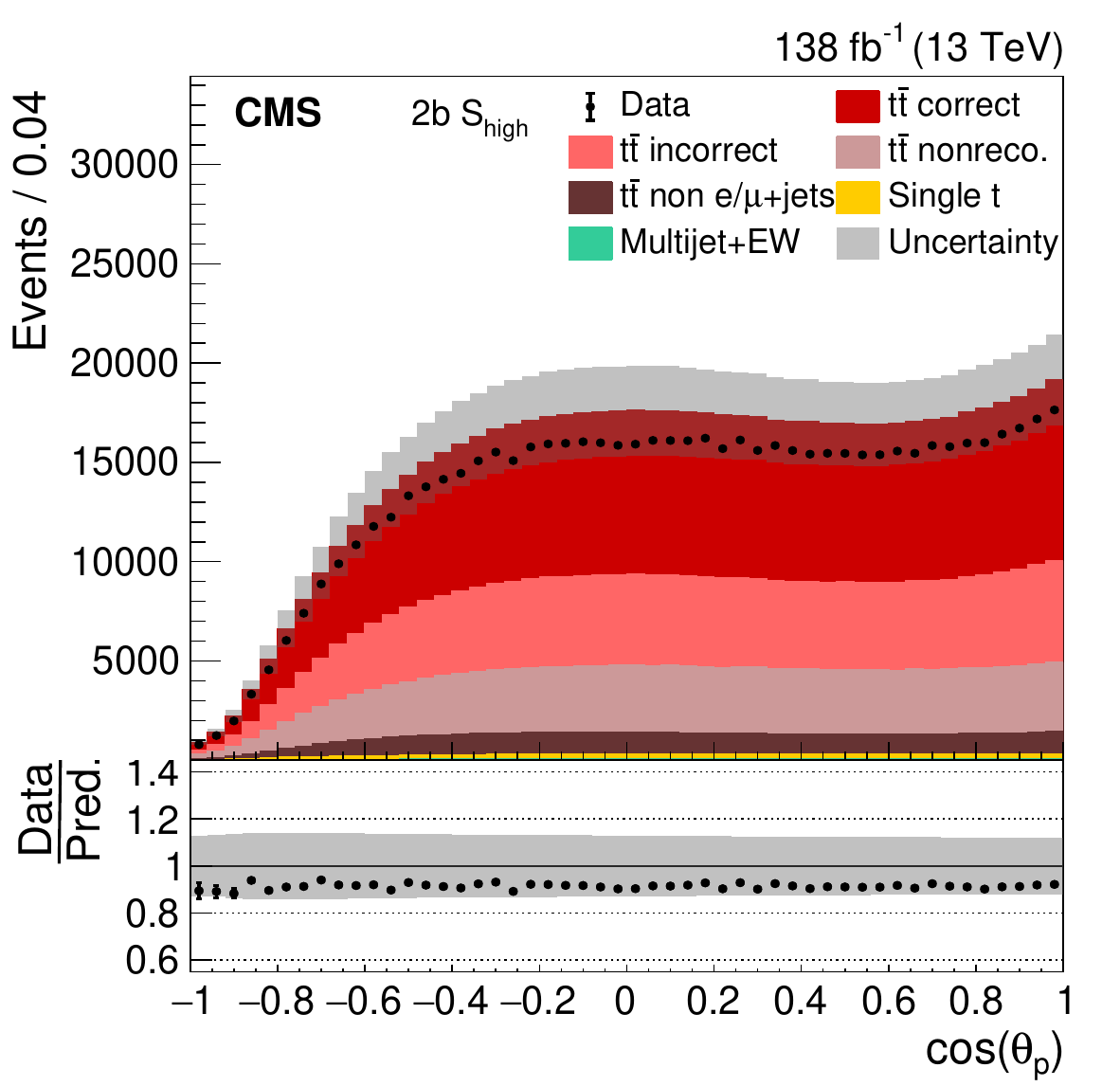}
\caption{Distribution of \costhetap in all four categories. The data (points) are compared to the prediction (stacked histograms). The \ttbar and single top quark contributions are taken from the simulation, while the multijet+EW background is obtained from the CR. The \ttbar contribution is split into the correctly and incorrectly reconstructed, ``nonreconstructable'', and non \lpj events. The gray uncertainty band indicates the combined statistical and systematic uncertainties in the prediction. The vertical bars on the points show the statistical uncertainty of the data. Ratios to the predicted yields are provided in the lower panels.}
\label{background_fig_3}
\end{figure*}

\begin{figure*}[tp!]
\centering
\includegraphics[width=0.48\textwidth]{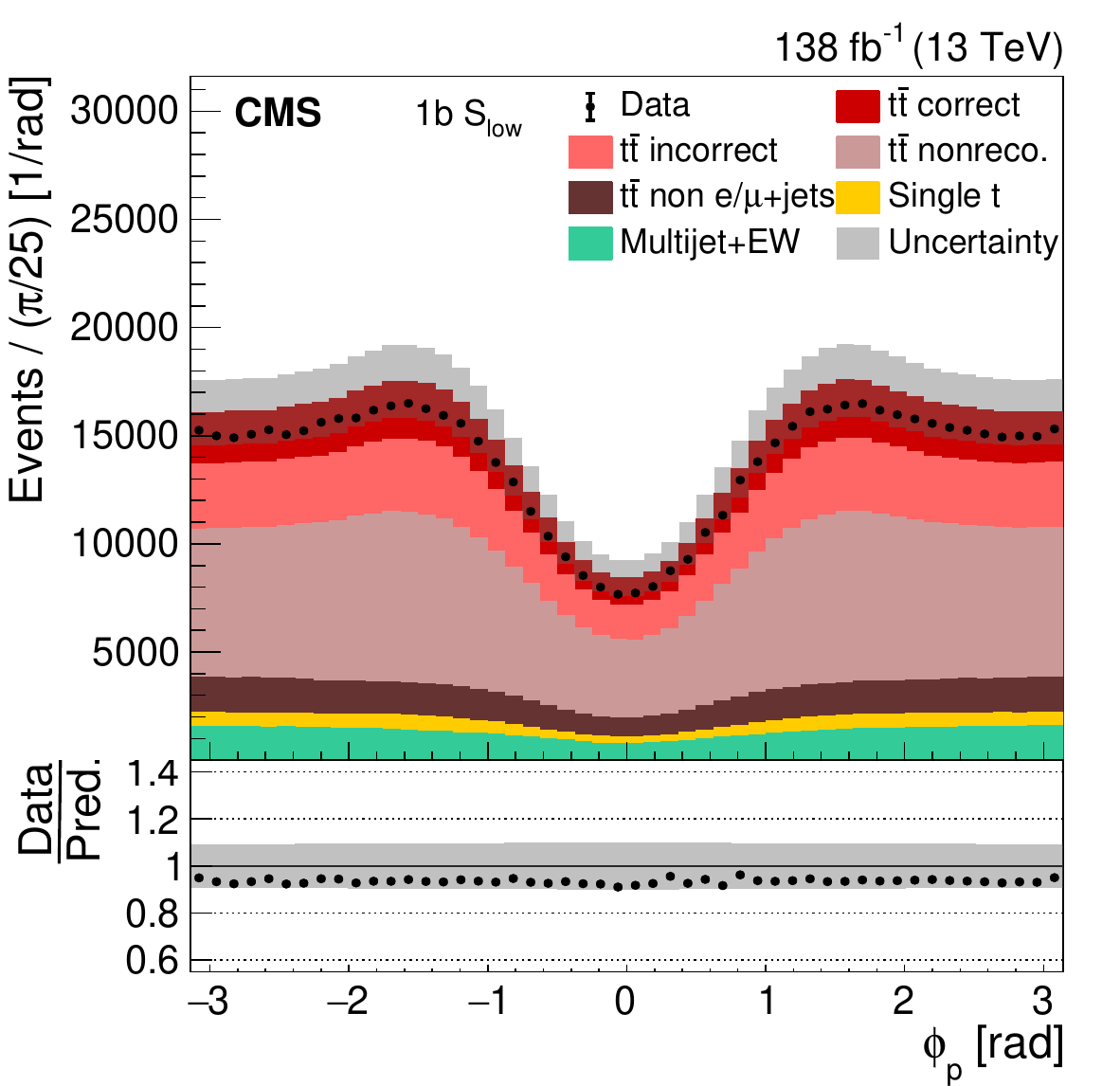}%
\hfill%
\includegraphics[width=0.48\textwidth]{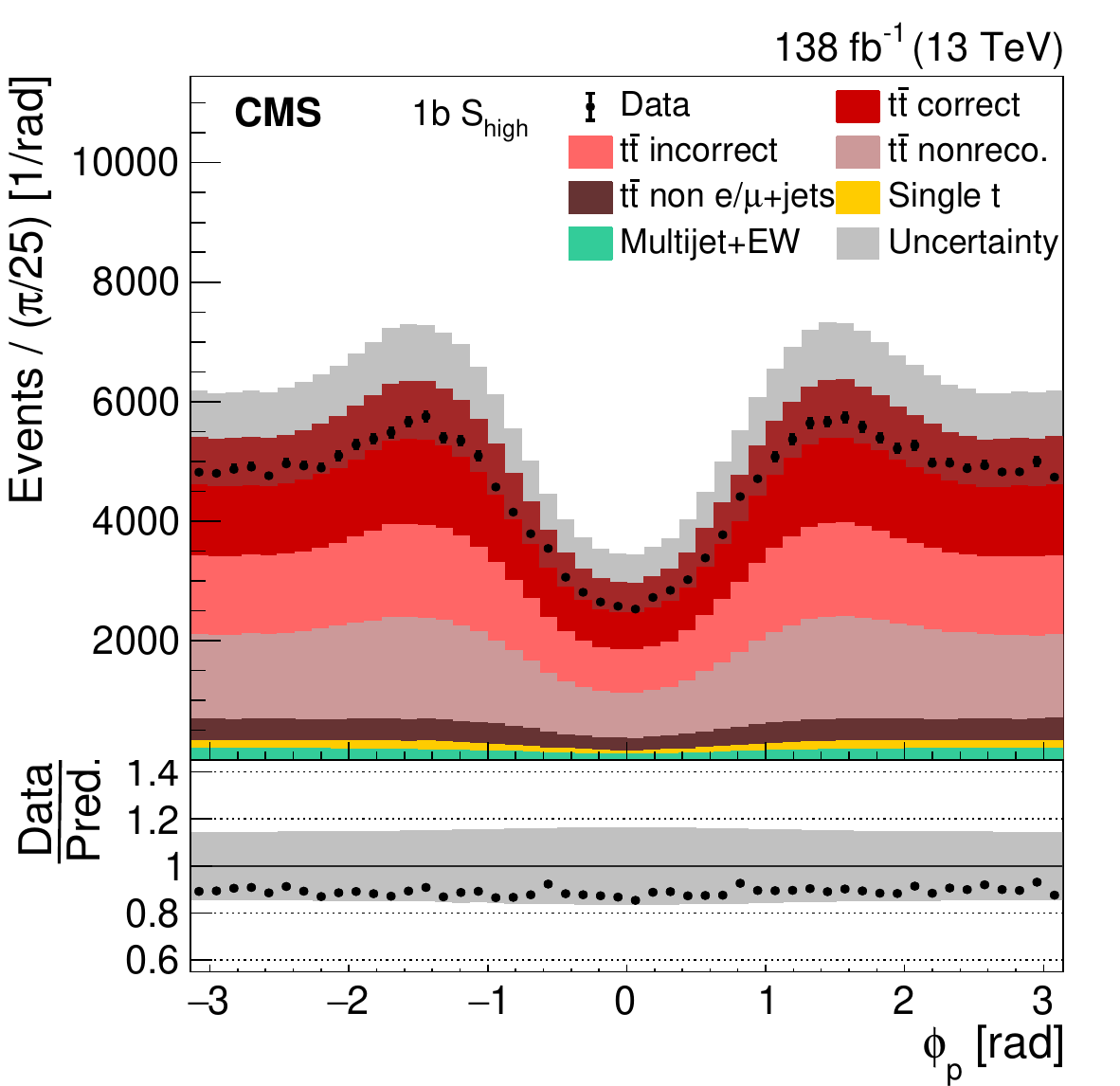} \\
\includegraphics[width=0.48\textwidth]{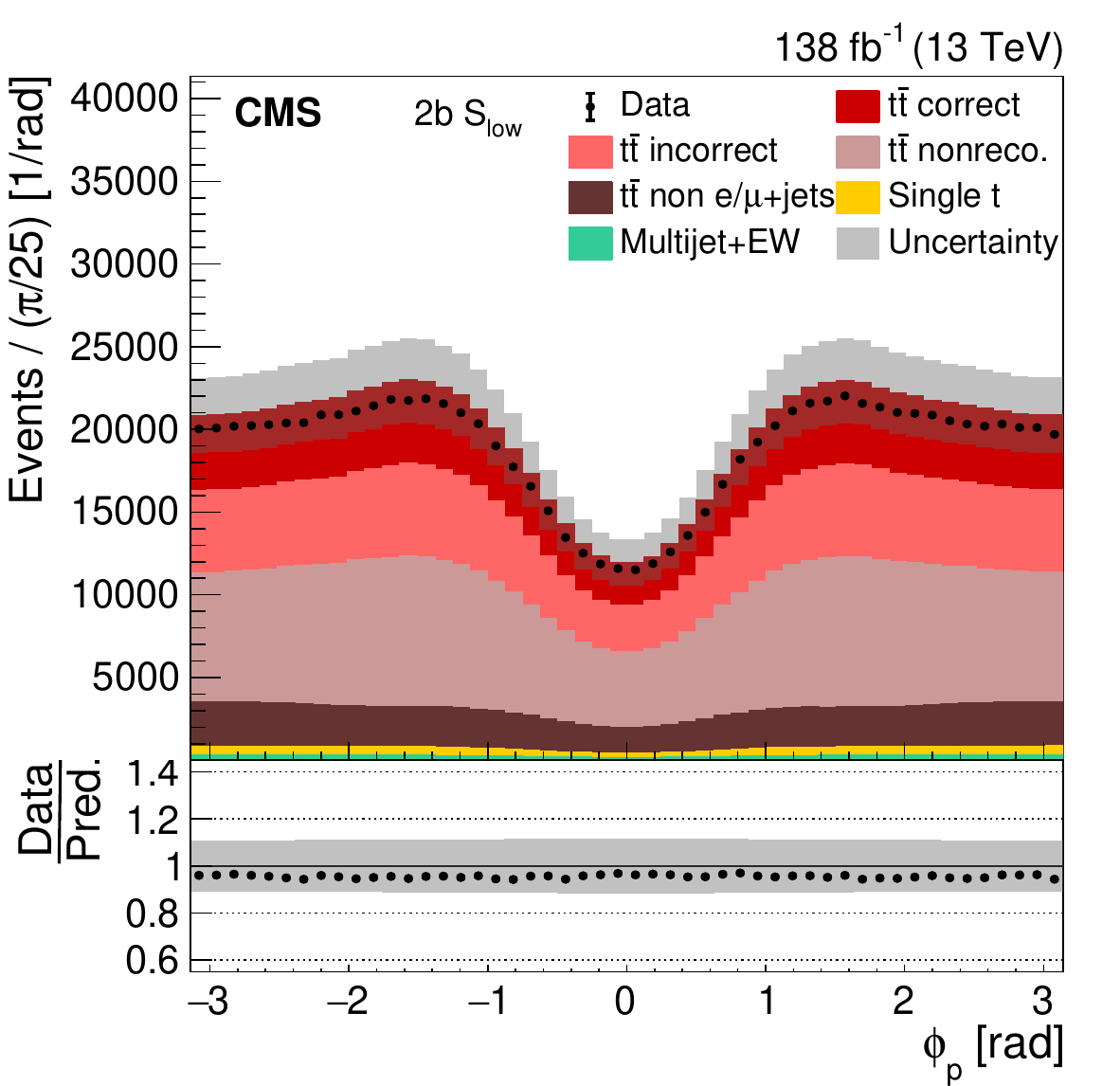}%
\hfill%
\includegraphics[width=0.48\textwidth]{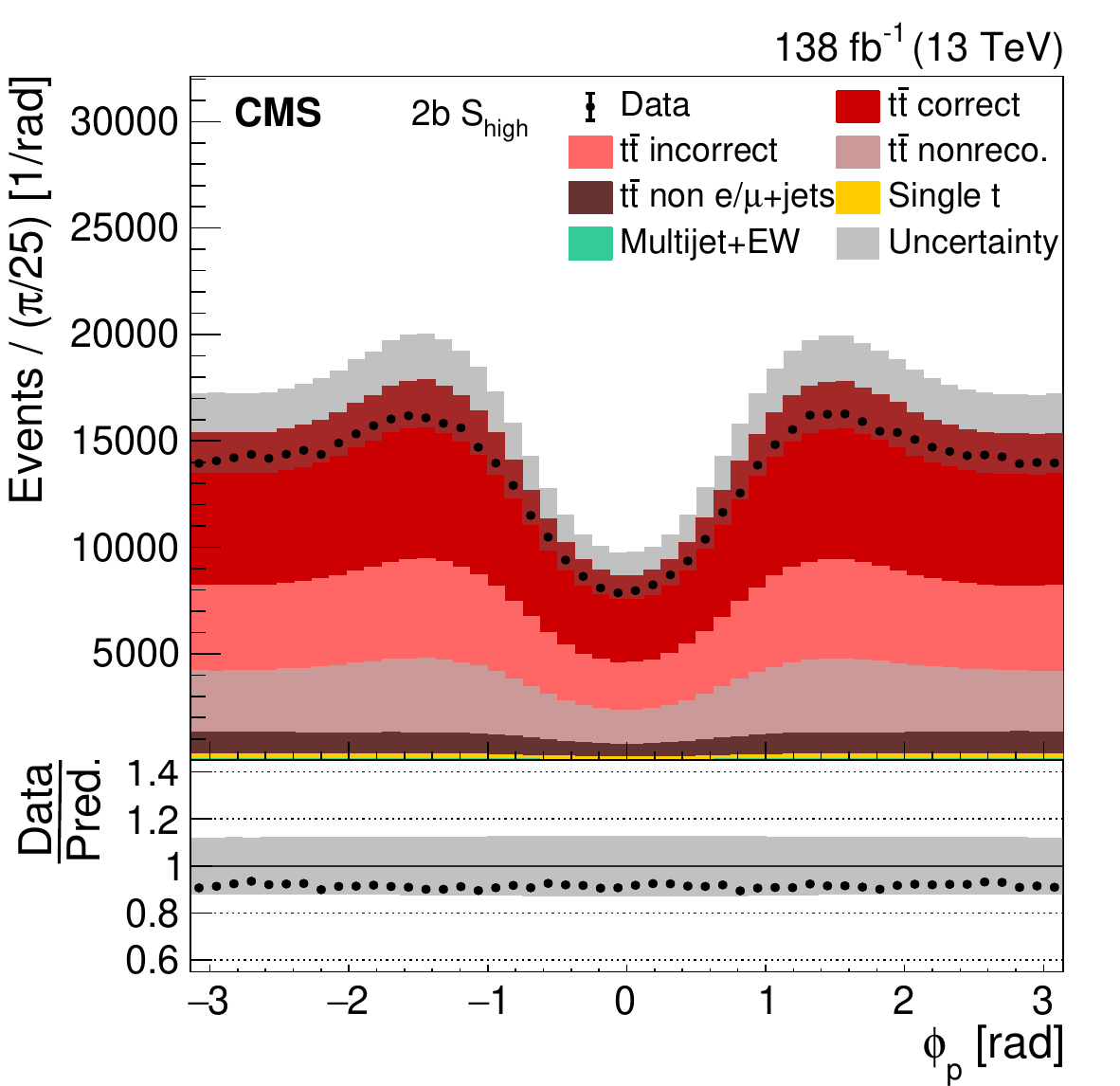}
\caption{Distribution of \phip in all four categories. The data (points) are compared to the prediction (stacked histograms). The \ttbar and single top quark contributions are taken from the simulation, while the multijet+EW background is obtained from the CR. The \ttbar contribution is split into the correctly and incorrectly reconstructed, ``nonreconstructable'', and non \lpj events. The gray uncertainty band indicates the combined statistical and systematic uncertainties in the prediction. The vertical bars on the points show the statistical uncertainty of the data. Ratios to the predicted yields are provided in the lower panels.}
\label{background_fig_4}
\end{figure*}

\begin{figure*}[p!]
\centering
\includegraphics[width=0.48\textwidth]{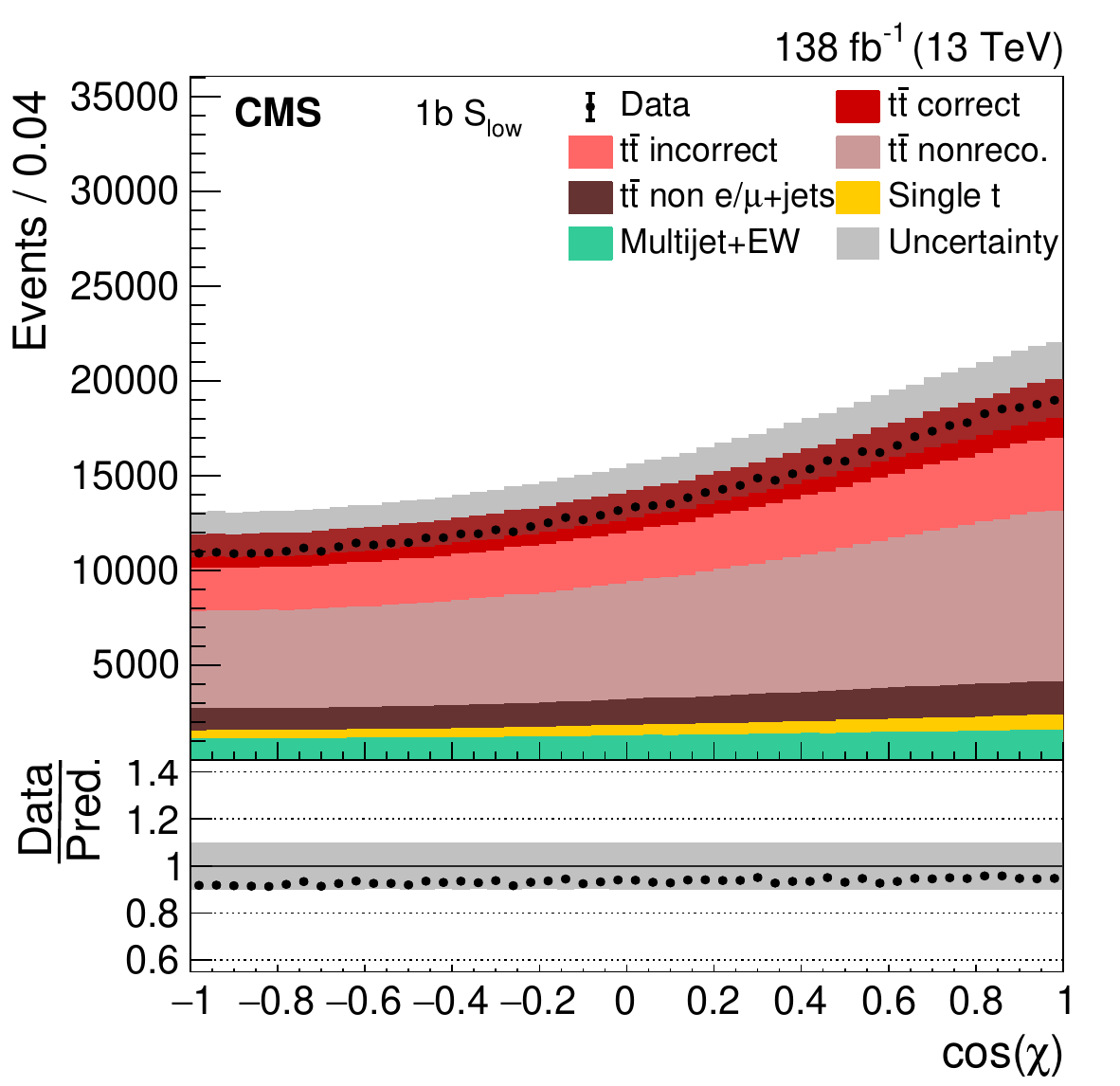}%
\hfill%
\includegraphics[width=0.48\textwidth]{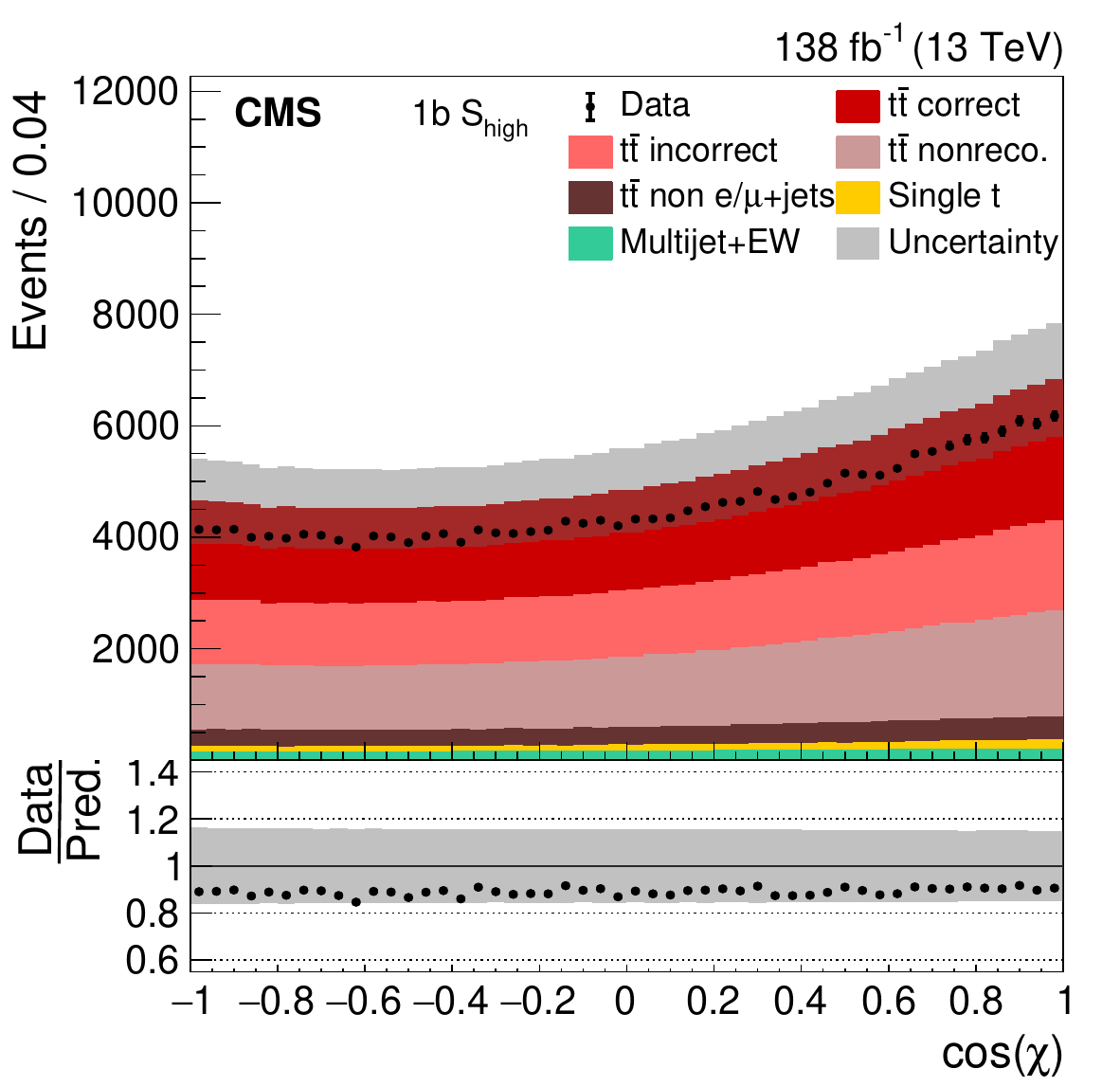} \\
\includegraphics[width=0.48\textwidth]{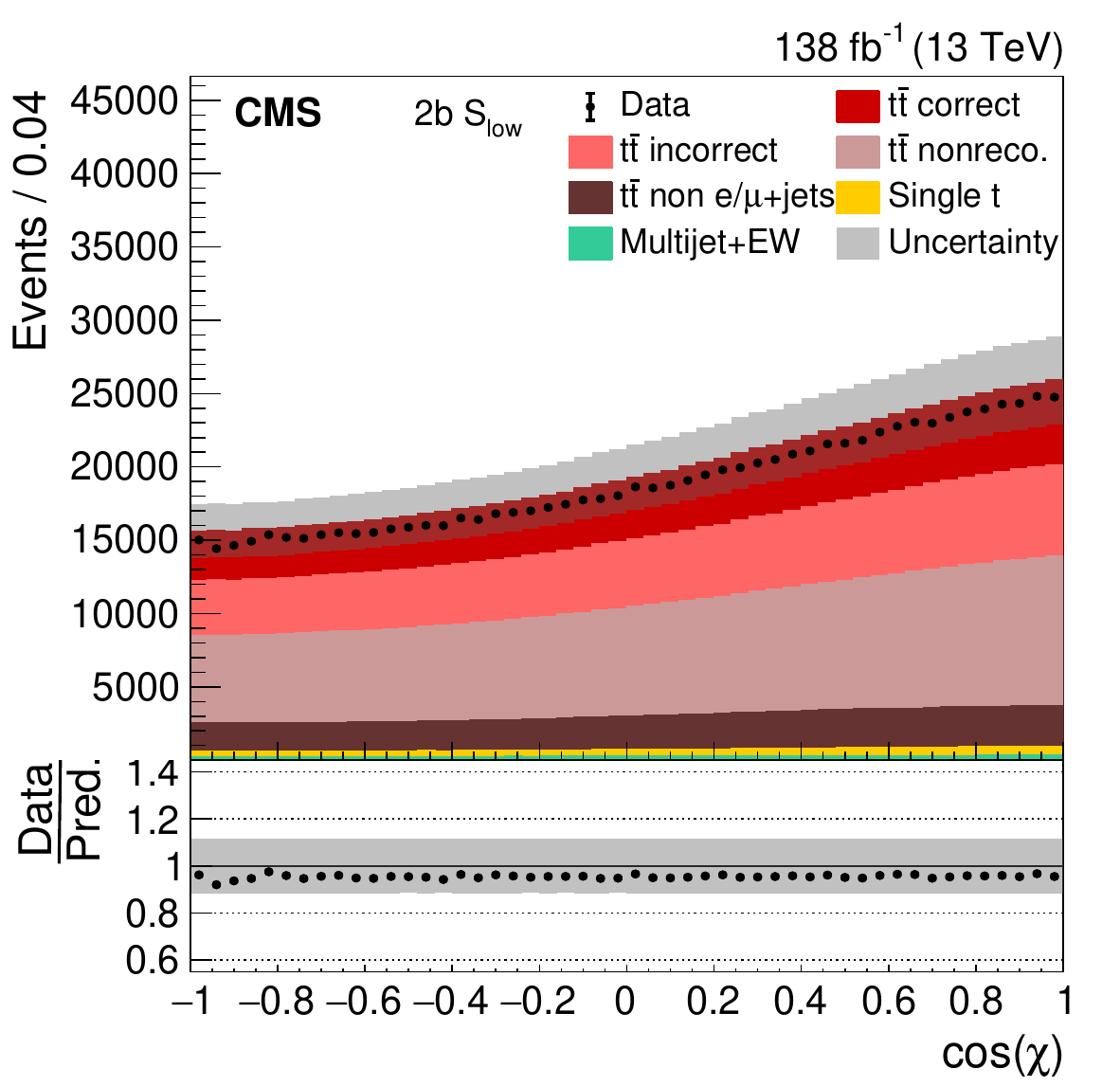}%
\hfill%
\includegraphics[width=0.48\textwidth]{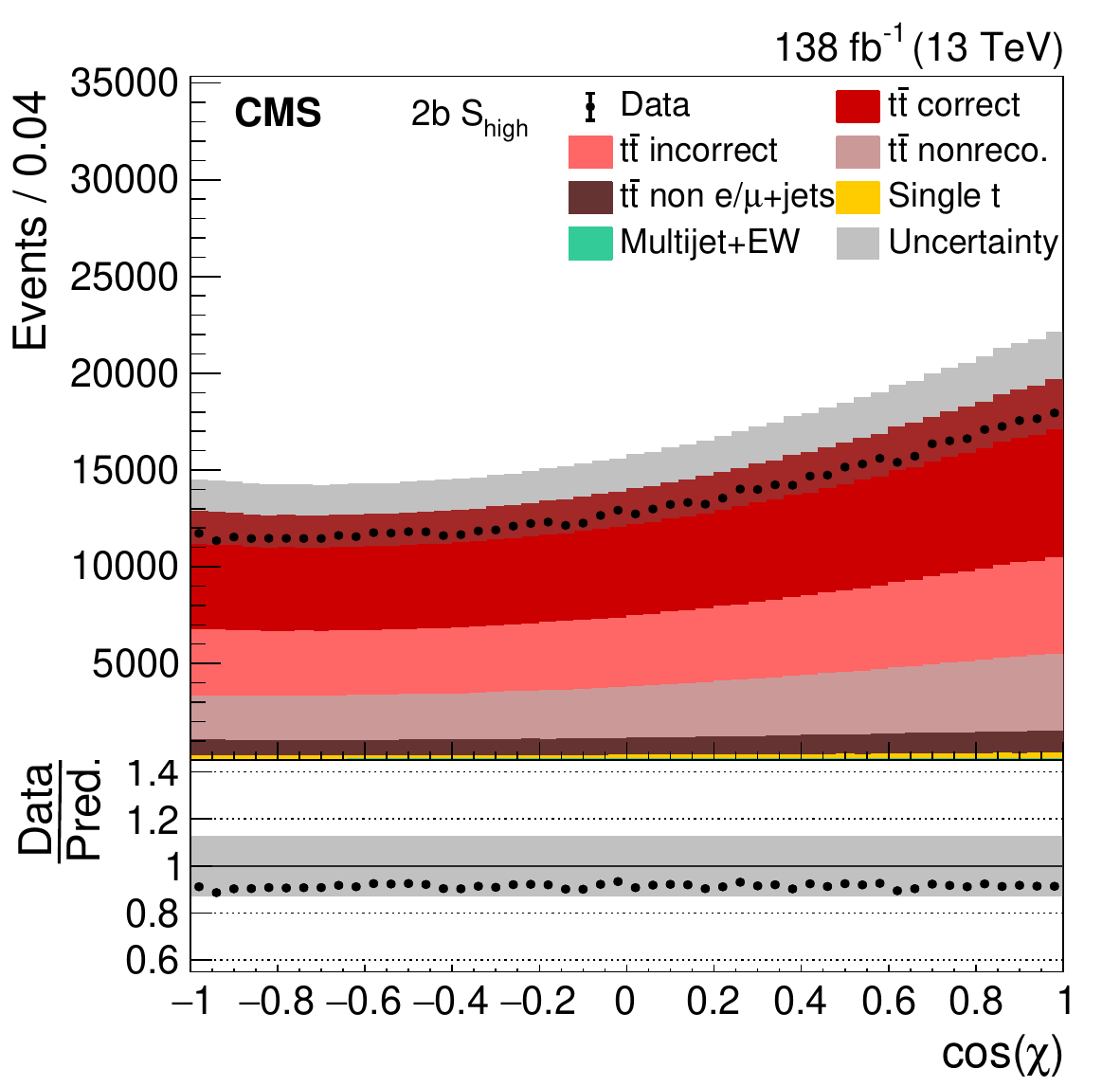}
\caption{Distribution of \coschi in all four categories. The data (points) are compared to the prediction (stacked histograms). The \ttbar and single top quark contributions are taken from the simulation, while the multijet+EW background is obtained from the CR. The \ttbar contribution is split into the correctly and incorrectly reconstructed, ``nonreconstructable'', and non \lpj events. The gray uncertainty band indicates the combined statistical and systematic uncertainties in the prediction. The vertical bars on the points show the statistical uncertainty of the data. Ratios to the predicted yields are provided in the lower panels.}
\label{background_fig_5}
\end{figure*}

\begin{figure*}[tp!]
\centering
\includegraphics[width=0.48\textwidth]{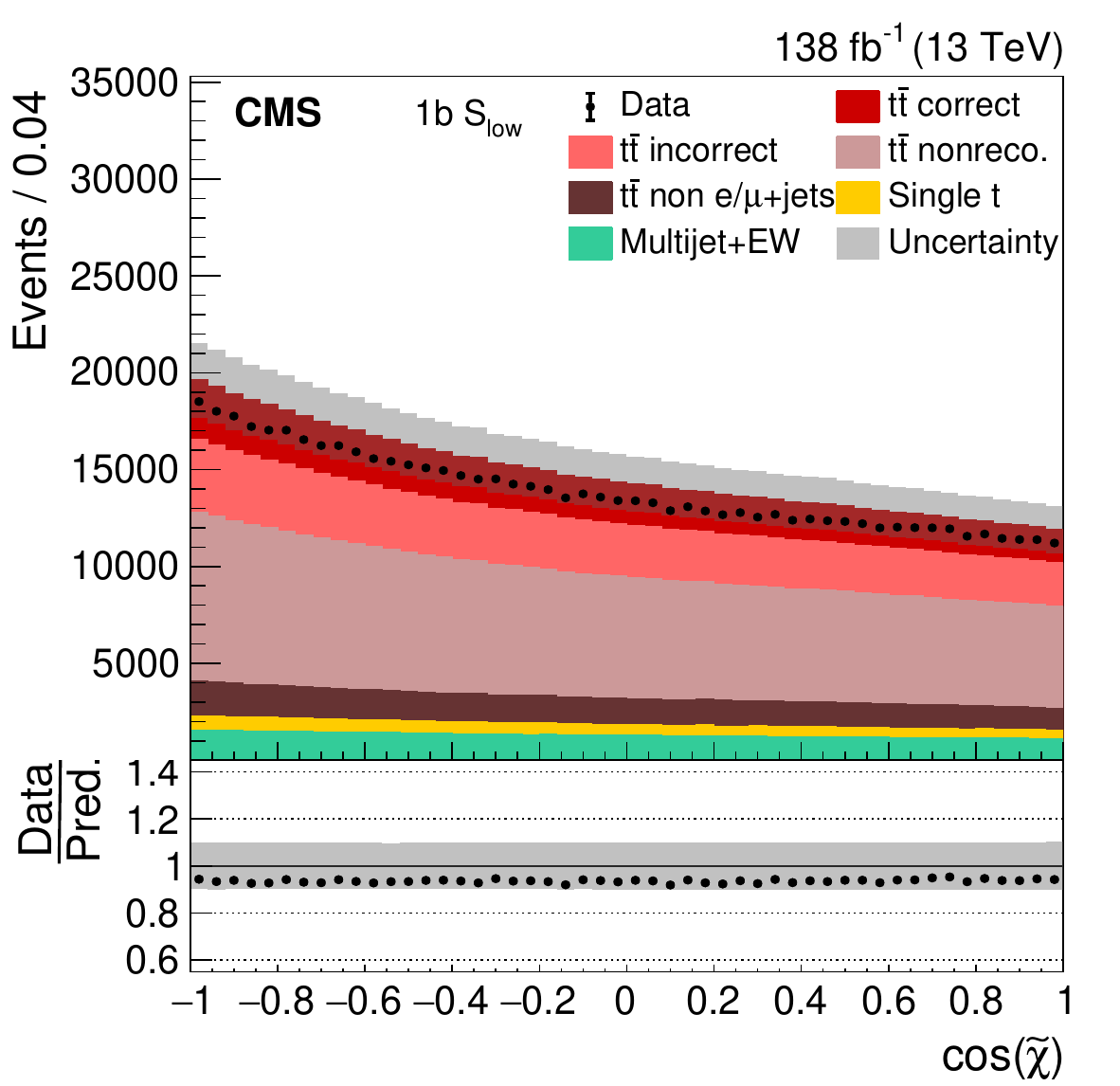}%
\hfill%
\includegraphics[width=0.48\textwidth]{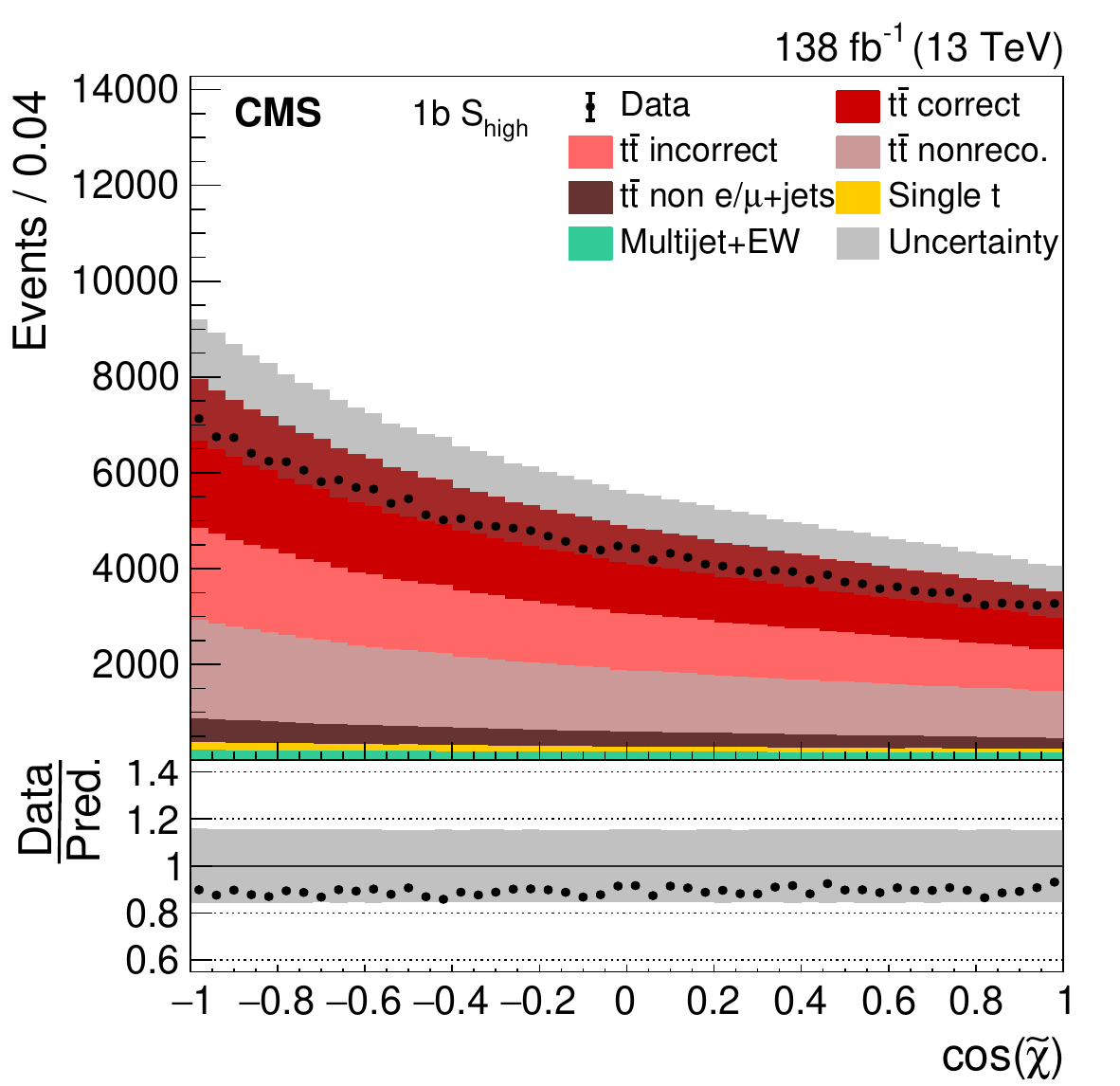} \\
\includegraphics[width=0.48\textwidth]{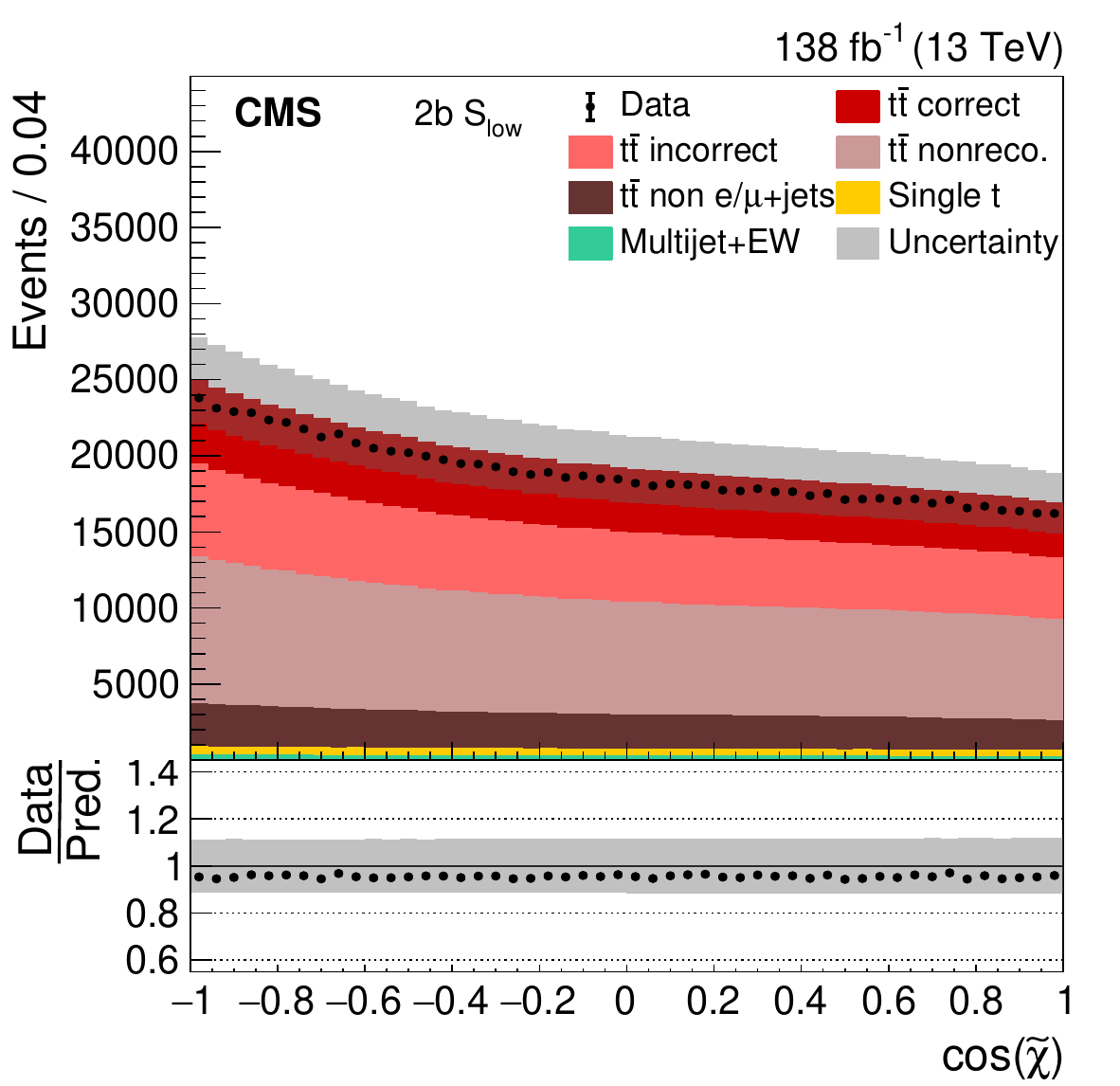}%
\hfill%
\includegraphics[width=0.48\textwidth]{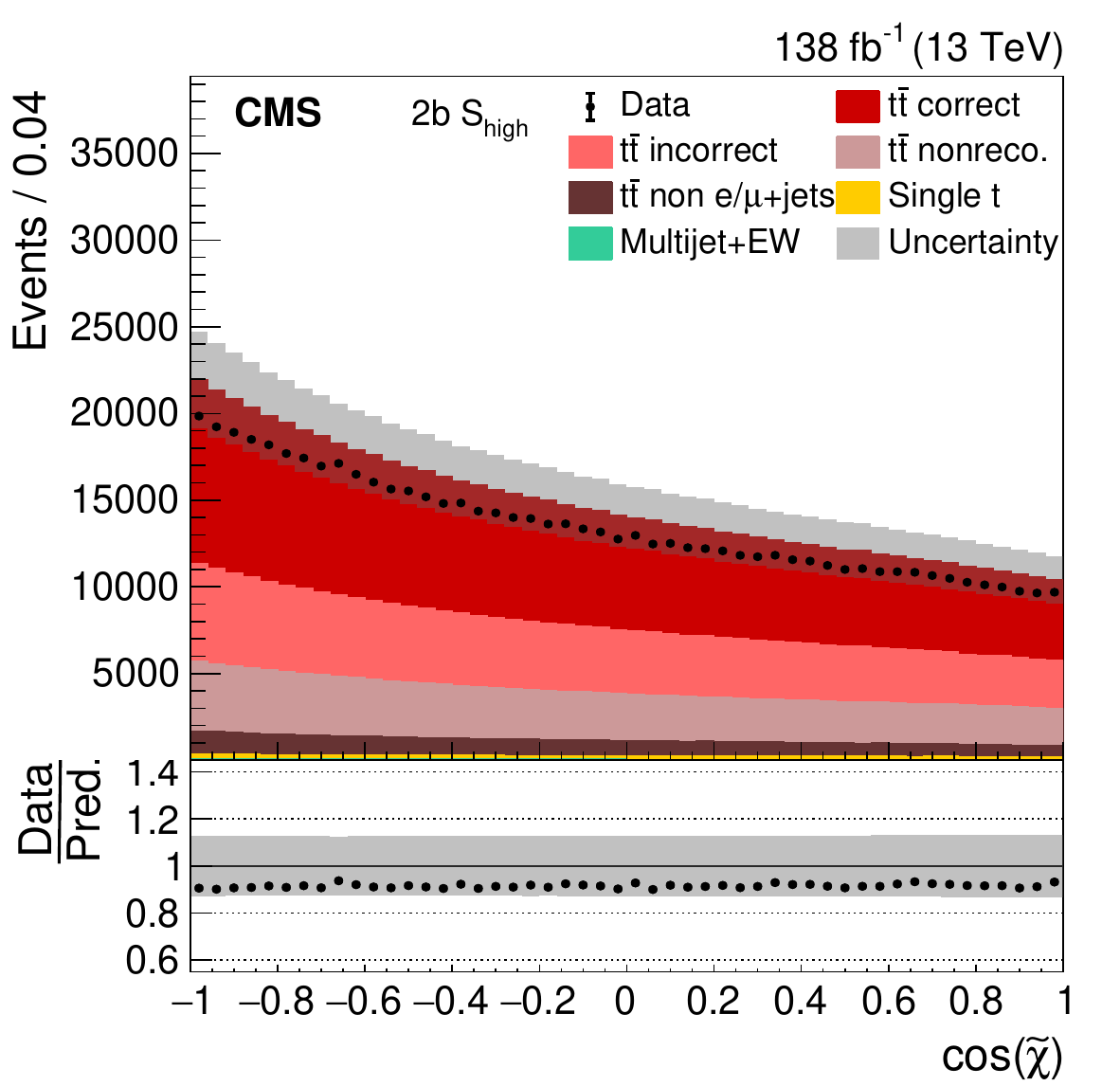}
\caption{Distribution of \coschit in all four categories. The data (points) are compared to the prediction (stacked histograms). The \ttbar and single top quark contributions are taken from the simulation, while the multijet+EW background is obtained from the CR. The \ttbar contribution is split into the correctly and incorrectly reconstructed, ``nonreconstructable'', and non \lpj events. The gray uncertainty band indicates the combined statistical and systematic uncertainties in the prediction. The vertical bars on the points show the statistical uncertainty of the data. Ratios to the predicted yields are provided in the lower panels.}
\label{background_fig_6}
\end{figure*}

In Figs.~\ref{background_fig_2}--\ref{background_fig_6}, the distributions of several observables in all signal categories are shown with the multijet+EW background estimation taken from the CR. The uncertainty bands include statistical uncertainties and all systematic uncertainties detailed in Section~\ref{uncertainties}, and are in general dominated by uncertainties in the jet energy scale, the \PQb tagging efficiencies, PS modeling, and ME scales, with the latter contributing the largest uncertainty in the overall normalization.

\section{Extraction of polarization and spin correlation coefficients}
\label{signalfit}

Following the formalism introduced in Eq.~\eqref{method_eq1}, the differential cross section \Sigtot can be written as a linear combination of functions \Sigm, which depend on the angles \phippb and \thetappb of the decay products of the top quark and antiquark:
\begin{equation}\label{signal_eq0}
    \Sigtot=\Sigma_0+\sum_{m=1}^{15}\Qm\Sigm.
\end{equation}
The spin analyzing powers $\kappa$ and the cross section \signorm are absorbed in the definitions of the functions \Sigm.

The values of the coefficients \Qm can be extracted by fitting \Sigtot with Eq.~\eqref{signal_eq0}. This approach is used at the generator level to obtain the \QmMC---the polarization and spin correlation of the partonic top quarks as predicted by each of the \ttbar simulations and their uncertainty variations. These fits are performed in bins of the additional observables \ttmcth and \tptcth. A binning in \ttm of $\{300,400,600,800,13000\}\GeV$ with the first \ttm bin including a few underflow events, and in \tpt of $\{0,100,200,300,6500\}\GeV$ is used. In both cases the bins are further divided into \absct bins with the boundaries $\{0,0.4,0.7,1\}$. As a result we obtain the average values of the \QmMC in each of these bins. The knowledge of the \QmMC facilitates the analytical calculation of \SigtotMC as a function of \phippb and \thetappb in each bin of the additional observables.

\begin{figure*}[!ph]
\centering
\includegraphics[width=0.48\textwidth]{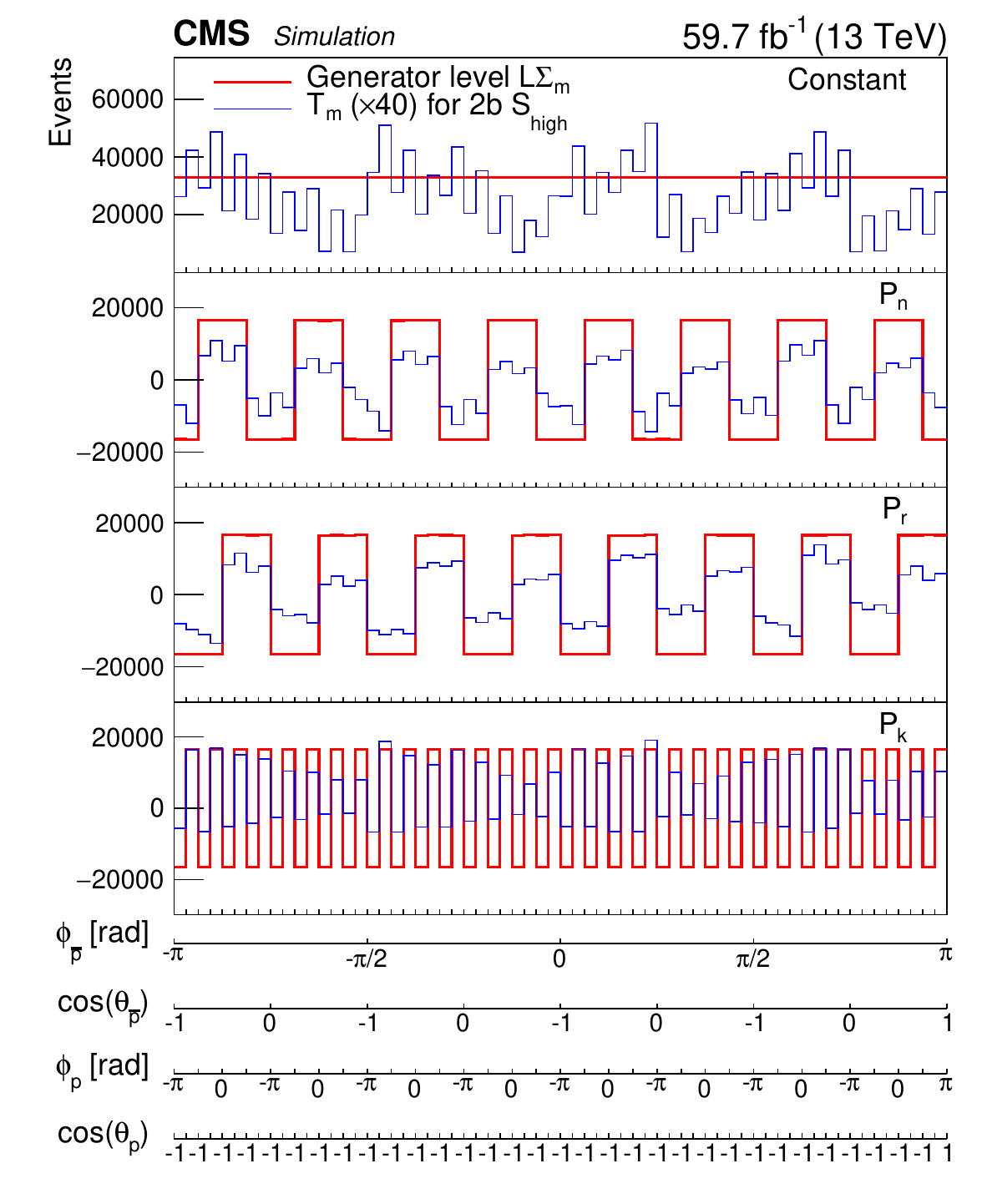}%
\hfill%
\includegraphics[width=0.48\textwidth]{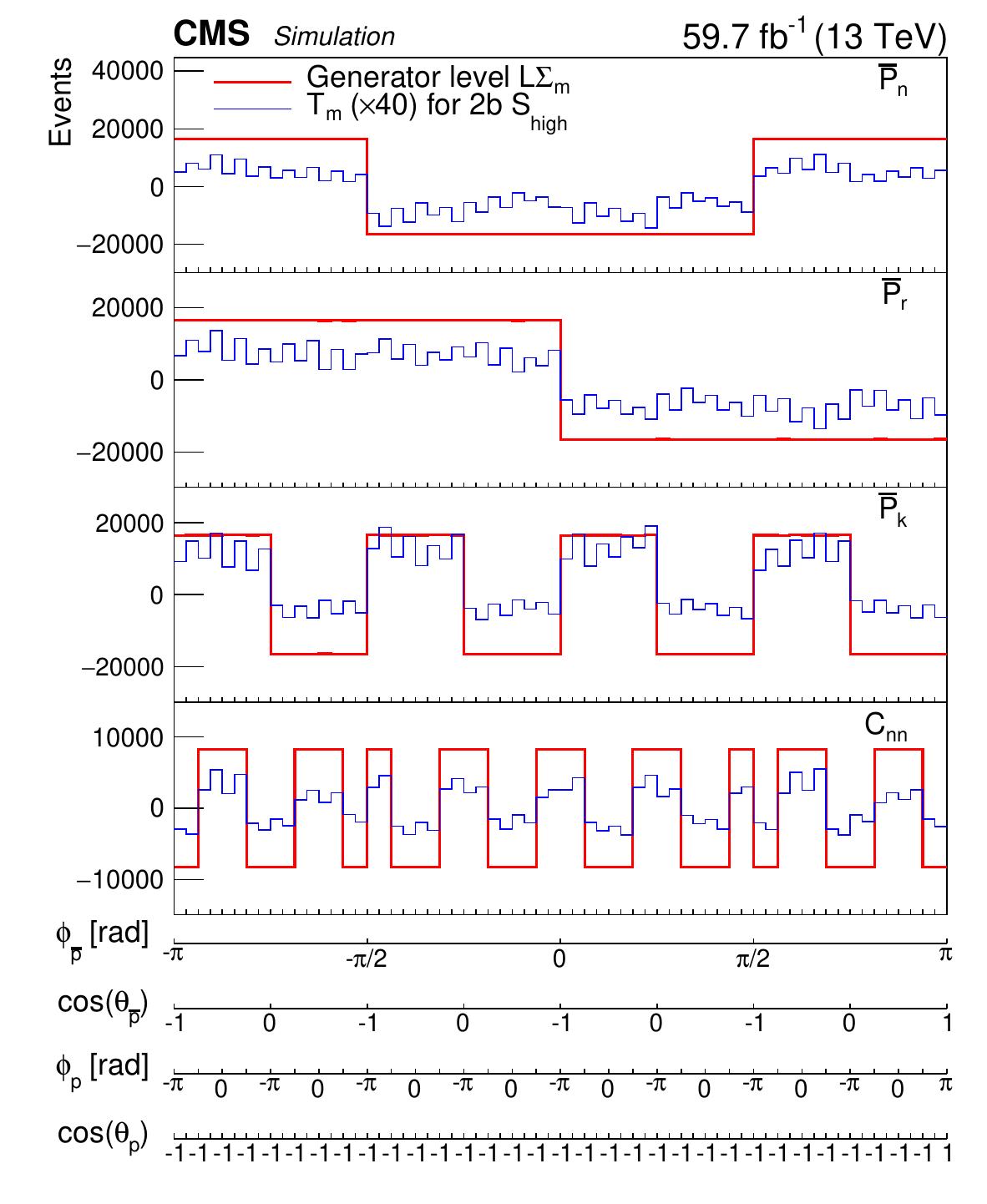} \\
\includegraphics[width=0.48\textwidth]{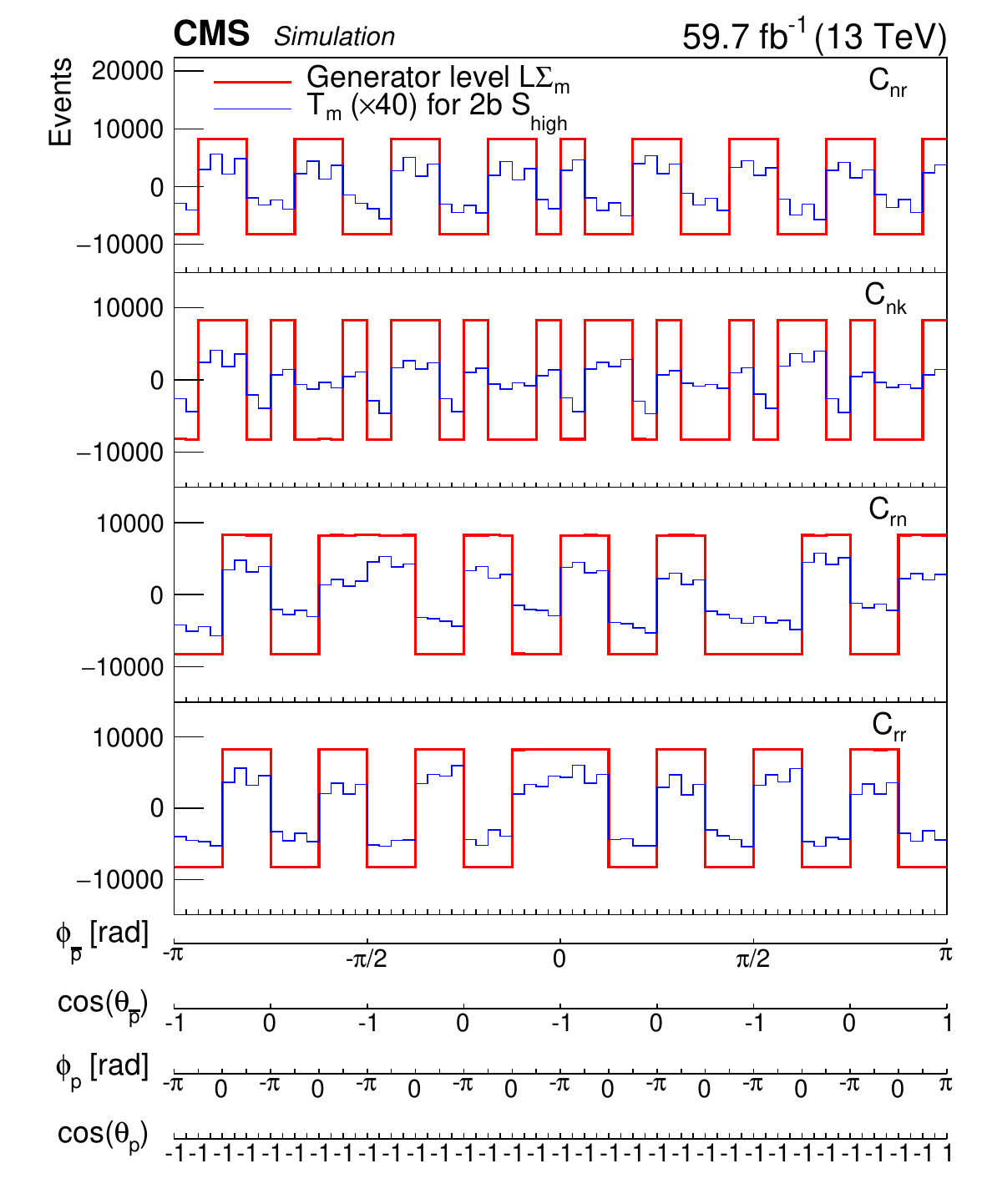}%
\hfill%
\includegraphics[width=0.48\textwidth]{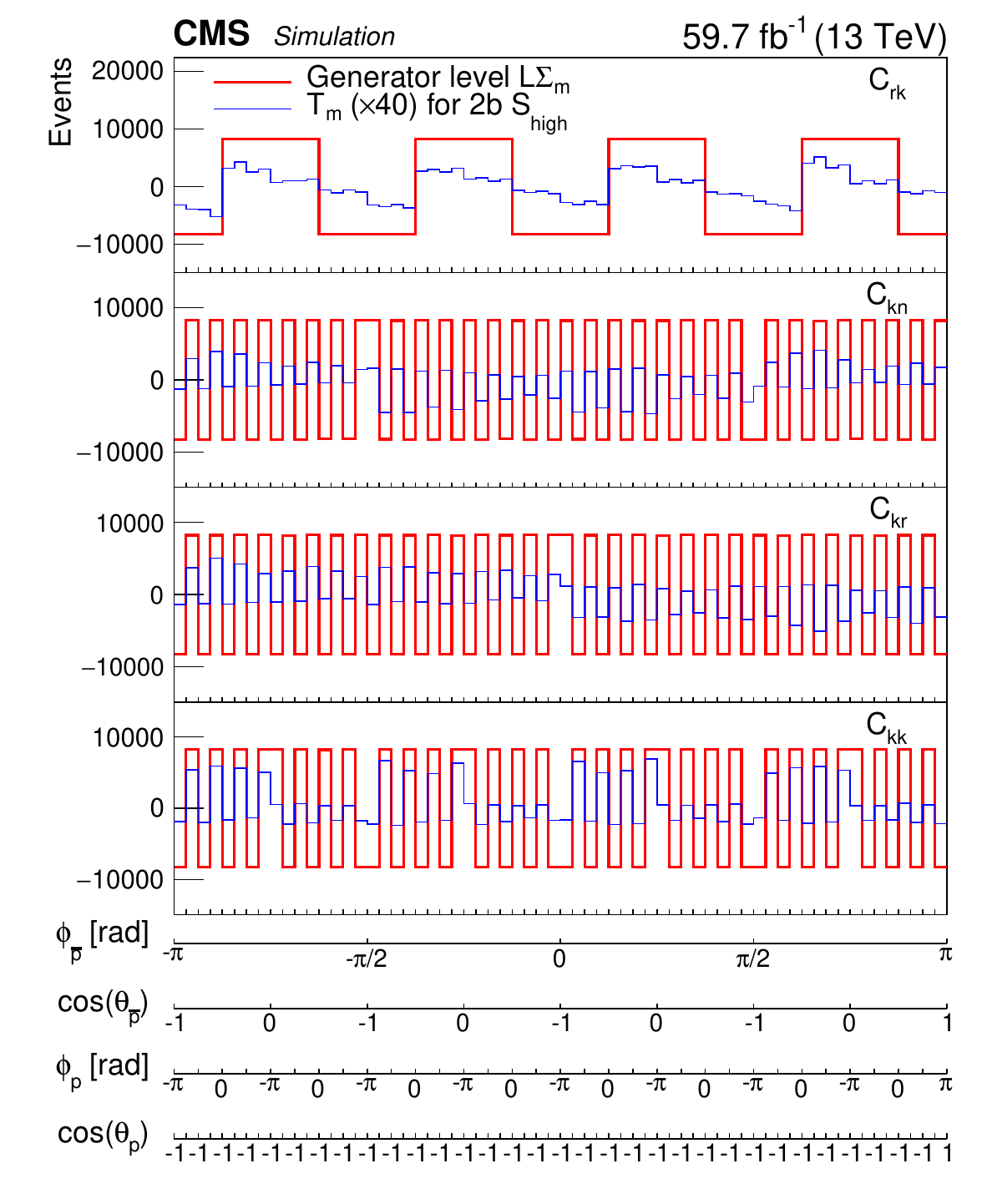}
\caption{Examples of unrolled 4-dimensional distributions \LSigm and \Tm as functions of \phippb and \thetappb for the individual coefficients of the polarization vectors and the spin correlation matrix for events with $400<\ttm<600\GeV$ and $\absct<0.4$. The \LSigm (red lines) are the distributions at the generator level in the full phase space, and the \Tm (blue lines) are the distributions in the \twobhigh category for the 2018 data. For the purpose of illustration, the events are required to be reconstructed and generated in the same \ttmcth bin. The detector-level distributions are enhanced by a factor of 40 to improve their visibility.}
\label{method_fig1}
\end{figure*}

For the measurement of the \Qm, we are interested in templates \Tm that can be fit to the data and describe only the effect of the corresponding coefficient. At the generator level, these templates are \LSigm, where $L$ is the integrated luminosity. Accordingly, the \Tm are the corresponding distributions of events at the detector level in the signal categories. The binnings at the generator and the detector levels are the same. The \Tm include all \ttbar events selected at the detector level, meaning that they describe polarization and spin correlation effects of \lpj and \ttbar background events, also referred as \ttbar non \lpj. To avoid the full simulation of \ttbar samples for each \Qm we use a reweighting technique to evaluate the \Tm. For this, each event is assigned a weight equal to $\Sigm/\SigtotMC$, which are evaluated for each event based on the generator-level values of \thetappb, \phippb, and the bin determined by the additional observables \ttmcth or \tptcth.
In this bin we know the average value of the \QmMC, as determined from the fits of the \Sigm at the generator level. The generator-level \Sigm and the \Tm for the \twobhigh category at the detector level are shown in Fig.~\ref{method_fig1}. Here, the $x$ axis shows the bin number of the unrolled 4-dimensional distribution of \phipb, \costhetapb, \phip, and \costhetap, listed from the outermost to the innermost variable, where $\cos(\thetappb)$ uses two bins: $\{-1,0,1\}$, and \phippb is divided into four bins: $\{-\pi,-\pi/2,0,\pi/2,\pi\}$, resulting in a total number of 64 bins.

In general, the \Qm are not constant within a bin. At the generator level, the functions \Sigm do not depend on the kinematic properties of the top quarks, so they factorize and the average values of the \Qm are fitted in each bin. However, at the detector level, the \Tm do change as a function of the top quark and antiquark kinematic properties due to selection requirements and detector effects. Therefore, it is important to perform the measurements in sufficiently small bins such that either the \Qm or the \Tm are approximately constant within each bin. If the \Tm vary significantly within a fitted bin, the measured \Qm could be biased if the values of the coefficients change within a bin in a different way than in the SM simulation. The binnings in \ttmcth and \tptcth were selected to minimize the bias due to nonconstant \Tm templates within the bins. The way the templates are constructed ensures that a template fit to the SM prediction extracts the correct \Qm, \ie, the bias is minimized for SM-like data.

The same reweighting procedure is used for the \Qone and \Dtilde measurements, but decomposing the distribution of \coschi and \coschit into constant and linear terms as given by Eq.~\eqref{method_eq4}. In these cases we use 10 equally sized bins in \coschi, and \coschit for all measurements. A finer binning in \ttm of $\{300,400,500,600,700,800,900,1000,13000\}\GeV$ and \tpt of $\{0,50,100,150,200,\allowbreak250,300,400,6500\}\GeV$ is selected for the \Qone and \Dtilde measurements.

{\tolerance=800
We perform a maximum likelihood fit combining the information of the four selections (\twobhigh, \twoblow, \onebhigh, \oneblow) in the four data-taking periods, for a total of 16 categories. The statistical model describes the total number of events in each bin
\ifthenelse{\boolean{cms@external}}{\begin{multline}
    N_{ij'n'}(\an,\setQmn,\setnuk)=S_{ij'n'}(\an,\setQmn,\setnuk) \\
    +B_{ij'n'}(\setnuk),
\end{multline}}{\begin{equation}
    N_{ij'n'}(\an,\setQmn,\setnuk)=S_{ij'n'}(\an,\setQmn,\setnuk)+B_{ij'n'}(\setnuk),
\end{equation}}}%
with $i$ denoting the category and $j'$ referring to a bin in the 1-dimensional concatenated detector-level distribution of \phippb and \thetappb. The index $n'$ ($n$) refers to a bin of the \ttmcth or \tptcth distribution at the detector (generator) level. The normalization parameters \an and the \Qmn (or \Qonen, \Dtilden) are defined separately for each bin $n$ and are free parameters in the fit. The $S_{ij'n'}$ and $B_{ij'n'}$ are the \ttbar and background contributions, respectively, and both can depend on the nuisance parameters \setnuk modeling the variation of the expected event yields due to systematic uncertainties. The \ttbar contribution takes the form
\ifthenelse{\boolean{cms@external}}{\begin{multline}
    S_{ij'n'}\big(\an,\setQmn,\setnuk\big)=\sum_n\an\Bigg(T_{i0nj'n'}\big(\setnuk\big) \\
    +\sum_{m=1}^{15}\Qmn T_{imnj'n'}\big(\setnuk\big)\Bigg),
\end{multline}}{\begin{equation}
    S_{ij'n'}\big(\an,\setQmn,\setnuk\big)=\sum_n\an\Bigg(T_{i0nj'n'}\big(\setnuk\big)+\sum_{m=1}^{15}\Qmn T_{imnj'n'}\big(\setnuk\big)\Bigg),
\end{equation}}
where $T_{imnj'n'}$ are the detector-level distributions (templates) obtained by reweighting the \ttbar simulation to the individual components proportional to the \Qmn and the templates $T_{i0nj'n'}$ correspond to the constant terms. Finally, the function
\ifthenelse{\boolean{cms@external}}{\begin{multline}
    -2\log\Big(L\big(\an,\setQmn,\setnuk\big)\Big)= \\
    -2\sum_{ij'n'}\Big[d_{ij'n'}\log\Big(N_{ij'n'}\big(\an,\setQmn,\setnuk\big)\Big) \\
    -N_{ij'n'}\big(\an,\setQmn,\setnuk\big)\Big] \\
    -2\log\Big(G\big(\setnuk\big)\Big)
\end{multline}}{\begin{multline}
    -2\log\Big(L\big(\an,\setQmn,\setnuk\big)\Big)=-2\sum_{ij'n'}\Big[d_{ij'n'}\log\Big(N_{ij'n'}\big(\an,\setQmn,\setnuk\big)\Big) \\
    -N_{ij'n'}\big(\an,\setQmn,\setnuk\big)\Big]-2\log\Big(G\big(\setnuk\big)\Big)
\end{multline}}
is minimized with respect to the value of the parameters \an, \Qmn, and \nuk. Here, $d_{ij'n'}$ are the observed event yields and $G(\setnuk)$ describes the Gaussian constraints of the \nuk. Goodness of fit tests indicate good agreement between the data and the fitted model with $p$-values~\cite{Gross:2010qma} of 0.80--0.95 for all fits. We tested that the Gaussian approximation can be used to describe the distributions of the uncertainties in the measured parameters. This allows us to use Gaussian error propagation when evaluating quantities derived from the measured parameters.

\begin{figure*}[!p]
\centering
\includegraphics[width=0.8\textwidth]{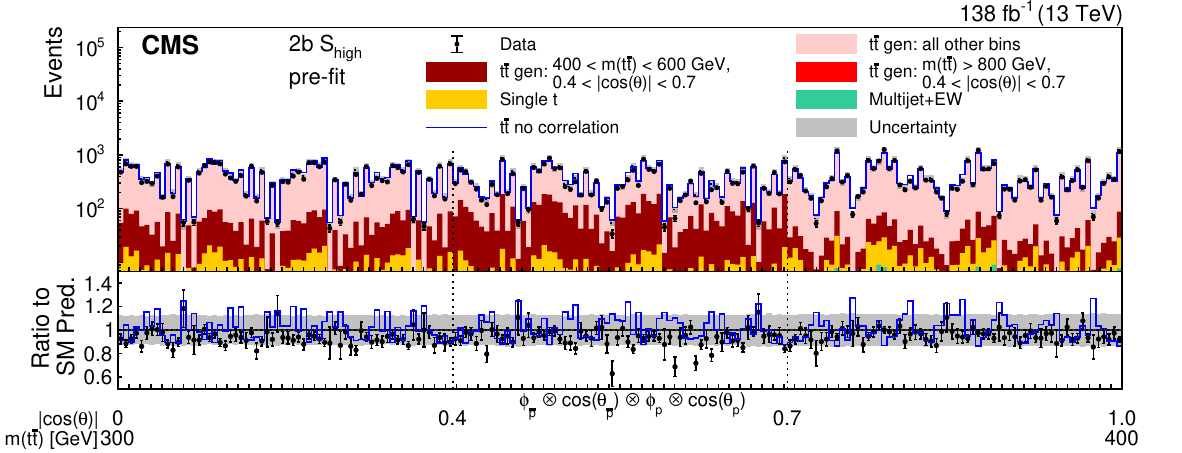}
\includegraphics[width=0.8\textwidth]{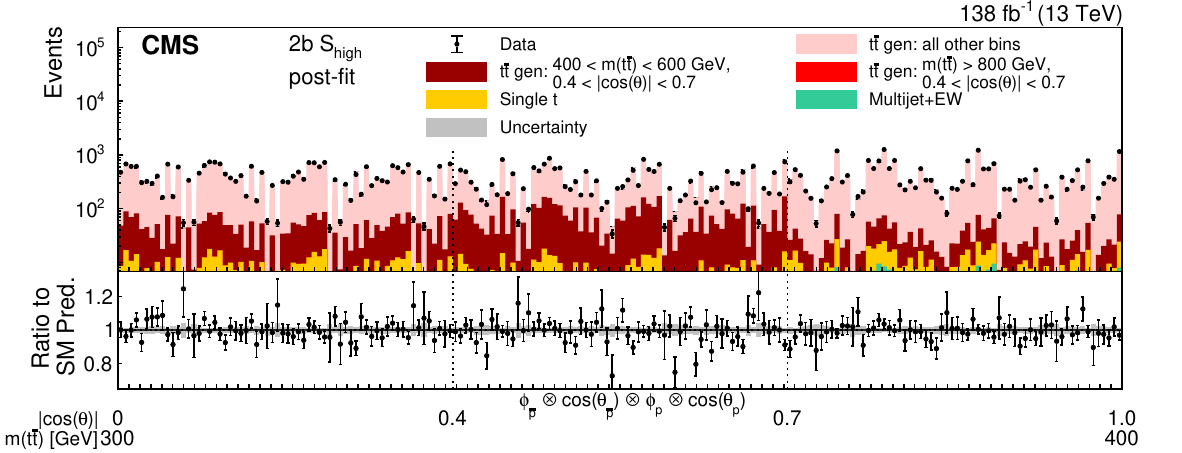}
\includegraphics[width=0.8\textwidth]{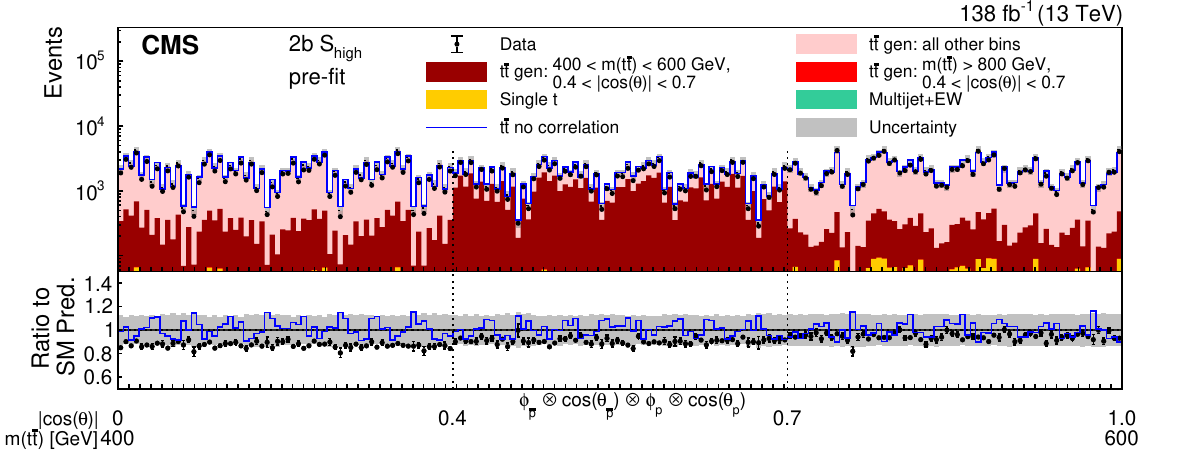}
\includegraphics[width=0.8\textwidth]{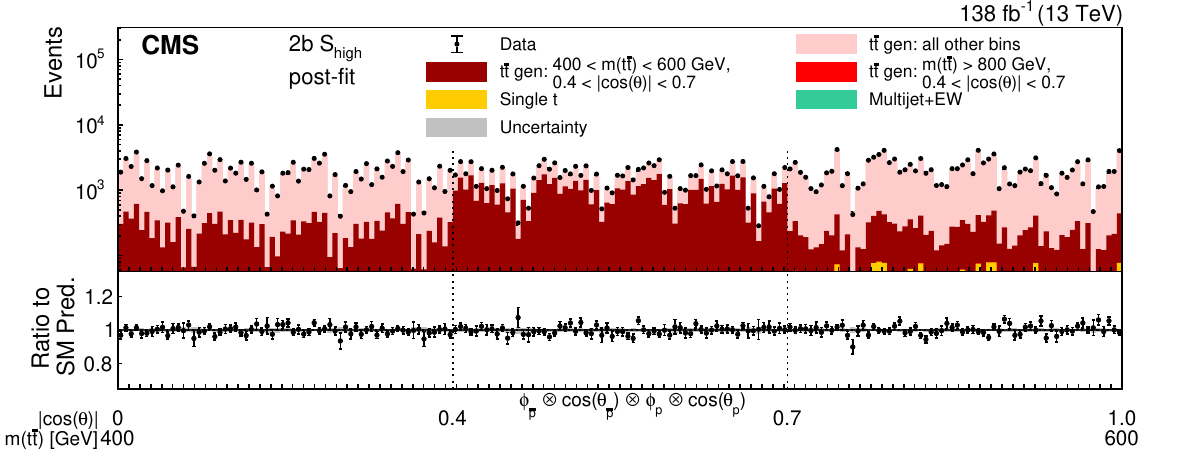}
\caption{Pre- and post-fit distributions comparing the data (points) to the \PWGPYT simulation (stacked histograms) for the full matrix measurement in bins of \ttmcth in the \twobhigh category. The $x$ axis shows the bins of the unrolled 4-dimensional distribution of \phipb, \costhetapb, \phip, and \costhetap, listed from the outermost to the innermost variable in each of the \ttmcth bins. The boundaries of the \absct and \ttm bins are labeled and indicated by dashed and solid lines, respectively. For the illustration of resolution effects, \ttbar events generated in two selected \ttmcth bins are shown in different shades of red. All other \ttbar contributions are shown in pink. A model without any polarization and spin correlation is shown as a blue line. The gray uncertainty band indicates the combined statistical and systematic uncertainties in the prediction. The vertical bars on the points show the statistical uncertainty. Ratios to the predicted yields are provided in the lower panels.}
\label{signalfit_fig_4a}
\end{figure*}

\begin{figure*}[!p]
\centering
\includegraphics[width=0.8\textwidth]{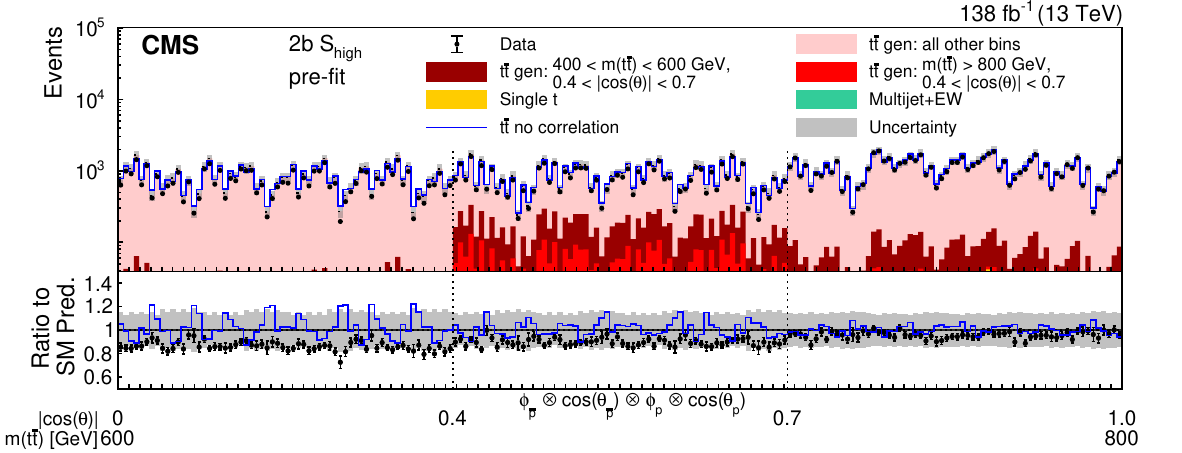}
\includegraphics[width=0.8\textwidth]{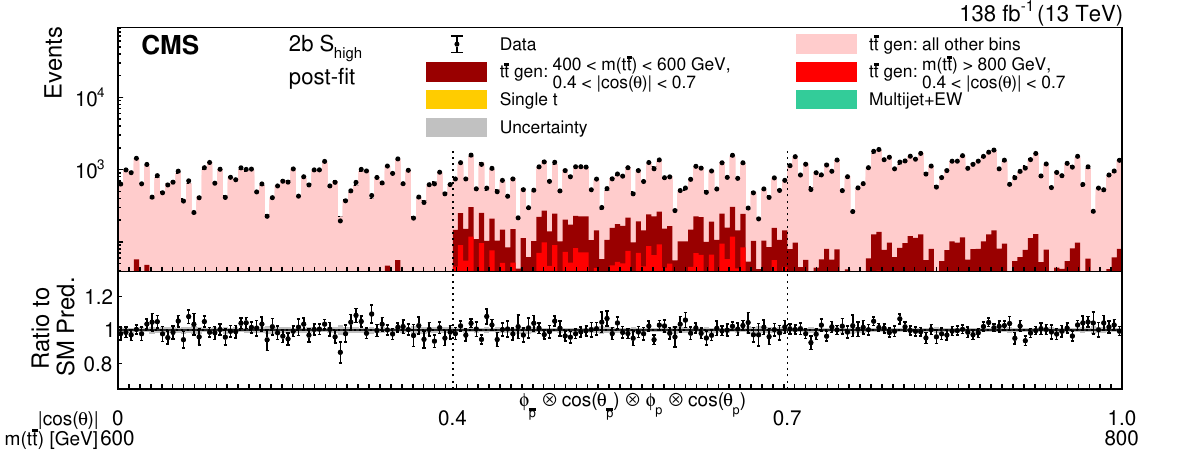}
\includegraphics[width=0.8\textwidth]{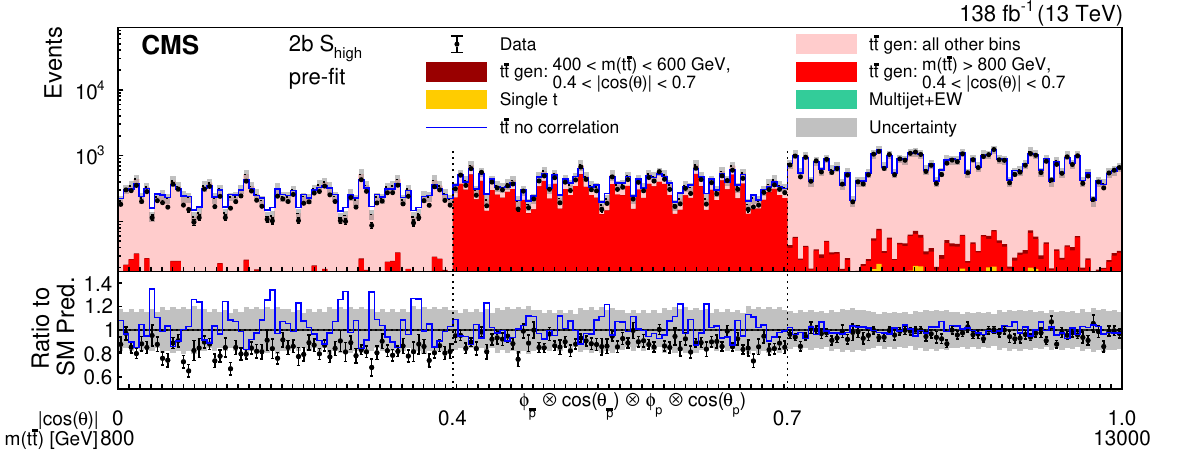}
\includegraphics[width=0.8\textwidth]{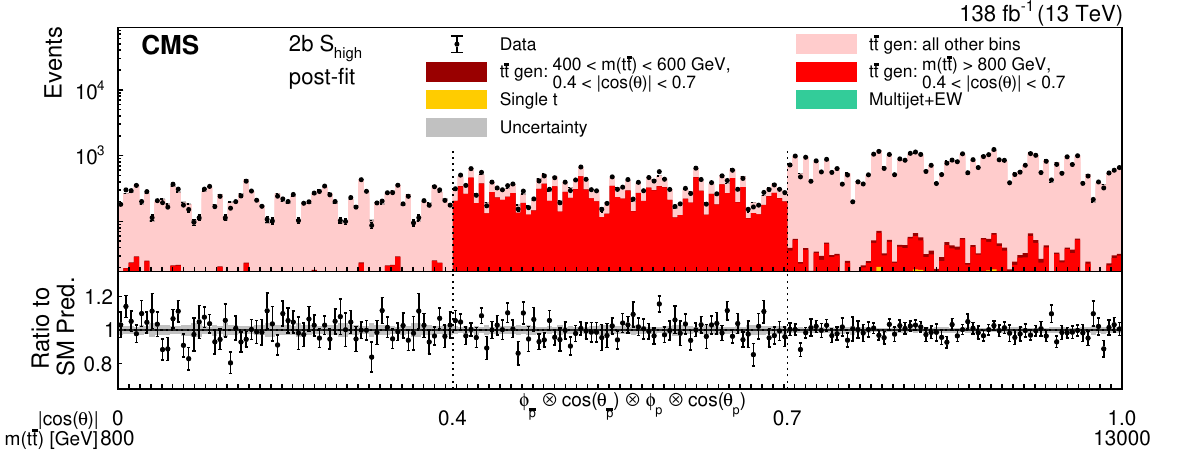}
\caption{Pre- and post-fit distributions comparing the data (points) to the \PWGPYT simulation (stacked histograms) for the full matrix measurement in bins of \ttmcth in the \twobhigh category. The $x$ axis shows the bins of the unrolled 4-dimensional distribution of \phipb, \costhetapb, \phip, and \costhetap, listed from the outermost to the innermost variable in each of the \ttmcth bins. The boundaries of the \absct and \ttm bins are labeled and indicated by dashed and solid lines, respectively. For the illustration of resolution effects, \ttbar events generated in two selected \ttmcth bins are shown in different shades of red. All other \ttbar contributions are shown in pink. A model without any polarization and spin correlation is shown as a blue line. The gray uncertainty band indicates the combined statistical and systematic uncertainties in the prediction. The vertical bars on the points show the statistical uncertainty. Ratios to the predicted yields are provided in the lower panels.}
\label{signalfit_fig_4b}
\end{figure*}

In Figs.~\ref{signalfit_fig_4a}--\ref{signalfit_fig_4b}, the pre- and post-fit distributions are shown for the full matrix measurement in the \twobhigh category. The pre-fit model uses the \PWGPYT predictions. In addition, a model without any polarization and spin correlation is shown as a blue line to demonstrate those effects. The \twobhigh category is shown here as an example since it is the category with the largest effects from polarization and spin correlation. The agreement between the data and the model is very similar in the other categories.

\begin{figure*}[!p]
\centering
\includegraphics[width=0.8\textwidth]{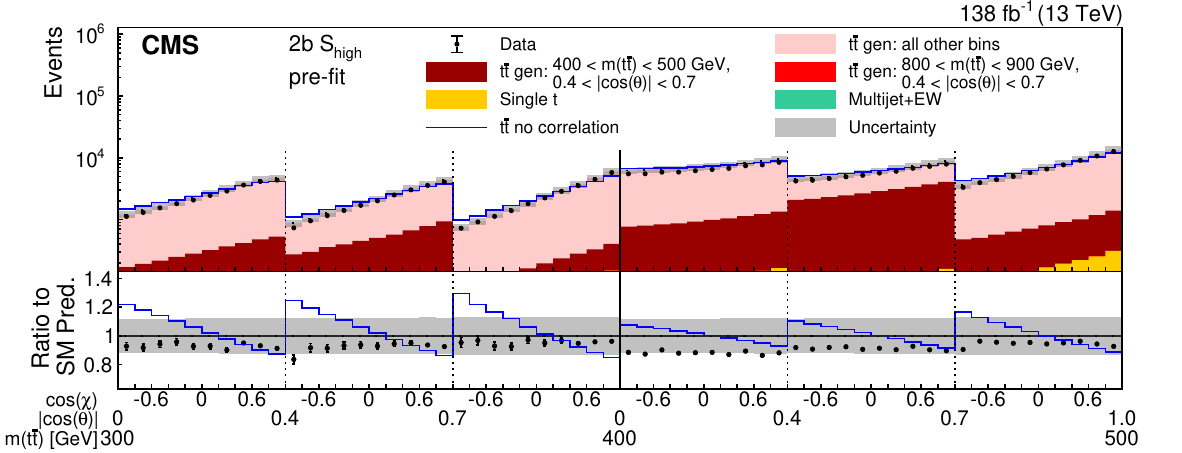}
\includegraphics[width=0.8\textwidth]{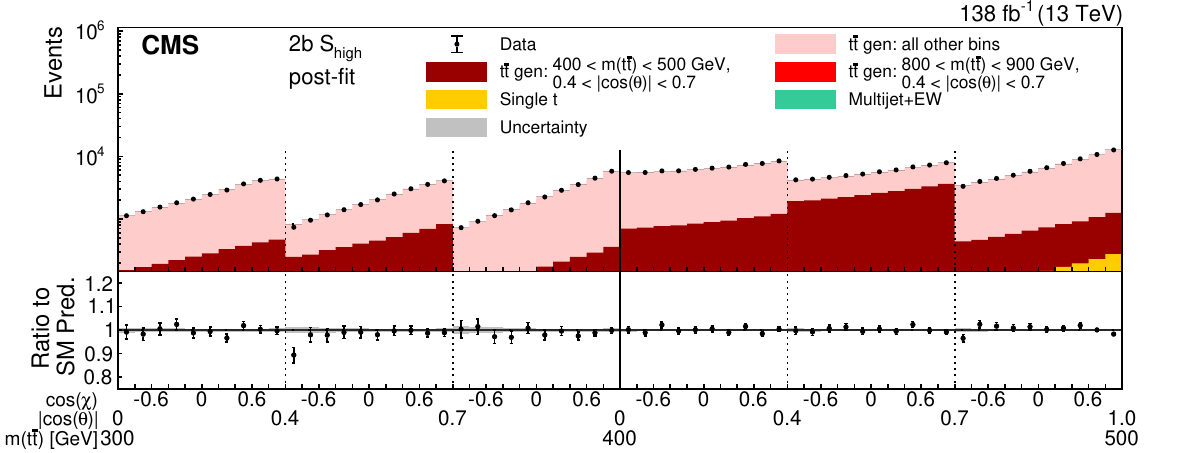}
\includegraphics[width=0.8\textwidth]{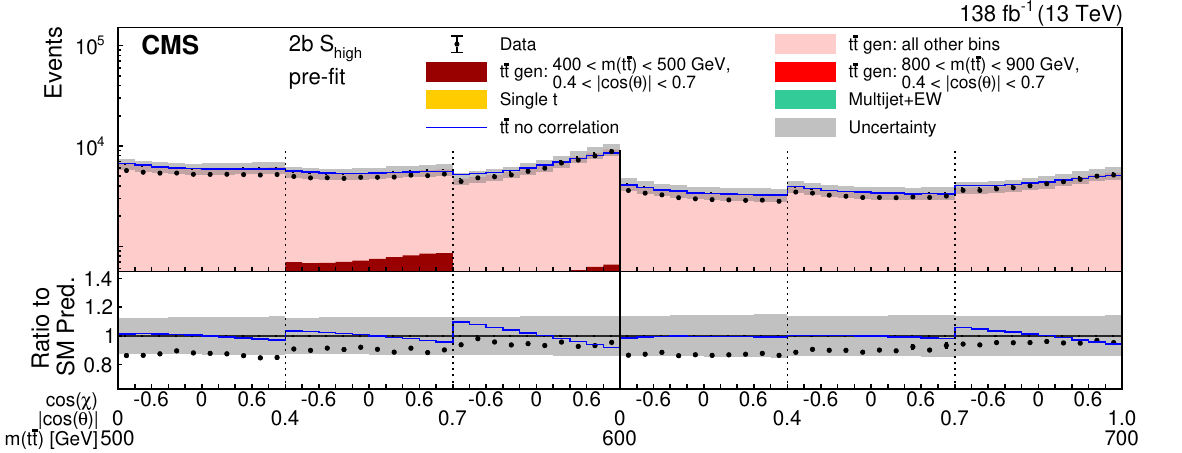}
\includegraphics[width=0.8\textwidth]{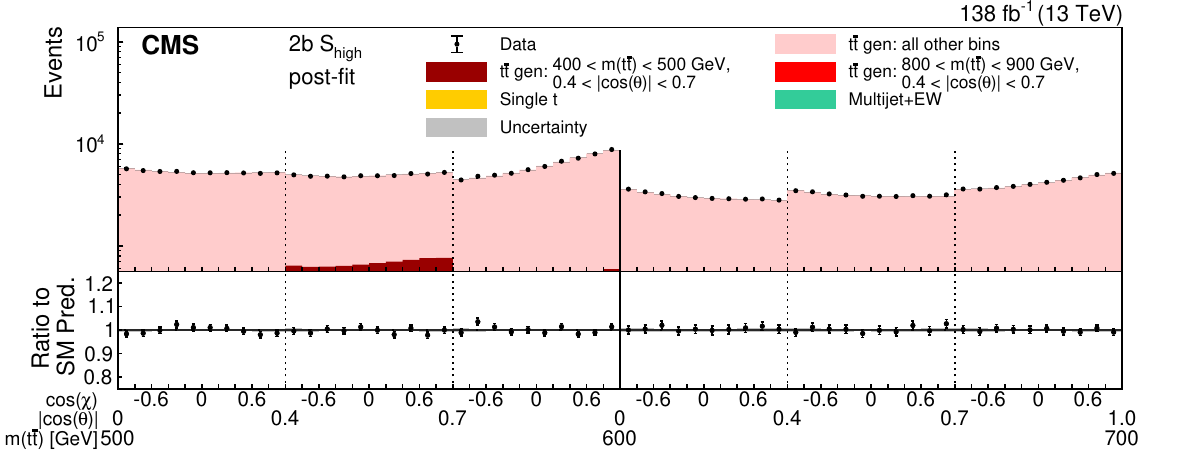}
\caption{Pre- and post-fit distributions of \coschi comparing the data (points) to the \PWGPYT simulation (stacked histograms) for the \Qone measurement in bins of \ttmcth in the \twobhigh category. The boundaries of the \absct and \ttm bins are labeled and indicated by dashed and solid lines, respectively. For the illustration of resolution effects, \ttbar events generated in two selected \ttmcth bins are shown in different shades of red. All other \ttbar contributions are shown in pink. A model without any polarization and spin correlation is shown as a blue line. The gray uncertainty band indicates the combined statistical and systematic uncertainties in the prediction. The vertical bars on the points show the statistical uncertainty. Ratios to the predicted yields are provided in the lower panels.}
\label{signalfit_fig_5a}
\end{figure*}

\begin{figure*}[!p]
\centering
\includegraphics[width=0.8\textwidth]{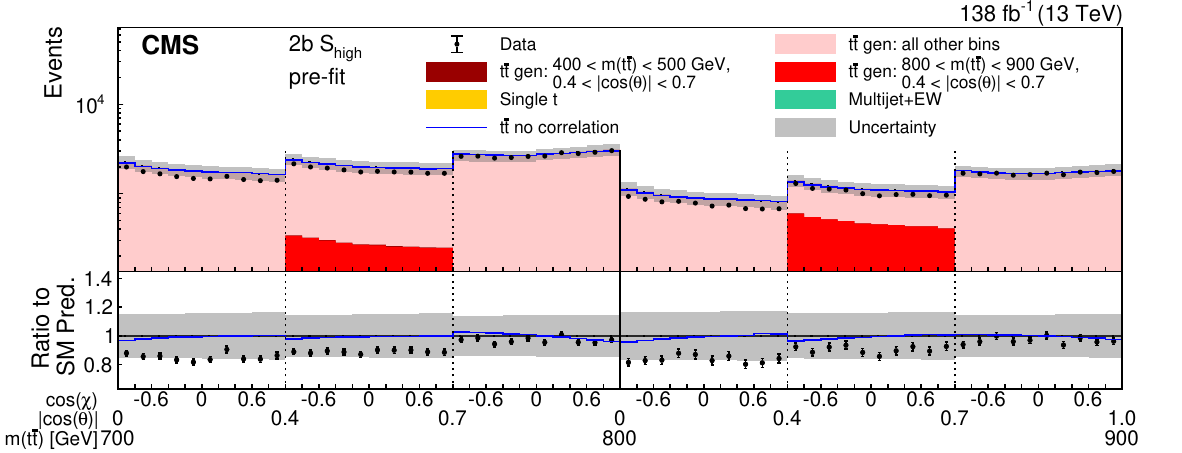}
\includegraphics[width=0.8\textwidth]{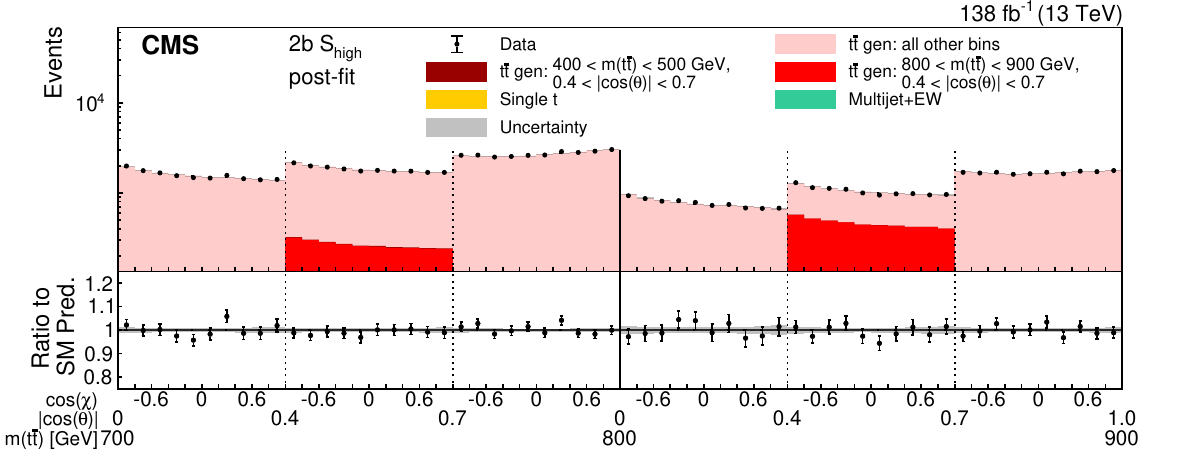}
\includegraphics[width=0.8\textwidth]{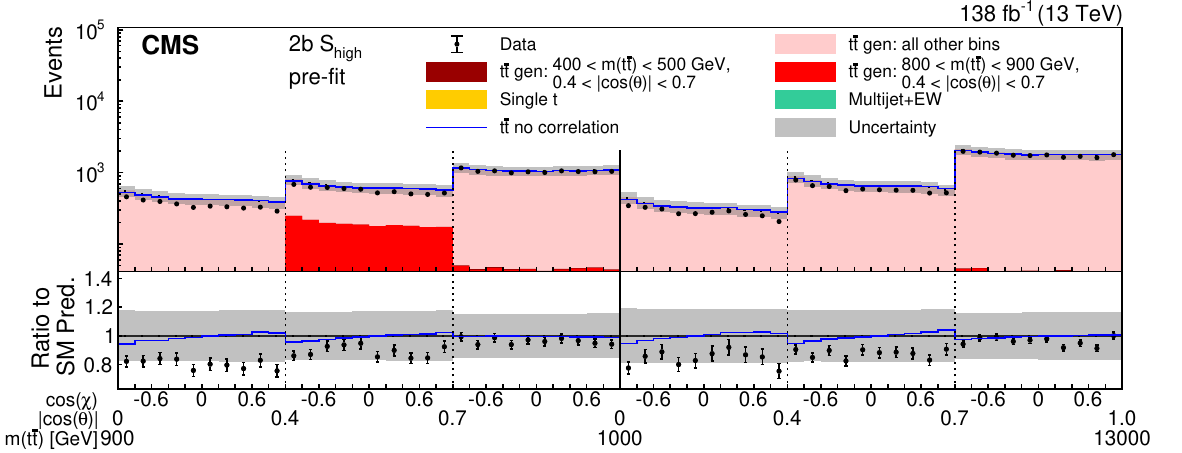}
\includegraphics[width=0.8\textwidth]{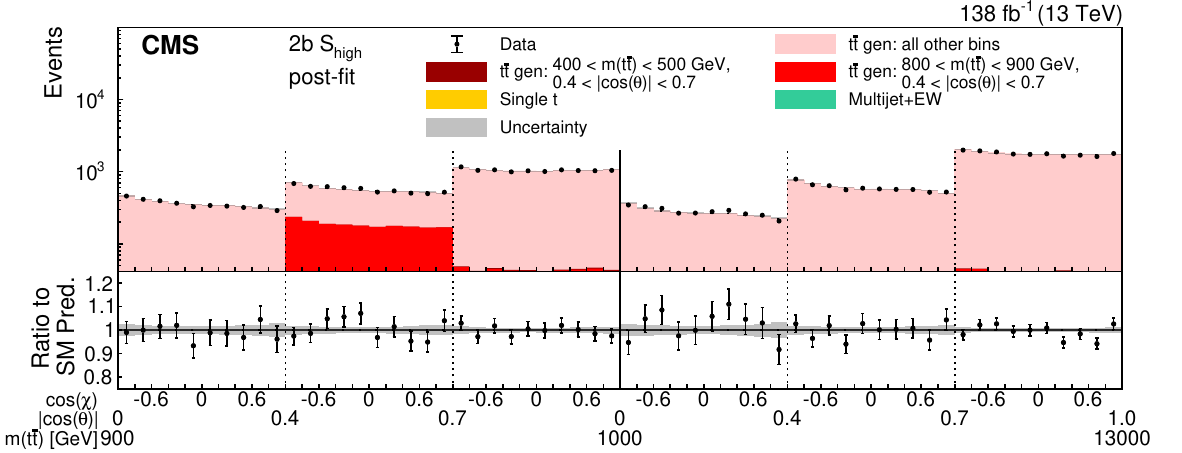}
\caption{Pre- and post-fit distributions of \coschi comparing the data (points) to the \PWGPYT simulation (stacked histograms) for the \Qone measurement in bins of \ttmcth in the \twobhigh category. The boundaries of the \absct and \ttm bins are labeled and indicated by dashed and solid lines, respectively. For the illustration of resolution effects, \ttbar events generated in two selected \ttmcth bins are shown in different shades of red. All other \ttbar contributions are shown in pink. A model without any polarization and spin correlation is shown as a blue line. The gray uncertainty band indicates the combined statistical and systematic uncertainties in the prediction. The vertical bars on the points show the statistical uncertainty. Ratios to the predicted yields are provided in the lower panels.}
\label{signalfit_fig_5b}
\end{figure*}

\begin{figure*}[!p]
\centering
\includegraphics[width=0.8\textwidth]{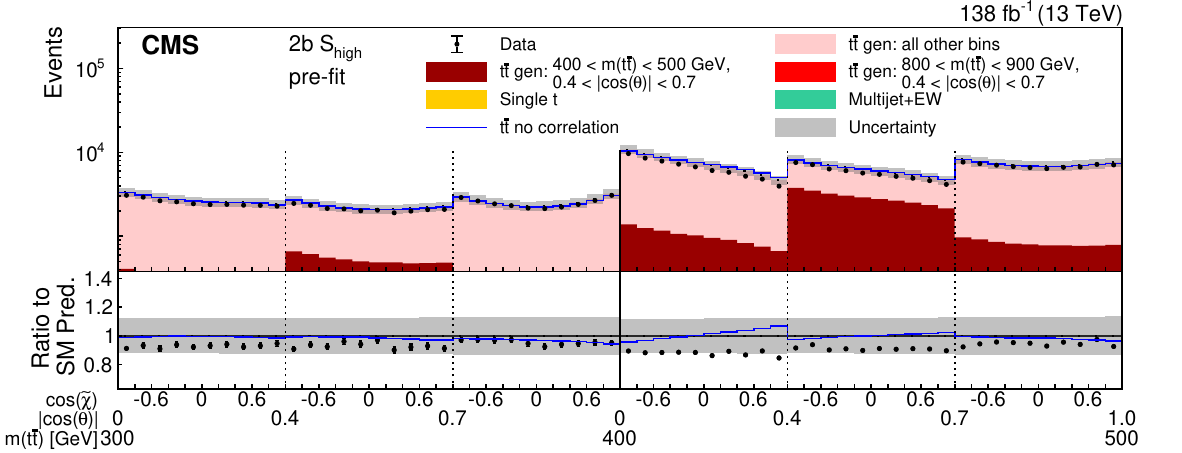}
\includegraphics[width=0.8\textwidth]{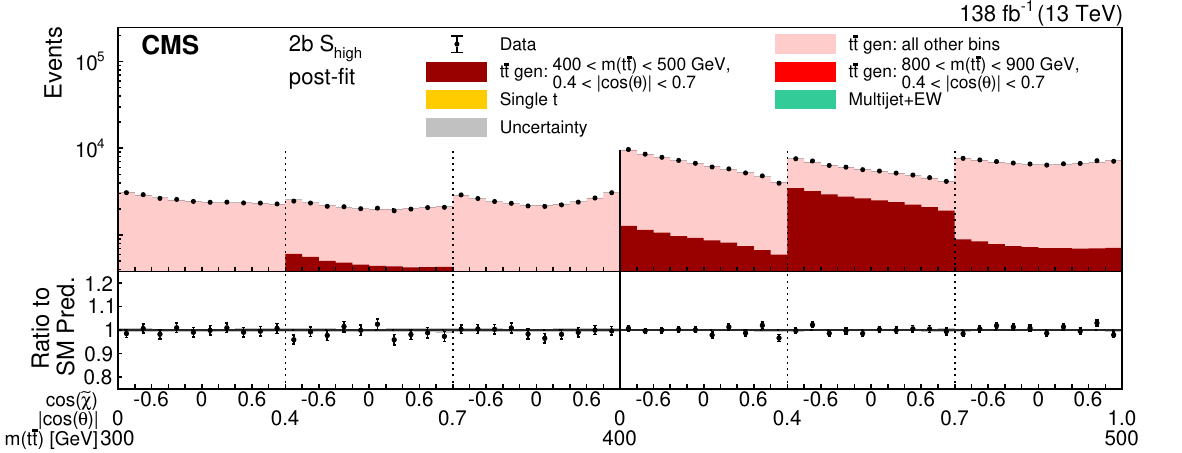}
\includegraphics[width=0.8\textwidth]{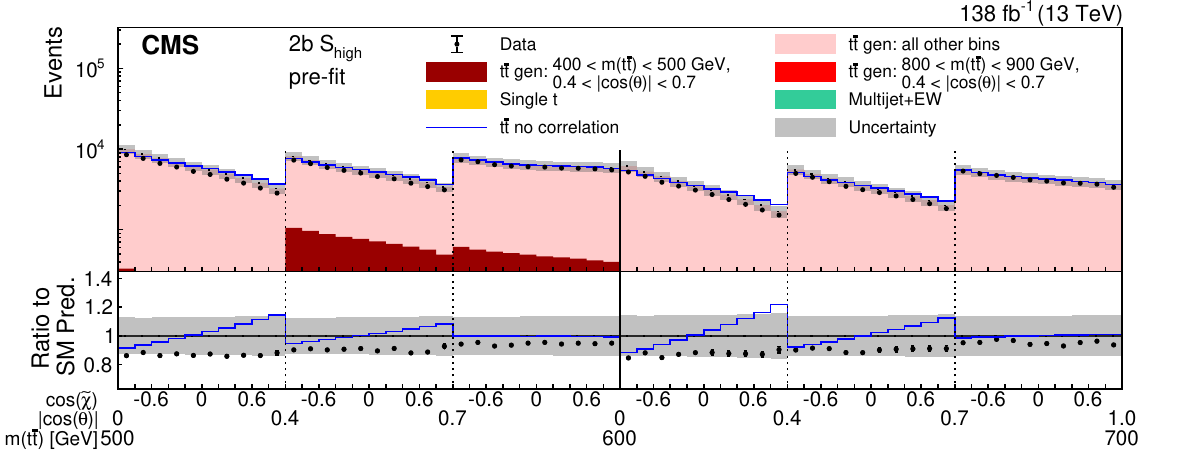}
\includegraphics[width=0.8\textwidth]{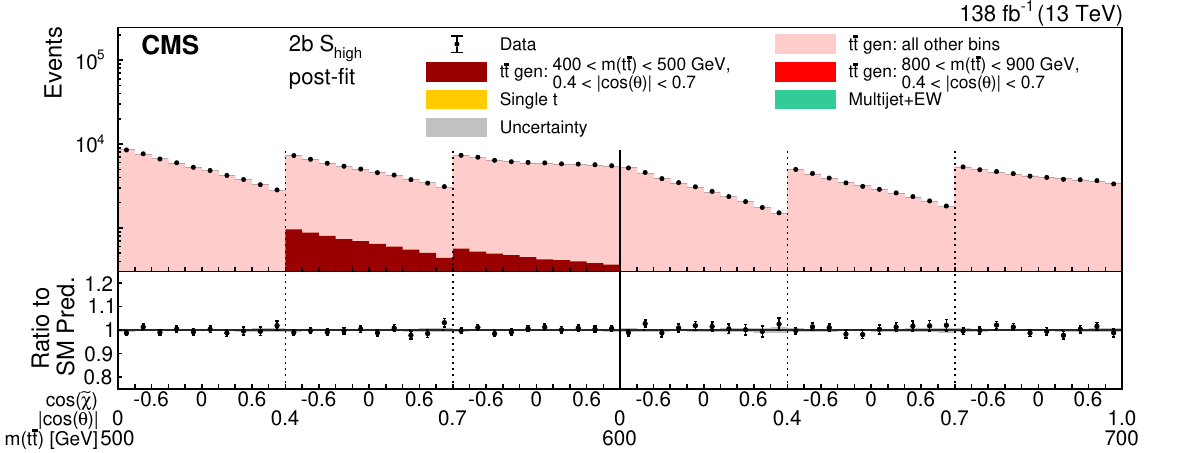}
\caption{Pre- and post-fit distributions of \coschit comparing the data (points) to the \PWGPYT simulation (stacked histograms) for the \Dtilde measurement in bins of \ttmcth in the \twobhigh category. The boundaries of the \absct and \ttm bins are labeled and indicated by dashed and solid lines, respectively. For the illustration of resolution effects, \ttbar events generated in two selected \ttmcth bins are shown in different shades of red. All other \ttbar contributions are shown in pink. A model without any polarization and spin correlation is shown as a blue line. The gray uncertainty band indicates the combined statistical and systematic uncertainties in the prediction. The vertical bars on the points show the statistical uncertainty. Ratios to the predicted yields are provided in the lower panels.}
\label{signalfit_fig_6a}
\end{figure*}

\begin{figure*}[!p]
\centering
\includegraphics[width=0.8\textwidth]{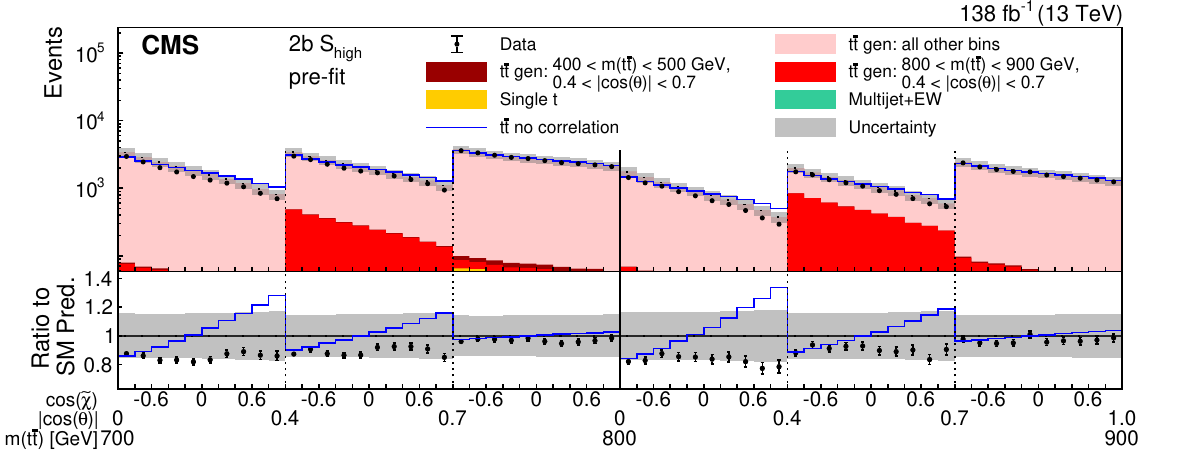}
\includegraphics[width=0.8\textwidth]{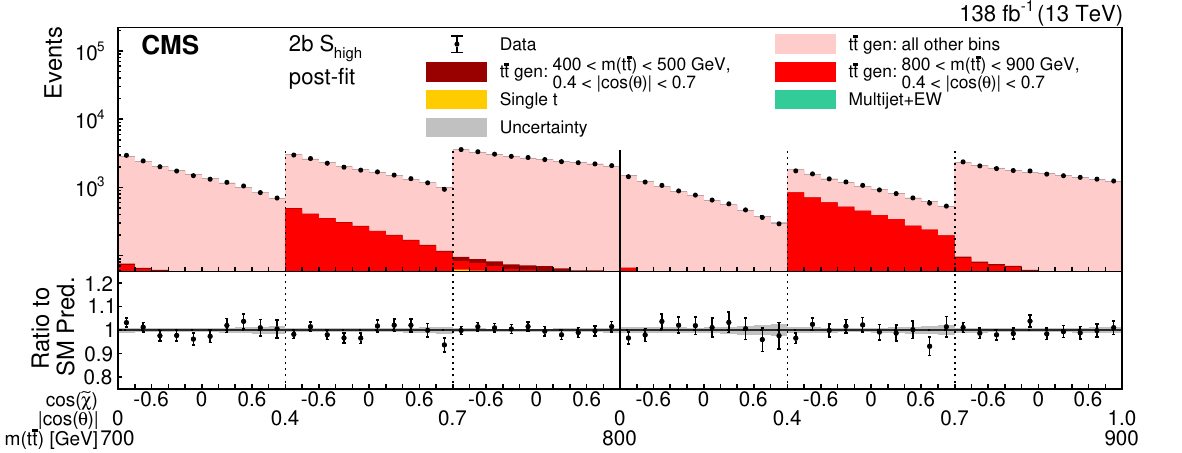}
\includegraphics[width=0.8\textwidth]{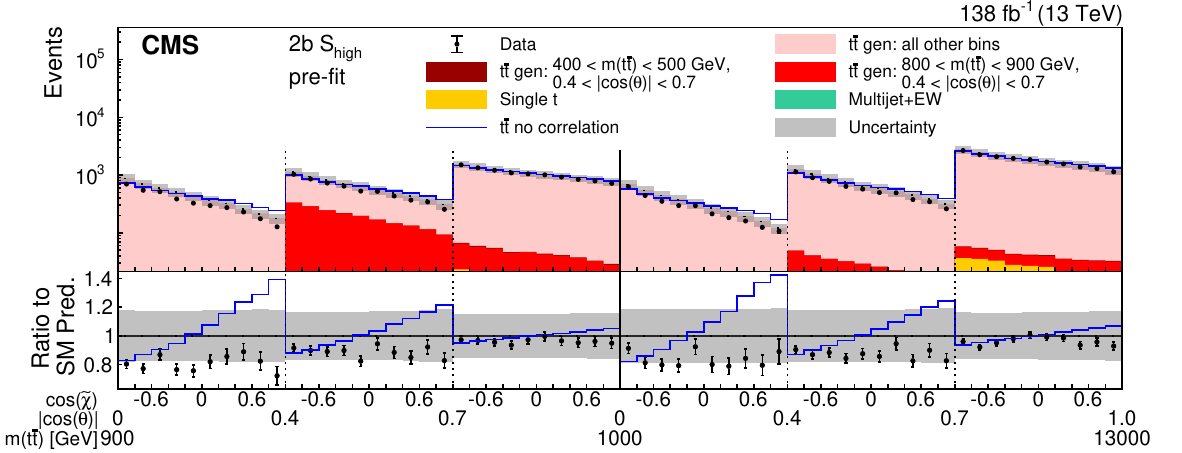}
\includegraphics[width=0.8\textwidth]{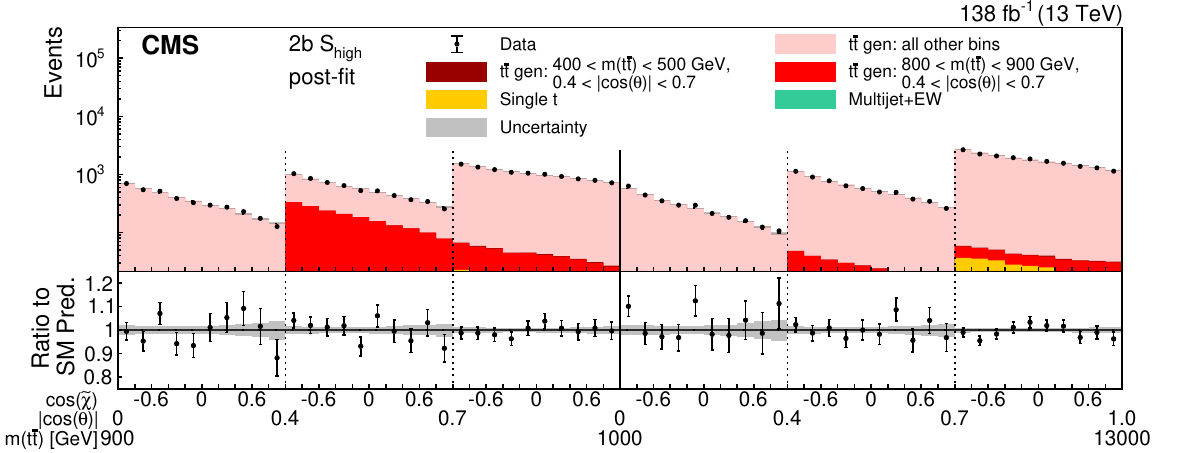}
\caption{Pre- and post-fit distributions of \coschit comparing the data (points) to the \PWGPYT simulation (stacked histograms) for the \Dtilde measurement in bins of \ttmcth in the \twobhigh category. The boundaries of the \absct and \ttm bins are labeled and indicated by dashed and solid lines, respectively. For the illustration of resolution effects, \ttbar events generated in two selected \ttmcth bins are shown in different shades of red. All other \ttbar contributions are shown in pink. A model without any polarization and spin correlation is shown as a blue line. The gray uncertainty band indicates the combined statistical and systematic uncertainties in the prediction. The vertical bars on the points show the statistical uncertainty. Ratios to the predicted yields are provided in the lower panels.}
\label{signalfit_fig_6b}
\end{figure*}

The same strategy is used to extract \Qonen (\Dtilden) directly. In this case, $j'$ represents a bin of \coschi (\coschit). The pre- and post-fit distributions in the \twobhigh category of the \Qone and \Dtilde measurements in bins of \ttmcth are presented in Figs.~\ref{signalfit_fig_5a}--\ref{signalfit_fig_6b}, respectively. The post-fit model describes the data well and no significant deviations are observed in any of the \ttmcth or \tptcth bins.

A possible bias in the measured \Qmn (\Qonen, \Dtilden) was estimated by performing fits on simulations with variations of the coefficients of up to $\pm0.3$, which exceeds the maximum observed difference between the expected and measured values. We found that any bias turns out to be negligible compared to the other uncertainties in the measurements. We also performed the fit using the \PWGHRW and \MGHPYT simulations as pseudo-data and found that we can extract the correct values of the coefficients.

\section{Systematic uncertainties}
\label{uncertainties}

Several theoretical and experimental systematic uncertainties affect the predicted event yields and are taken into account for the extraction of the \Qmn (\Qonen, \Dtilden). Their templates are obtained from alternative or reweighted simulations corresponding to variations in a specific uncertainty source, usually by one standard deviation. We take into account the following theoretical uncertainties:
\begin{itemize}
\item The effect of higher-order contributions to the ME calculation is estimated by varying the renormalization \mur and factorization \muf scales up and down by a factor of two. Distributions for these variations are obtained using event weights in the \PWGPYT simulation. The variations of \mur and \muf are parameterized in the fit by two independent nuisance parameters. The ME scales of \ttbar and single top quark production are treated separately.
\item  The difference in the \tpt spectrum between the \PWGPYT NLO and the NNLO calculations, obtained with \POWHEG \MINPYT exceeds what is  expected based on the \mur and \muf variations. Therefore, an additional uncertainty is introduced, whose $+1$ standard deviation variation corresponds to the reweighting of the NLO to the NNLO simulation using a NN-based method~\cite{mlreweighting, CMS:MLG-24-001}. The NN-based approach is used to reduce the statistical fluctuations in this uncertainty.
\item First order virtual electroweak corrections are calculated with \HATHOR and applied to LO \ttbar events obtained with \MGHPYT. The ratio to the LO simulation without electroweak corrections is determined as a function of \ttm and \absct. These ratios are used as event weights to correct the \POWHEG simulation. The differences between the electroweak corrected and the default \POWHEG simulation are taken into account as uncertainty.
\item The 100 Hessian variations of the NNPDF3.1 set plus the variation of strong coupling \alpS are taken into account as PDF uncertainties. For each of these variations a nuisance parameter is added. The distributions are obtained using the corresponding event weights. The PDF variations are correlated for \ttbar and single top quark production.
\item The uncertainties in the initial- and final-state parton showers are estimated by varying the scales for the different splittings $\Pg\to\qqbar$, $\Pg\to\gluglu$, $\PQq\to\PQq\Pg$, and $\PQb\to\PQb\Pg$ by a factor of two, resulting in a total of eight independent variations. The corresponding distributions are obtained using event weights. The parton shower scale variations are correlated for \ttbar and single top quark production.
\item The scale that separates the phase space of the first QCD emission into soft and hard parts is controlled by the $h_{\text{damp}}$ parameter for \POWHEG simulations. The values used for the CP5 tune are $1.38^{+0.92}_{-0.51}\,\Mtop$. Separate samples produced with the different values of $h_{\text{damp}}$ are used to obtain the corresponding distributions. To reduce the impact from statistical fluctuations we employ the NN based approach~\cite{mlreweighting} to determine the weights applied to the central simulation based on the kinematic properties of top quarks determined at the generator level.
\item To estimate the effect of the uncertainty in \Mtop, a variation of 0.5\GeV~\cite{CMS:2015lbj} is taken into account. For the evaluation of the expected event yields, we use the \Mtop-dependent \ttbar production cross sections of 843 (820)\unit{pb} for $\Mtop=172.0$ (173.0)\GeV~\cite{mitov}.
\item The uncertainties in the underlying event modeling are estimated using separate samples that represent an envelope of the uncertainties in the \PYTHIA CP5 tune~\cite{CP5}.
\item The fraction of leptonically decaying \PQb hadrons is changed according to the known uncertainty in the branching fraction using event-based reweighting~\cite{ParticleDataGroup:2024}.
\item The uncertainty in the color reconnection is assessed using an alternative model where the reconnection of colored particles from resonant decays is activated in \PYTHIA, while this is deactivated in the default tune. Other variations use the gluon move and the QCD-inspired models~\cite{Argyropoulos:2014zoa, Christiansen:2015yqa}. The differences between these three and the default samples are added as symmetric uncertainties.
{\tolerance=1600
\item At the \ttbar production threshold, theoretical calculations based on nonrelativistic QCD~\cite{Ju:2020otc} predict \ttbar bound states and other effects not included in the \PWGPYT simulation. To estimate their effects on the measurement, we mimic the theoretical calculation by adding a pseudoscalar particle \etat with a mass of 343\GeV and a width of twice the top quark width. It is produced in \gluglu fusion and decays as $\etat\to\PW\PQb\PW\PAQb$. This is calculated using \MGHPYT and normalized using the cross section of 6.43\unit{pb} from Ref.~\cite{Fuks:2021xje}. The difference observed using this model is used to estimate the uncertainty due to bound-state effects.
\par}
\end{itemize}
All theoretical uncertainties affect all data-taking periods in the same way and the corresponding nuisance parameters are fully correlated between them.

We take into account the following experimental uncertainties:
\begin{itemize}
\item The integrated luminosities for the 2016, 2017, and 2018 data-taking years have 1.2--2.5\% individual uncertainties~\cite{CMS-LUM-17-003,CMS-PAS-LUM-17-004,CMS-PAS-LUM-18-002}, while the overall uncertainty for the 2016--2018 period is 1.6\%.
\item The prediction of the number of pileup interactions in simulation is assuming a total inelastic proton-proton cross section of 69.2\unit{mb}~\cite{CMS:2018mlc}. Changes in the simulated pileup multiplicity are estimated by varying the total inelastic cross section by $\pm4.6\%$. Templates with enhanced and reduced pileup are obtained by applying event weights. This uncertainty is treated as fully correlated between the data-taking periods.
\item The jet energy scale uncertainties are split into 19 different sources~\cite{JET}. The combined uncertainties are \pt- and $\eta$-dependent, with a magnitude that varies between 0.3 and 1.8\% for the relevant jets. In addition, variations are applied depending on the true generated type to \PQb jets, \PQc jets, $\PQu\PQd\PQs$ jets, and gluon jets. The correlations among the years are evaluated for each source. The differences in the distributions are obtained by rescaling the jet momenta in the simulation.
\item Separate uncertainties for the jet energy resolution are taken into account for jets in the endcap and barrel regions by varying the resolution corrections within their uncertainties. These uncertainties are uncorrelated among the years.
\item The dominant uncertainty in the \ptmiss is due to the jet energy calibration. Therefore, the \ptmiss is also recalculated whenever the jet momenta are rescaled. An additional contribution to the uncertainty due to particles that do not belong to the selected jets is estimated~\cite{CMS:2019ctu}. This uncertainty is uncorrelated among years.
\item Uncertainties in the electron and muon reconstruction and trigger efficiencies are determined~\cite{TNP}. For each flavor a statistical and systematic uncertainty in the derived scale factors are taken into account, where the statistical component is uncorrelated and the systematic component is correlated among the years. In addition, an overall normalization uncertainty of 0.5\% is used to account for the differences in DY and \ttbar events.
\item Since the analysis uses three \PQb tagging categories as input to the NN, we allow for separate variations of the uncertainties in the scale factors for the tight, medium, and loose \PQb tagging selections~\cite{Bols_2020, CMS-DP-2023-005}. The variations are performed by recalculating an event probability using all jets and their true type. The uncertainties in the correction factors for \PQb and \PQc jets are split into several sources such as statistical, jet energy correction, and pileup uncertainties. The statistical uncertainty is uncorrelated among the years, while the rest is treated as correlated. The uncertainties in the correction factors for the light jet flavors are split into a correlated and an uncorrelated component.
\item During the data-taking, a gradual shift in the timing of the inputs of the ECAL L1 trigger in the region $\abs{\eta}>2.0$ and of the muon trigger caused a specific trigger inefficiency. Correction factors as functions of \pt, $\eta$, and time are computed from data and applied to the simulation. The statistical uncertainties in these correction factors are taken into account.
\item The uncertainties in the background estimations are detailed in Section~\ref{background}.
\end{itemize}
For each bin, an additional nuisance parameter is added~\cite{BARLOW1993219} whose variation corresponds to the statistical uncertainty in the central templates. It is known that statistical fluctuations are also important for the systematic variations. In particular, if the variations are evaluated based on statistically independent simulations, the statistical effects can easily reach or even exceed the magnitude of the systematic effect. Therefore, it is often helpful to require a certain smoothness of the relative systematic effects. This reduces unphysical constraints of the related nuisance parameters. A 6(3)-dimensional smoothing~\cite{LOWESS} is applied for the full matrix (\Qone and \Dtilde) measurements.

\section{Results}
\label{results}

From the fits to the data we obtain the values of \Qmn (\Qonen, \Dtilden) in bins of \ttmcth or \tptcth. In the following, we concentrate on regions of the phase space that are of special interest, \eg, where a higher level of entanglement is expected. Most of the presented results are obtained from the combination of several \ttmcth or \tptcth bins.

With the \ttbar event yields of the post-fit model at the generator level $Y_n$, and the normalization factors \an, we obtain the total fitted event yields $K_n=Y_n\an$. The result in a combined bin $g$ is then obtained by averaging the measurements from bins $\{n\}$ using
\begin{equation}
    \hat{Q}_{mg}=\frac{1}{\sum\limits_{n\in g}K_n}\,\sum_{n\in g}K_n\Qmn.
\end{equation}
The uncertainties in $\hat{Q}_{mg}$ are calculated using error propagation taking into account the uncertainties and their correlations in the event yields and the \Qmn as obtained from the fit.
The combined normalization factors are obtained based on the $K_n$ sum
\begin{equation}
    c_g=\frac{1}{\sum\limits_{n\in g}Y^\prime_n}\,\sum_{n\in g}K_n,
\end{equation}
where $Y^\prime_n$ are the \ttbar yields of the pre-fit model at the generator level.

Following Refs.~\cite{Bernreuther:2015yna, CMS:2019spincorrdilep}, we use the new observables $C^\pm_{nr}=C_{nr}\pm C_{rn}$, $C^\pm_{rk}=C_{rk}\pm C_{kr}$, and $C^\pm_{nk}=C_{nk}\pm C_{kn}$, where for the calculation the covariances are taken into account. These are either even or odd under parity ($P$) and charge-parity ($CP$) transformations.

\begin{figure*}[tp!]
\centering
\includegraphics[width=\textwidth]{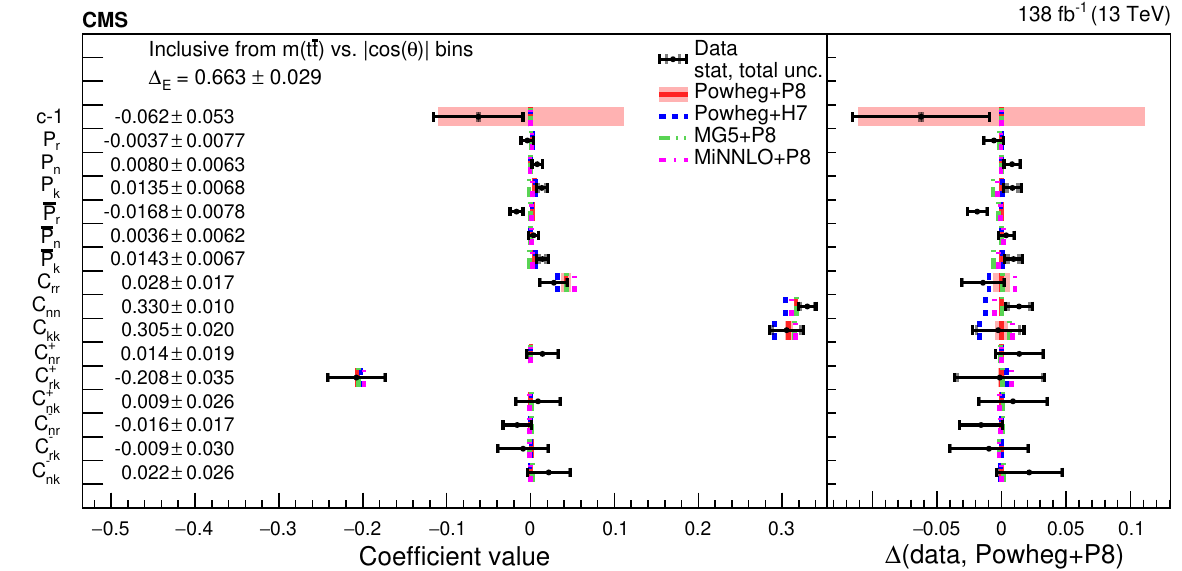}
\includegraphics[width=\textwidth]{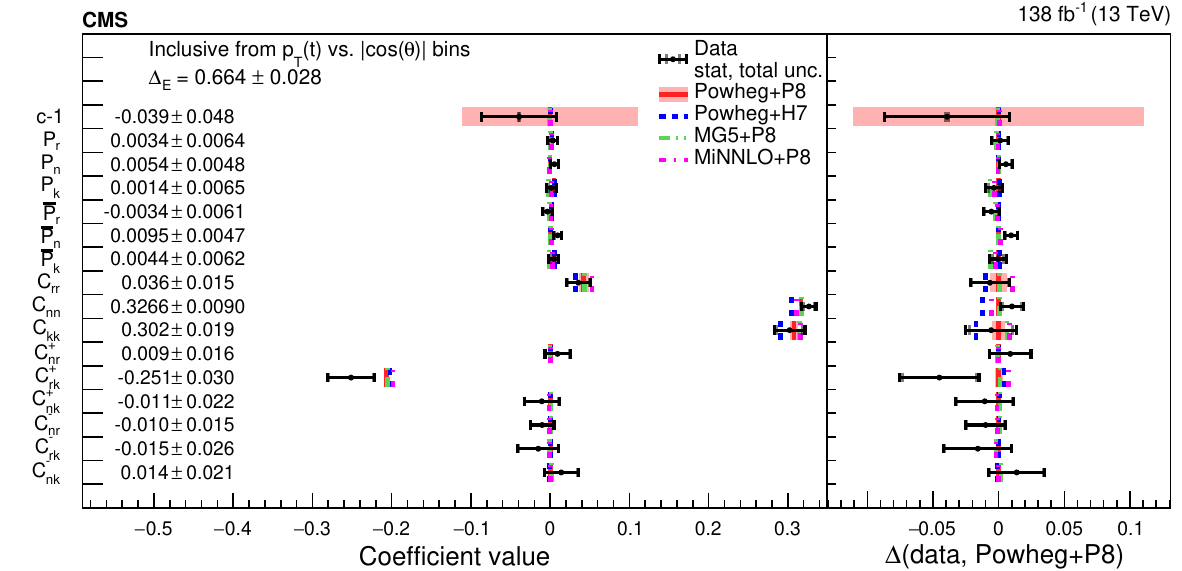}
\caption{Results of the inclusive full matrix measurement obtained by combining the bins of the \ttmcth (upper) and \tptcth (lower) measurements. The measurements (markers) are shown with the statistical uncertainty (inner error bars) and total uncertainty (outer error bars) and compared to the predictions of \PWGPYT, \PWGHRW, \MGHPYT and \MINPYT. In the right panels, results are presented with the \PWGPYT predictions subtracted. The \PWGPYT prediction is displayed with ME scale and PDF uncertainties. The values of \DE are displayed for each measurement.}
\label{results_fig_1}
\end{figure*}

The inclusive full matrix measurements based on the \ttmcth and \tptcth fits are obtained from the combination of all bins in the additional observables, and are shown in Fig.~\ref{results_fig_1}. The displayed values of \DE are calculated following Eq.~\eqref{method_eq3}.

\begin{figure*}[tp!]
\centering
\includegraphics[width=\textwidth]{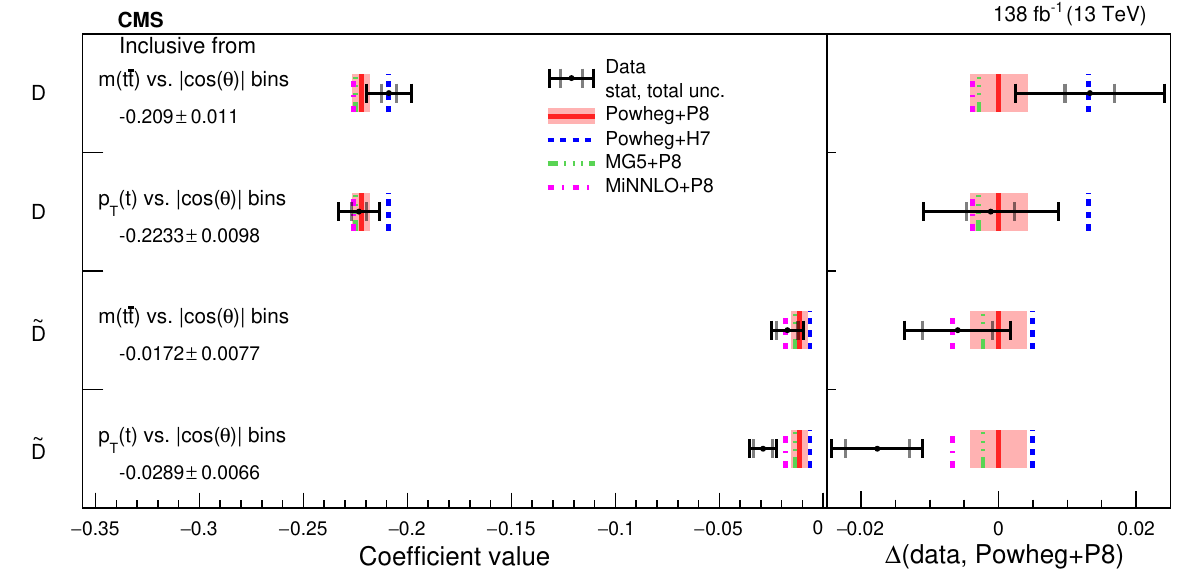}
\caption{Results of the inclusive \Qone and \Dtilde measurement obtained by combining the bins of the \ttmcth and \tptcth measurements. The measurements (markers) are shown with the statistical uncertainty (inner error bars) and total uncertainty (outer error bars) and compared to the predictions of \PWGPYT, \PWGHRW, \MGHPYT and \MINPYT. In the right panel, results are presented with the \PWGPYT predictions subtracted. The \PWGPYT prediction is displayed with ME scale and PDF uncertainties.}
\label{results_fig_1_1}
\end{figure*}

{\tolerance=800
As expected, both binnings lead to consistent results, where the \tptcth binning has a higher expected and observed precision. For the inclusive coefficients, the values predicted by \PWGPYT, \PWGHRW, \MGHPYT, and \MINPYT are similar. The measured coefficients are in agreement with the predictions and consistent with the previous measurement in the \ttbar dilepton channel~\cite{CMS:2019spincorrdilep}. The measured polarizations are all compatible with zero. Only the diagonal elements of $C$ differ from zero with the exception of $C^+_{rk}$, which is the only even coefficient under $P$ and $CP$ transformation.  In Fig.~\ref{results_fig_1_1} the results of the inclusive \Qone and \Dtilde measurements obtained with the \ttmcth and \tptcth binnings are shown.
\par}

\begin{figure*}[htp!]
\centering
\includegraphics[width=\textwidth]{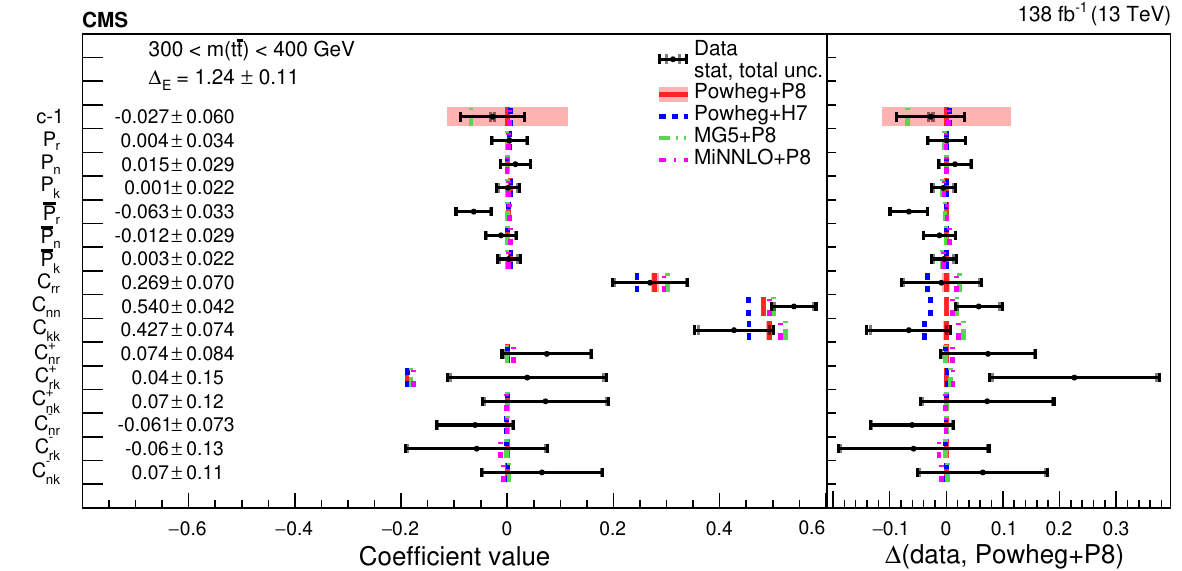}
\includegraphics[width=\textwidth]{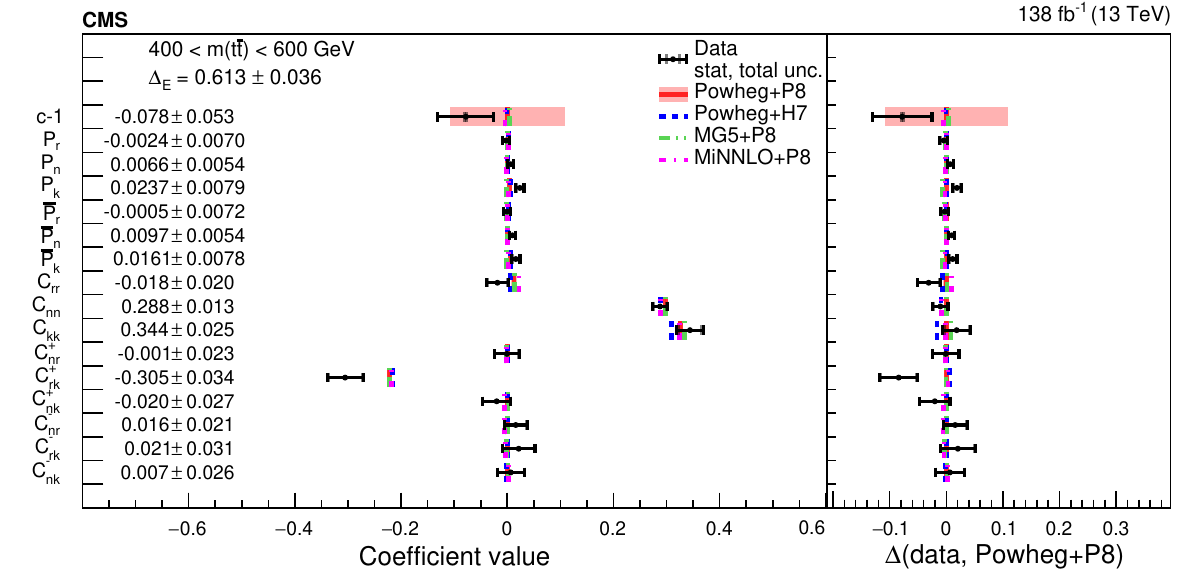}
\caption{Results of the full matrix measurement in bins of \ttm. The measurements (markers) are shown with the statistical uncertainty (inner error bars) and total uncertainty (outer error bars) and compared to the predictions of \PWGPYT, \PWGHRW, \MGHPYT and \MINPYT. In the right panels, results are presented with the \PWGPYT predictions subtracted. The \PWGPYT prediction is displayed with ME scale and PDF uncertainties. The values of \DE are displayed for each measurement.}
\label{results_fig_4}
\end{figure*}

\begin{figure*}[htp!]
\centering
\includegraphics[width=\textwidth]{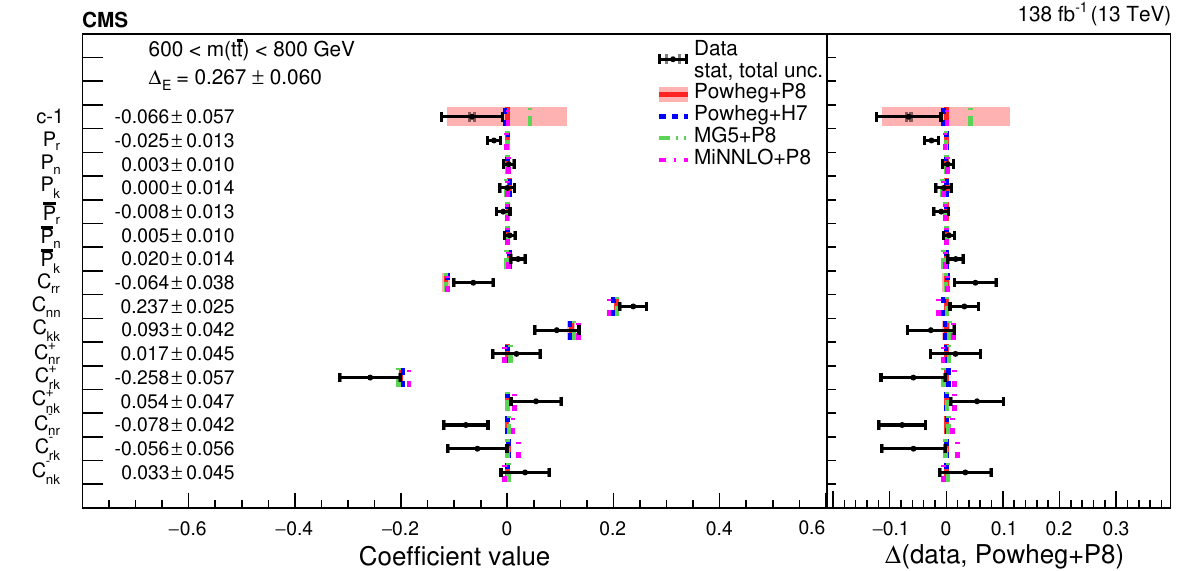}
\includegraphics[width=\textwidth]{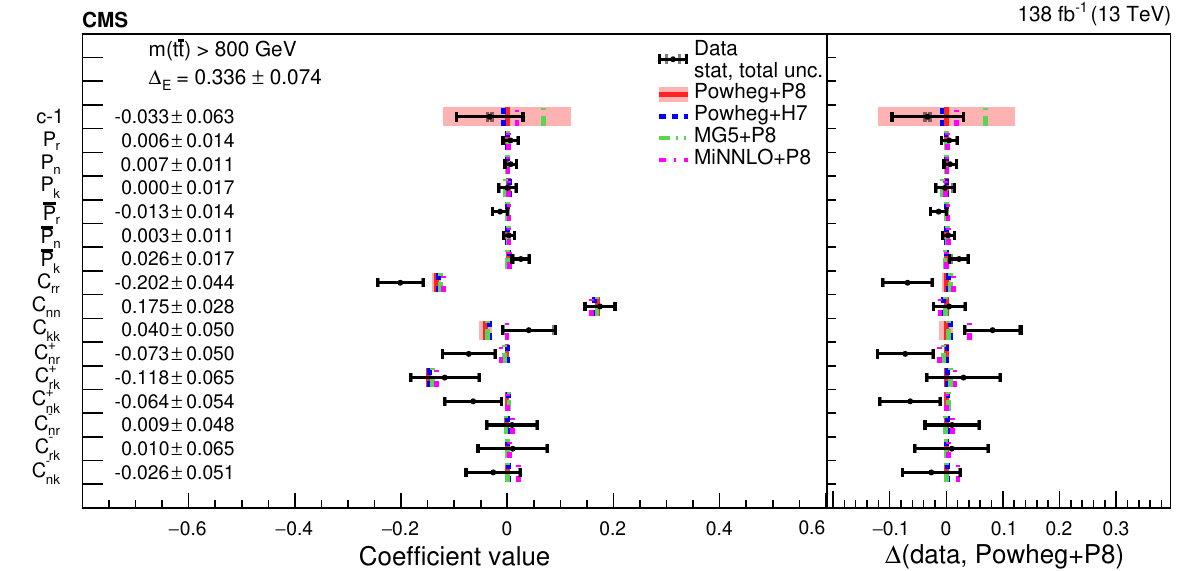}
\caption{Results of the full matrix measurement in bins of \ttm. The measurements (markers) are shown with the statistical uncertainty (inner error bars) and total uncertainty (outer error bars) and compared to the predictions of \PWGPYT, \PWGHRW, \MGHPYT and \MINPYT. In the right panels, results are presented with the \PWGPYT predictions subtracted. The \PWGPYT prediction is displayed with ME scale and PDF uncertainties. The values of \DE are displayed for each measurement.}
\label{results_fig_4_2}
\end{figure*}

\begin{figure*}[htp!]
\centering
\includegraphics[width=\textwidth]{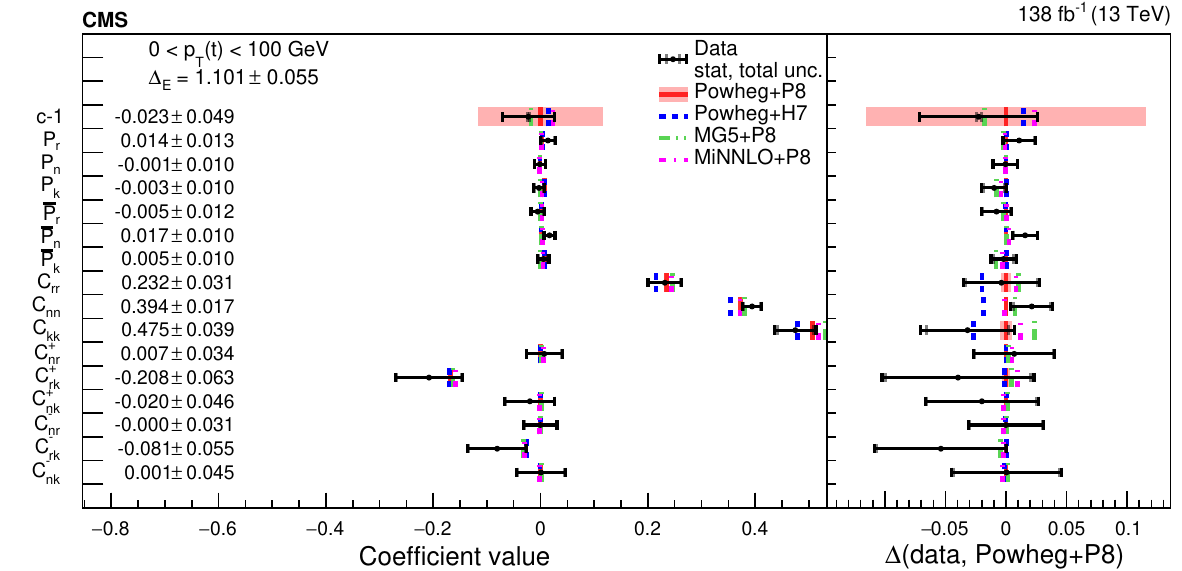}
\includegraphics[width=\textwidth]{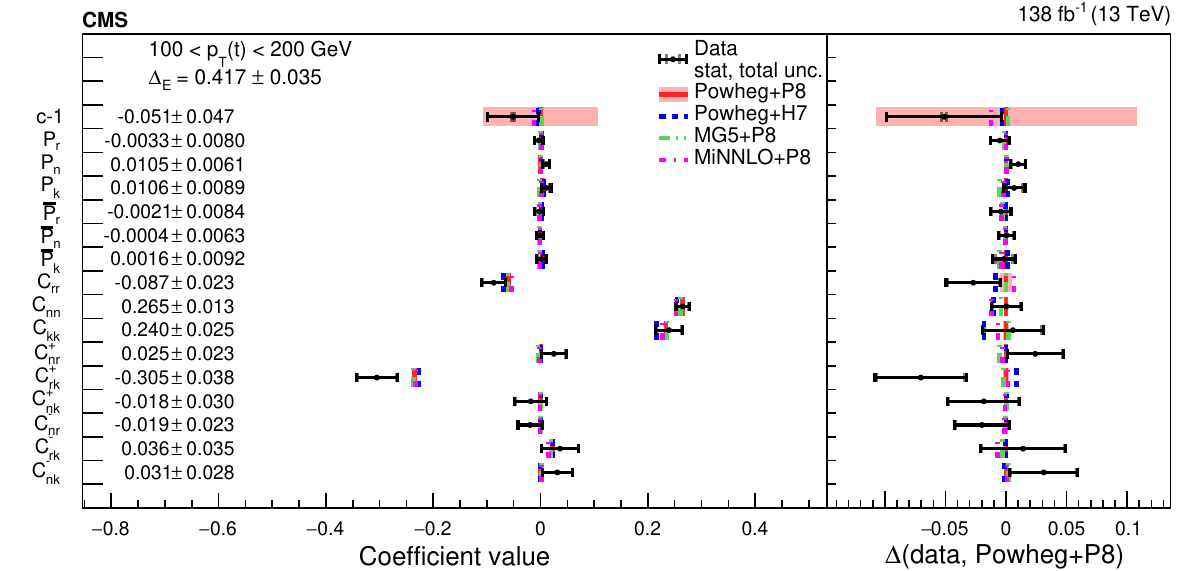}
\caption{Results of the full matrix measurement in bins of \tpt. The measurements (markers) are shown with the statistical uncertainty (inner error bars) and total uncertainty (outer error bars) and compared to the predictions of \PWGPYT, \PWGHRW, \MGHPYT and \MINPYT. In the right panels, results are presented with the \PWGPYT predictions subtracted. The \PWGPYT prediction is displayed with ME scale and PDF uncertainties. The values of \DE are displayed for each measurement.}
\label{results_fig_5_0}
\end{figure*}

\begin{figure*}[htp!]
\centering
\includegraphics[width=\textwidth]{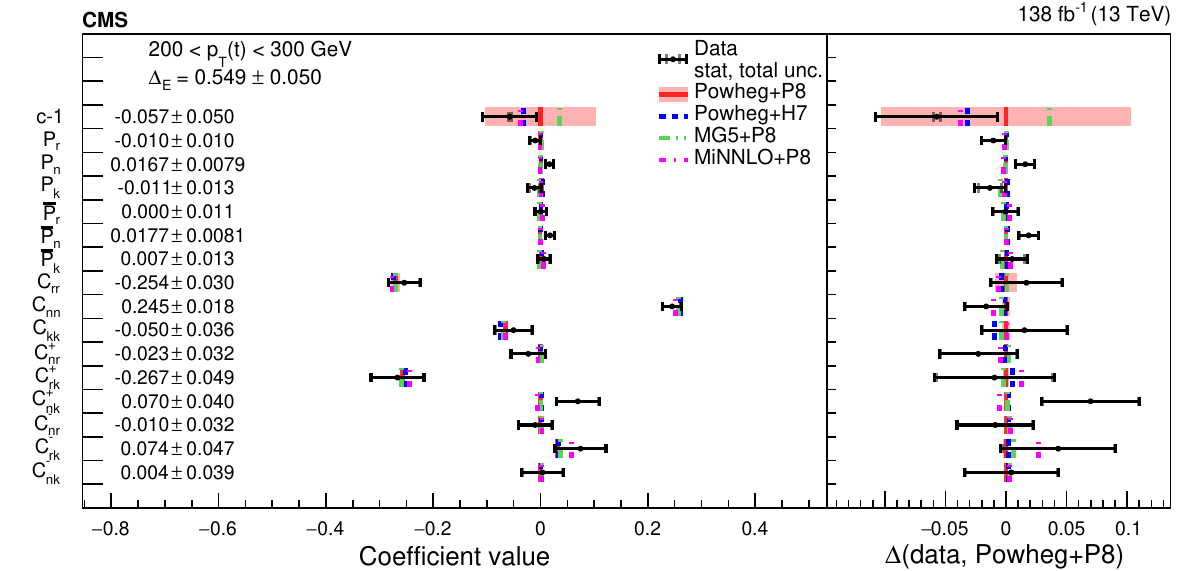}
\includegraphics[width=\textwidth]{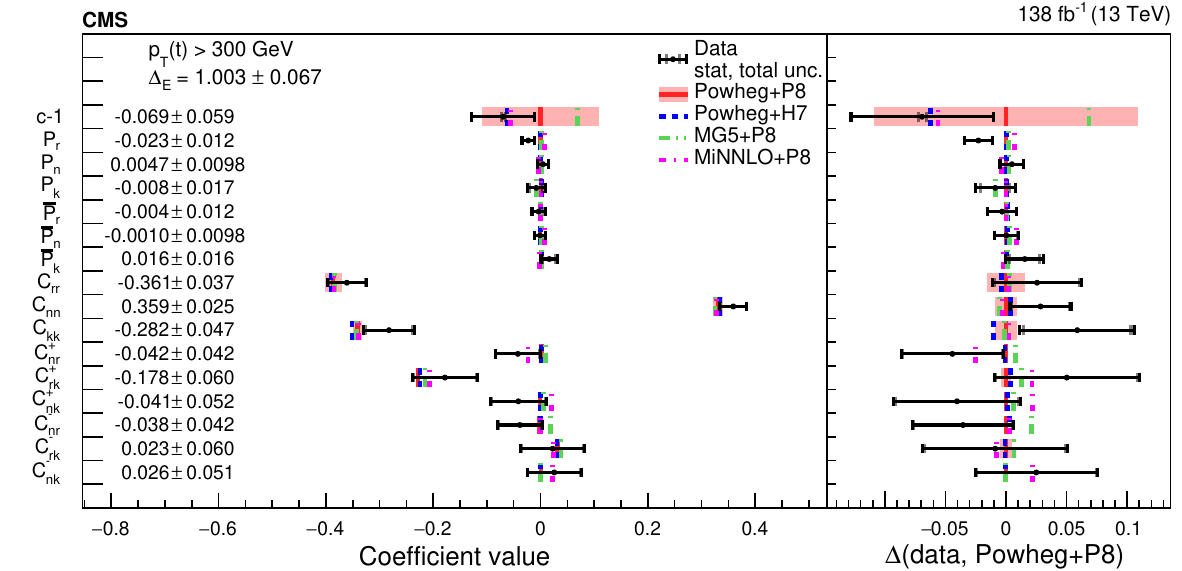}
\caption{Results of the full matrix measurement in bins of \tpt. The measurements (markers) are shown with the statistical uncertainty (inner error bars) and total uncertainty (outer error bars) and compared to the predictions of \PWGPYT, \PWGHRW, \MGHPYT and \MINPYT. In the right panels, results are presented with the \PWGPYT predictions subtracted. The \PWGPYT prediction is displayed with ME scale and PDF uncertainties. The values of \DE are displayed for each measurement.}
\label{results_fig_5}
\end{figure*}

Entangled quantum states are expected at the threshold of \ttbar production and also at high \ttm and \tpt in events with low \absct~\cite{Dong:2023xiw}. Therefore, taking advantage of the binning in the additional observables, we can obtain the spin correlation coefficients and thus test the entanglement criterion in several \ttm and \tpt regions and its dependence on \absct. Figures~\ref{results_fig_4}--\ref{results_fig_5} provide the results of the full matrix measurements including the \DE values in bins of \ttm and \tpt, \ie, we combine the \absct bins in each of these regions. While \PWGHRW predicts a slightly smaller spin correlation than the other simulations, the measured coefficients are compatible with all predictions. With these measurements, the differences in the spin correlation for various kinematic regions become clearly visible. In particular, for the measurement in \tpt, we observe the signs of \Crr and \Ckk changing from positive to negative with increasing \tpt, indicating the transition from the spin-singlet to the spin-triplet as the dominant state.

At the \ttbar production threshold, the most significant results for entanglement using the full matrix measurement, based on the criterion $\DE>1$, are obtained for $\ttm<400\GeV$ and for $\tpt<100\GeV$. We evaluate the significance of the deviation from the separable state hypothesis with $\DE=1$ based on the uncertainties in the measured values of \DE. However, the observed significance for entanglement does not exceed 2 standard deviations.

\begin{figure*}[htp!]
\centering
\includegraphics[width=0.48\textwidth]{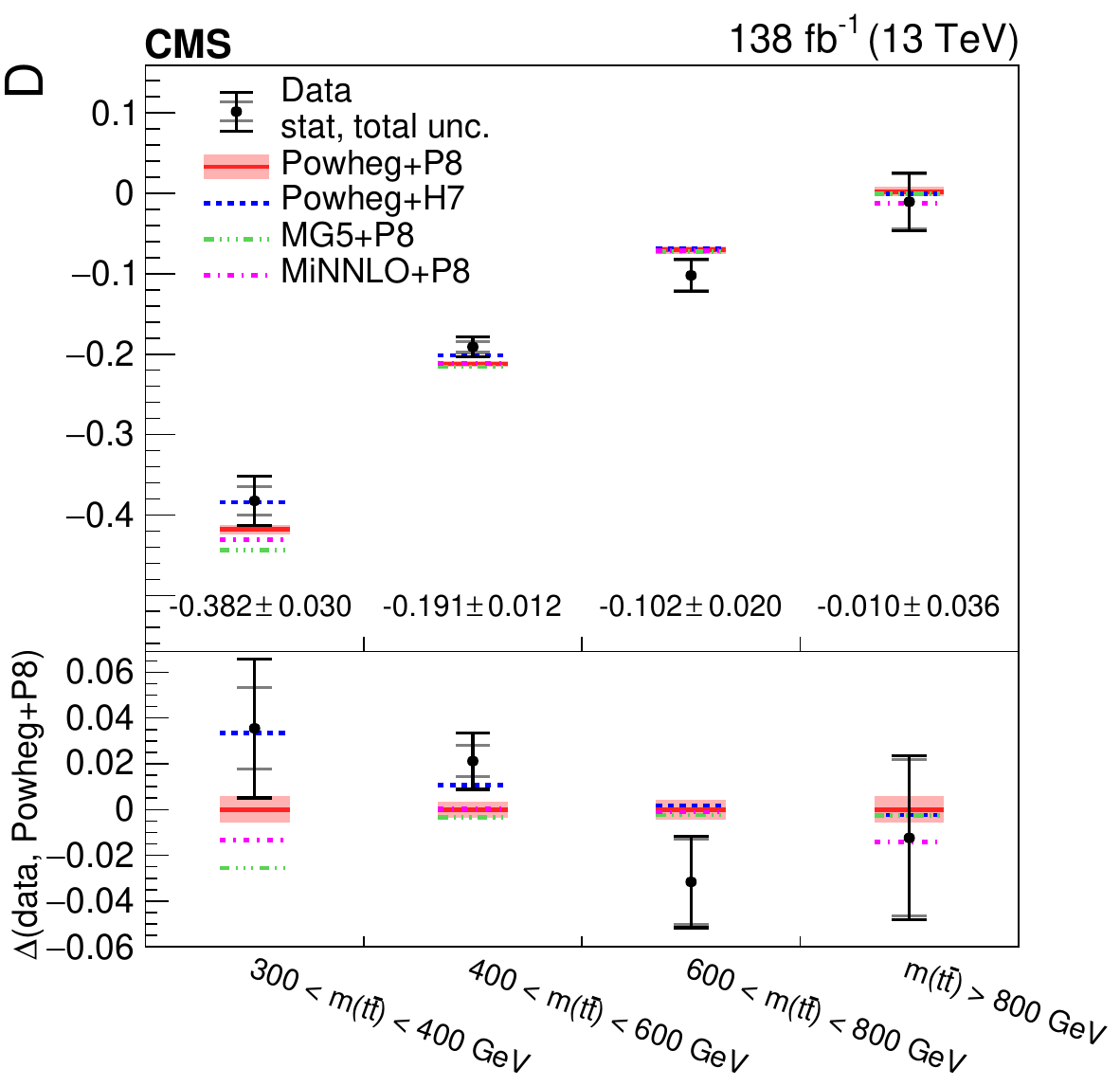}%
\hfill%
\includegraphics[width=0.48\textwidth]{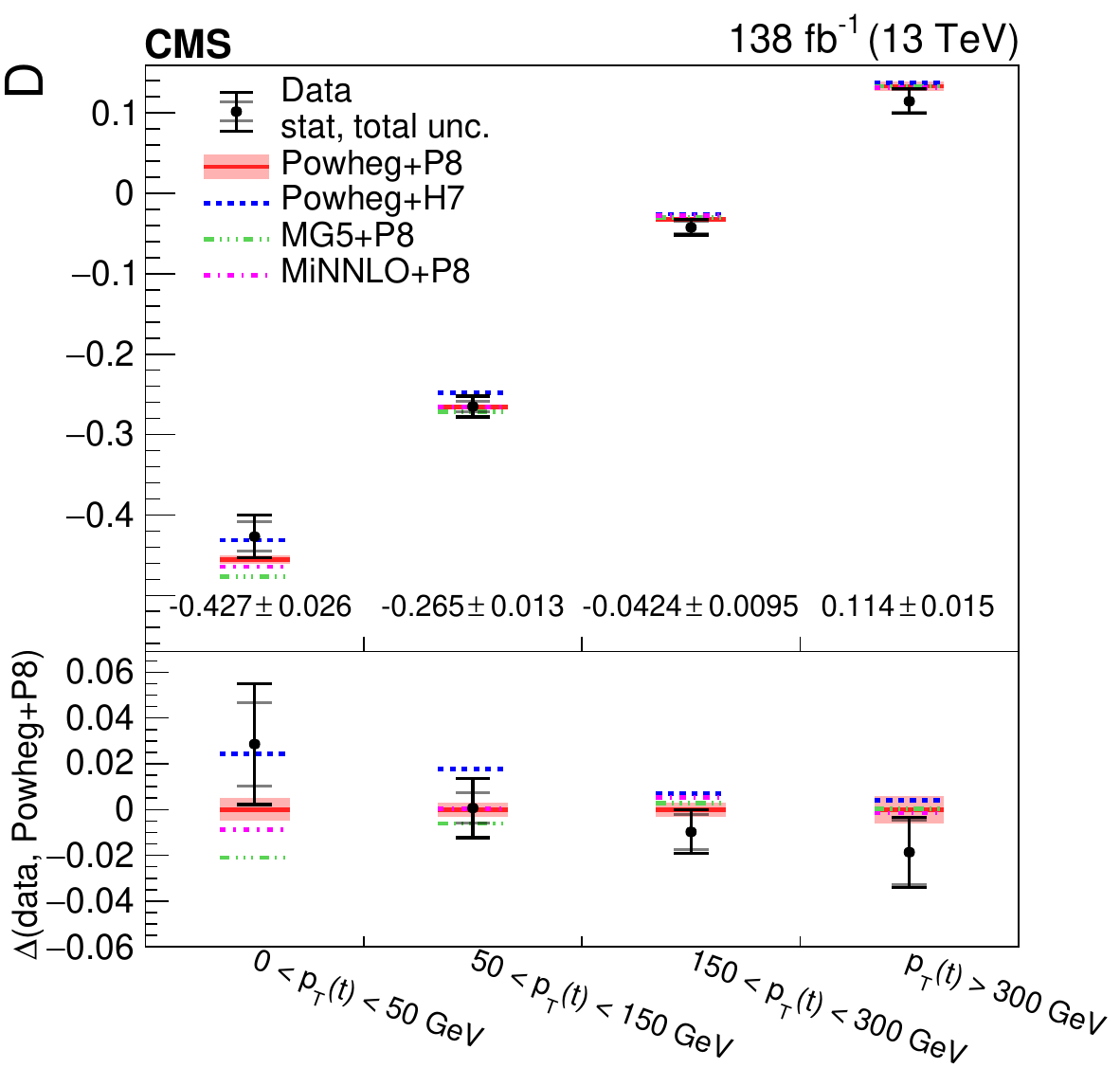}\\
\includegraphics[width=0.48\textwidth]{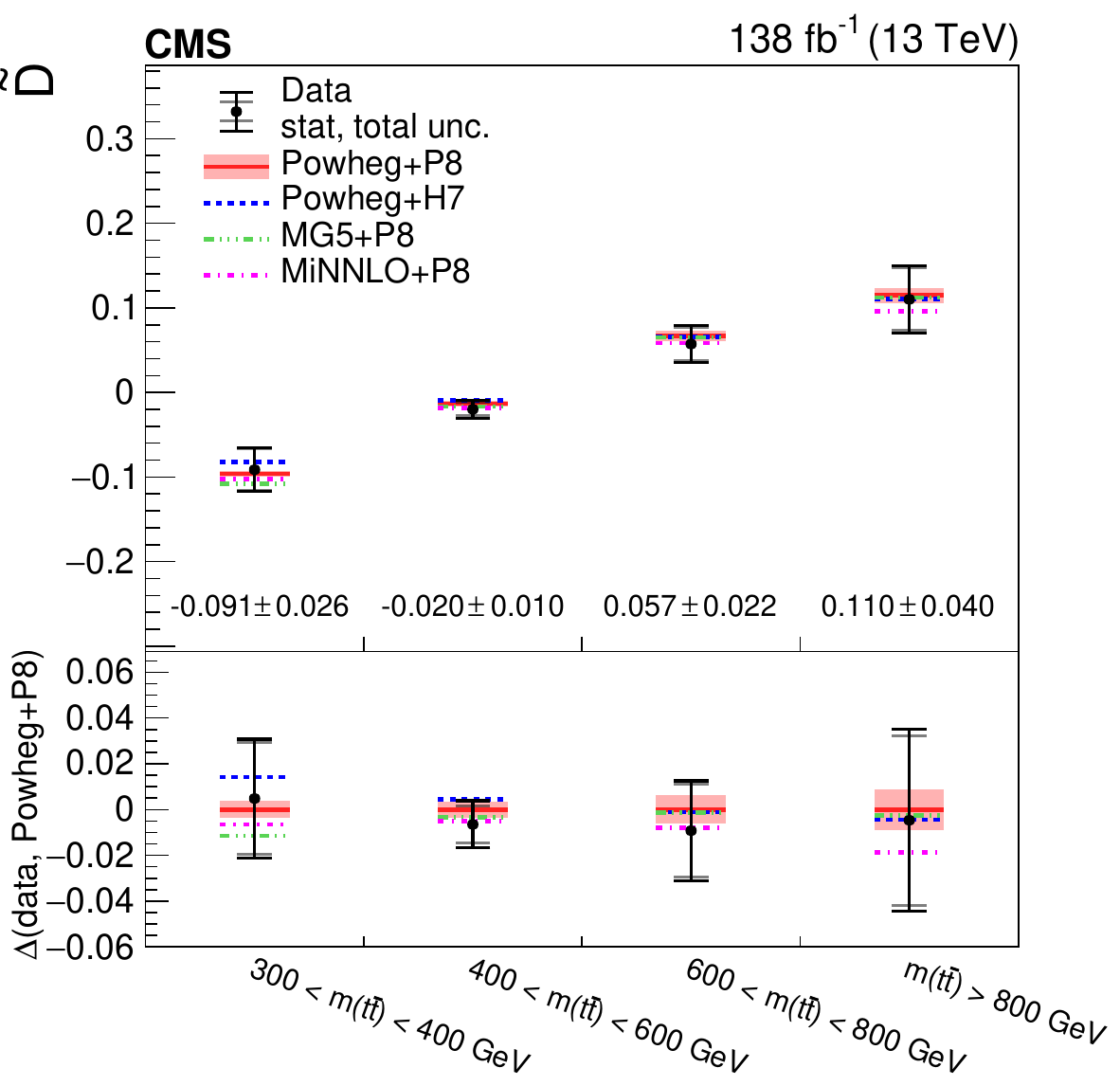}%
\hfill%
\includegraphics[width=0.48\textwidth]{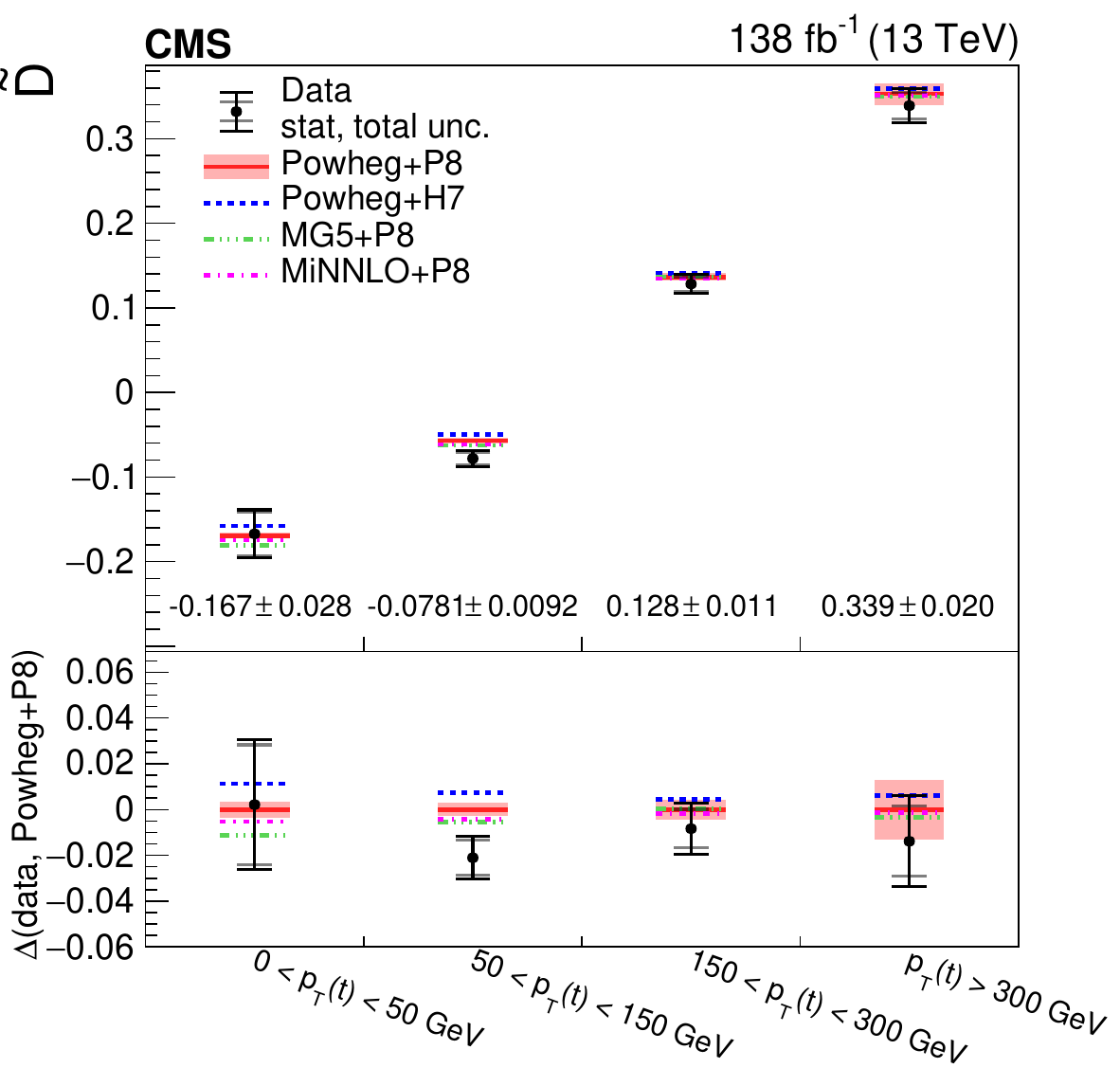}
\caption{Results of the \Qone and \Dtilde measurements in bins of \ttm and \tpt. The measurements (markers) are shown with the statistical uncertainty (inner error bars) and total uncertainty (outer error bars) and compared to the predictions of \PWGPYT, \PWGHRW, \MGHPYT and \MINPYT. In the lower panel results are presented with the \PWGPYT predictions subtracted. The \PWGPYT prediction is displayed with ME scale and PDF uncertainties.}
\label{results_fig_4_1}
\end{figure*}

The \Qone and \Dtilde measurements are presented in bins of \ttm and \tpt in Fig.~\ref{results_fig_4_1}. The finer \tpt binning in the \Qone measurement allows studying events with $\tpt<50\GeV$, where a significance for entanglement of 3.5 (4.4) standard deviations is observed (expected). A similar analysis performed in the dilepton channel~\cite{CMS-PAS-TOP-23-001} has higher sensitivity for entanglement at the \ttbar production threshold. Overall \PWGHRW predicts slightly higher values of \Qone in the threshold region, but the measured coefficients remain compatible with all predictions.

\begin{figure*}[htp!]
\centering
\includegraphics[width=0.48\textwidth]{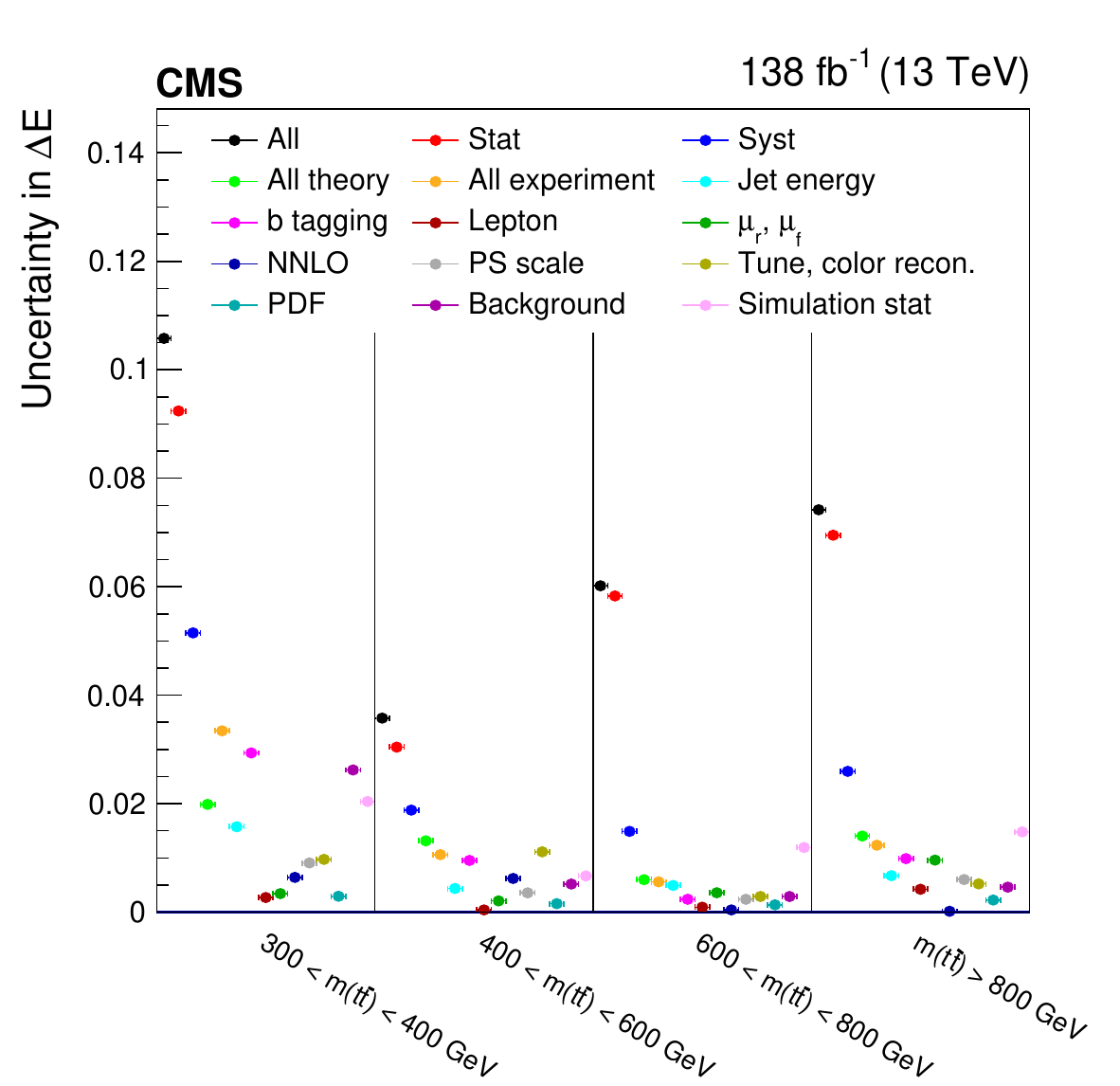}%
\hfill%
\includegraphics[width=0.48\textwidth]{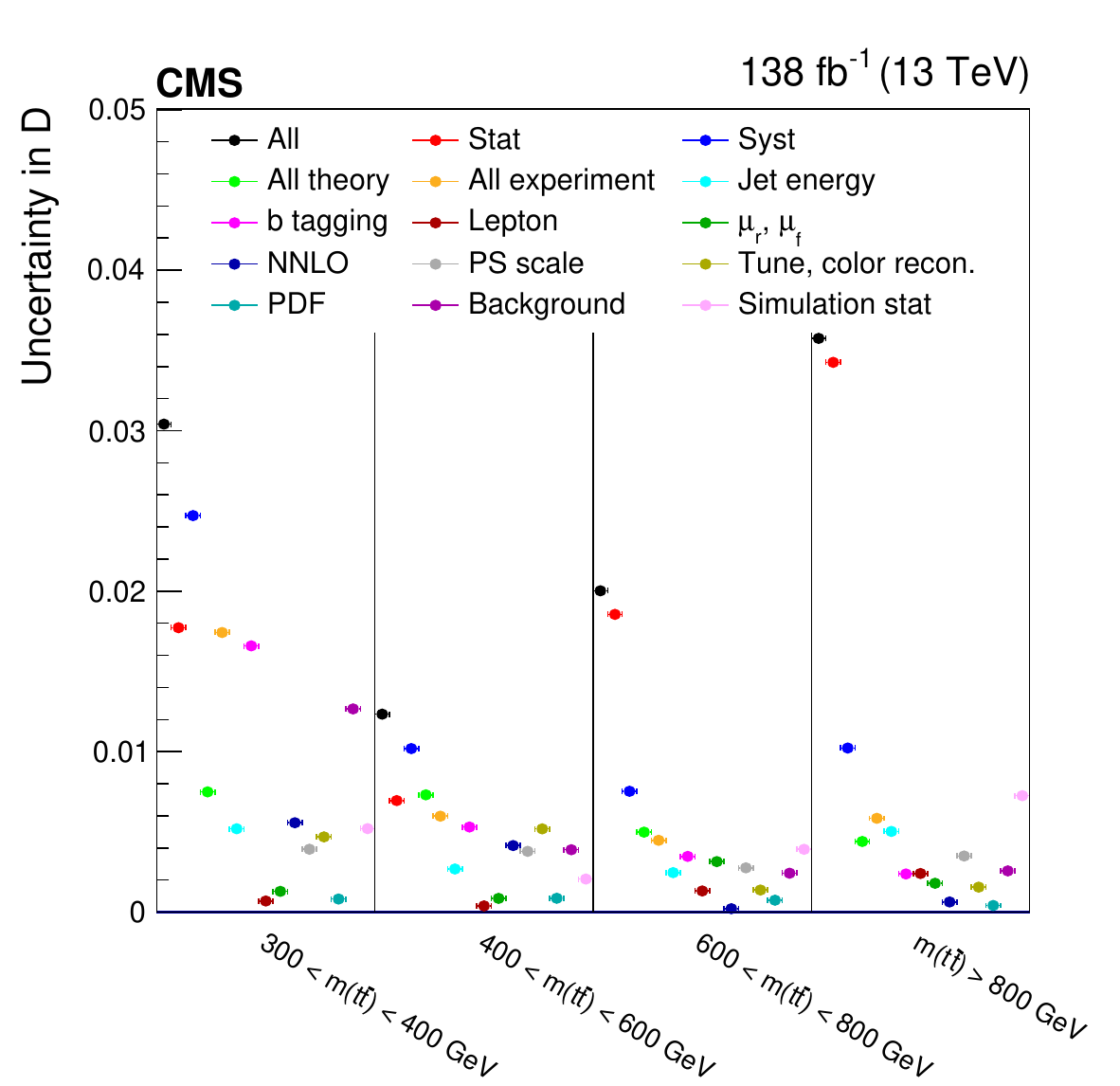} \\
\includegraphics[width=0.48\textwidth]{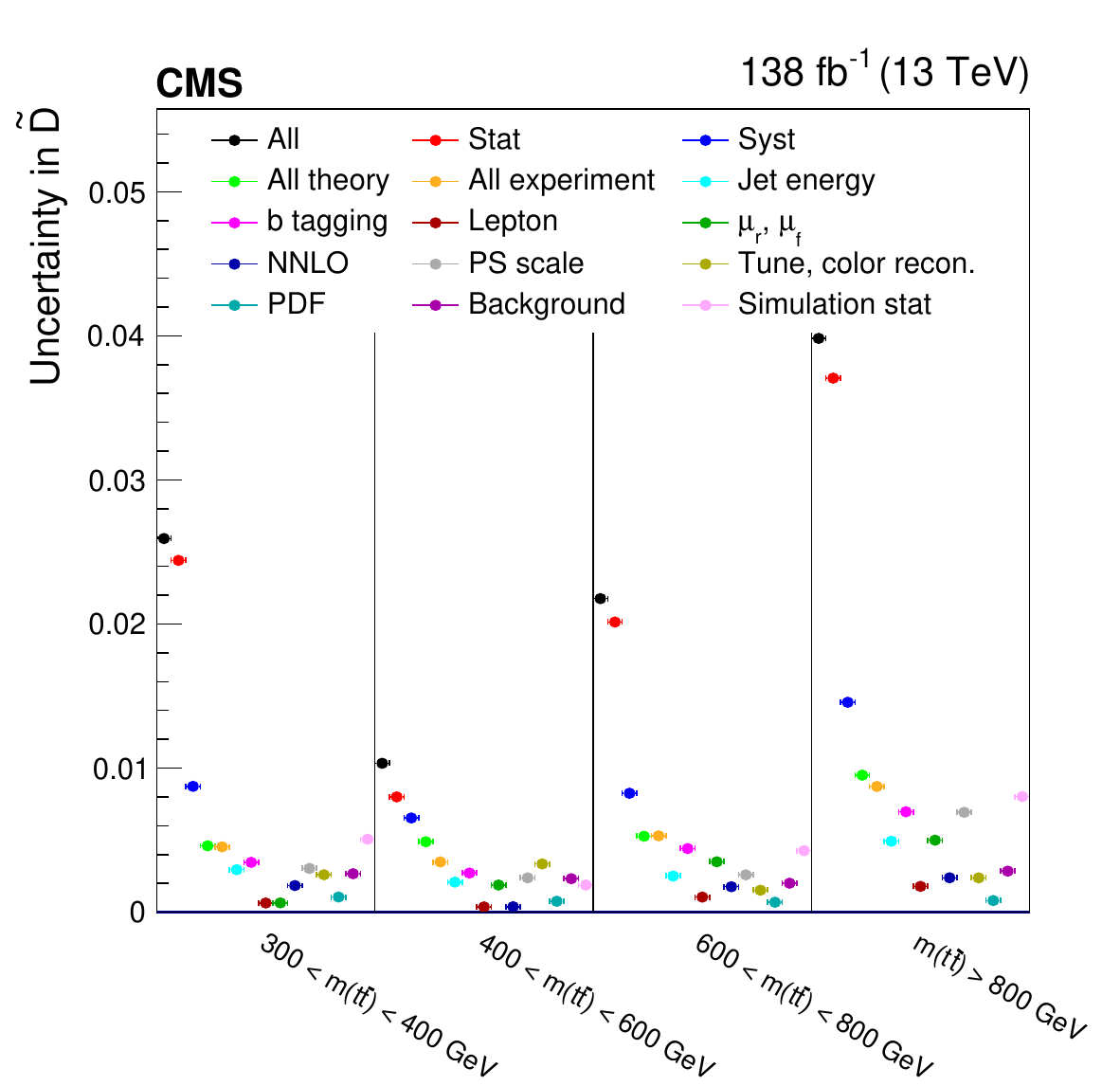}
\caption{Contribution of individual uncertainty sources or groups of uncertainties in the measured \DE, \Qone, and \Dtilde in bins of \ttm.}
\label{results_fig_uncssource_1}
\end{figure*}

Figure~\ref{results_fig_uncssource_1} shows the contributions of various uncertainty sources to the uncertainties in the measured \DE, \Qone, and \Dtilde in bins of \ttm.
In general, the uncertainties in the results are dominated by the statistical contribution, with the exception of the \Qone measurement, where systematic effects are more important at low \ttm. The uncertainty in the \PQb tagging calibration is the dominant source, but depending on the bin, theoretical uncertainties can be of similar sizes, in particular ME and PS scales.

\begin{figure*}[htp!]
\centering
\includegraphics[width=\textwidth]{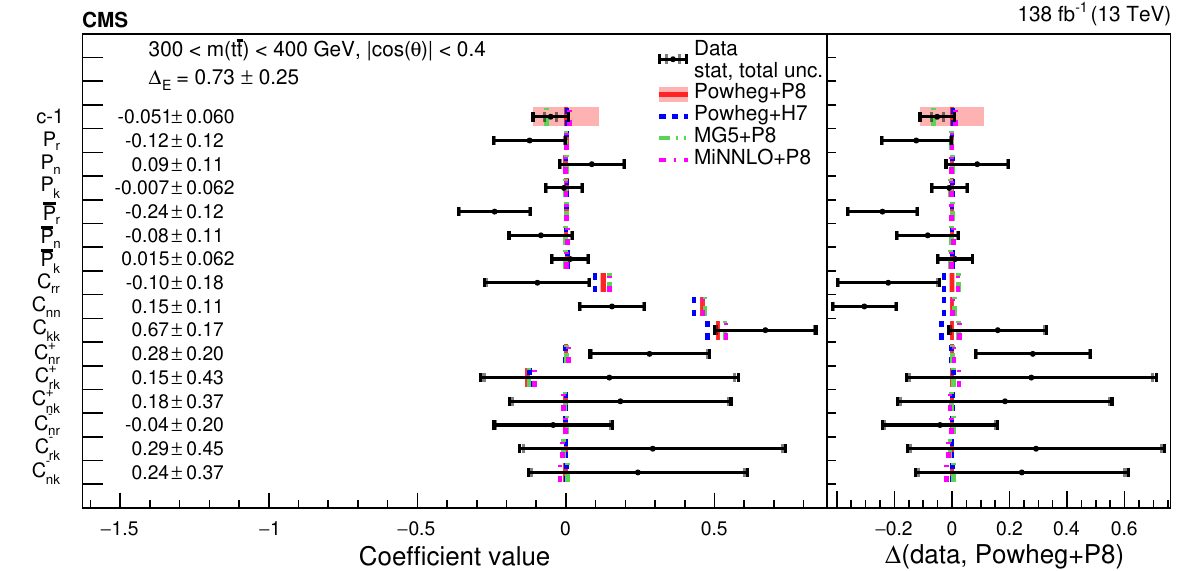}
\includegraphics[width=\textwidth]{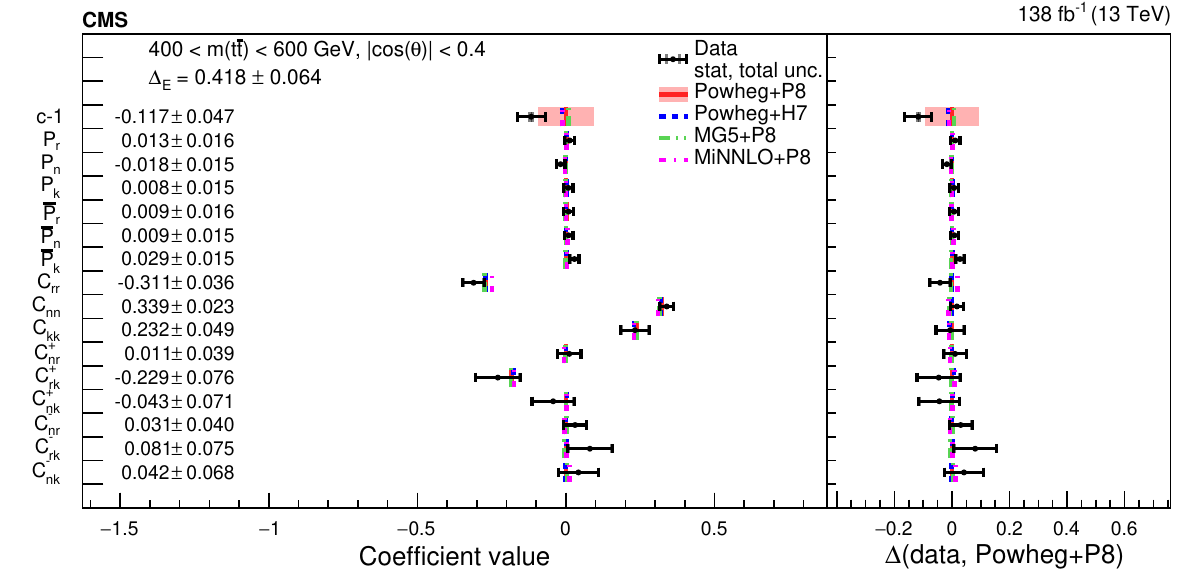}
\caption{Results of the full matrix measurement in bins of \ttm for $\absct<0.4$. The measurements (markers) are shown with the statistical uncertainty (inner error bars) and total uncertainty (outer error bars) and compared to the predictions of \PWGPYT, \PWGHRW, \MGHPYT and \MINPYT. In the right panels, results are presented with the \PWGPYT predictions subtracted. The \PWGPYT prediction is displayed with ME scale and PDF uncertainties. The values of \DE are displayed for each measurement.}
\label{results_fig_6}
\end{figure*}

\begin{figure*}[htp!]
\centering
\includegraphics[width=\textwidth]{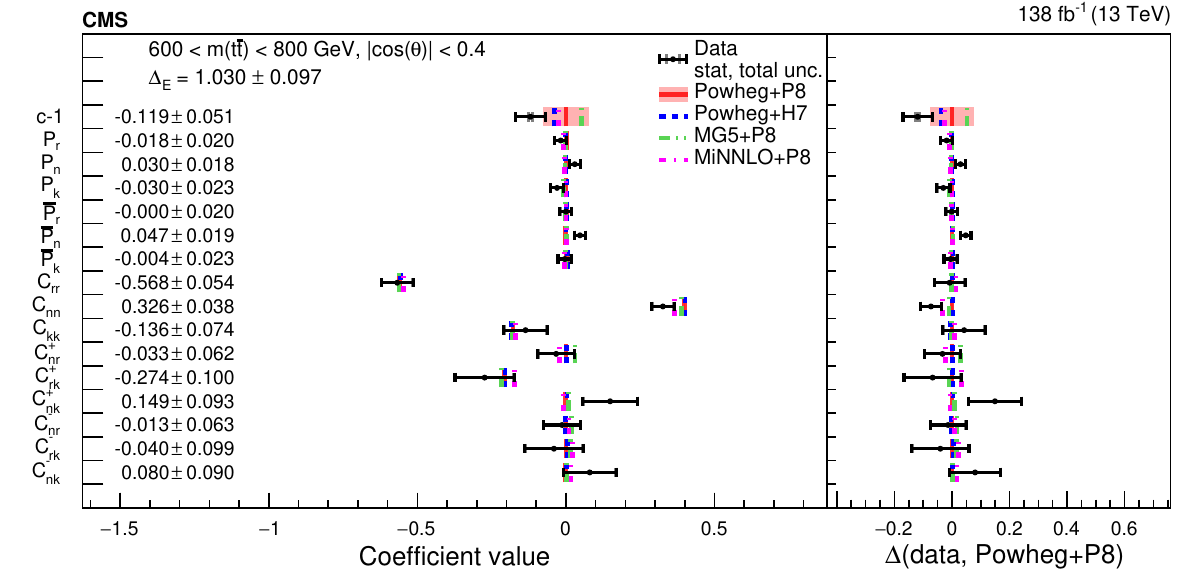}
\includegraphics[width=\textwidth]{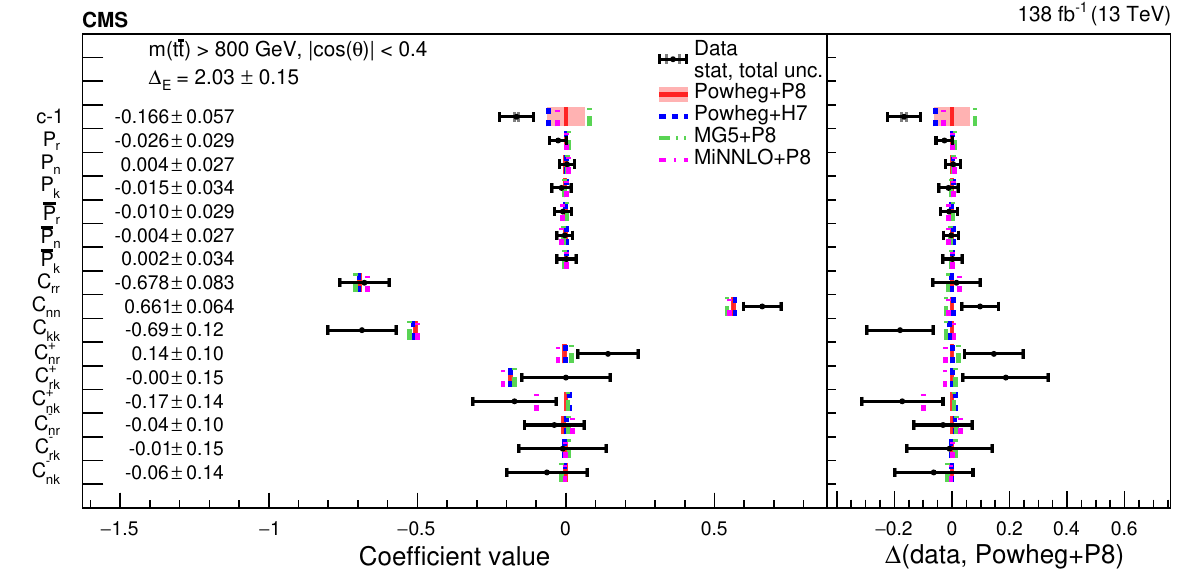}
\caption{Results of the full matrix measurement in bins of \ttm for $\absct<0.4$. The measurements (markers) are shown with the statistical uncertainty (inner error bars) and total uncertainty (outer error bars) and compared to the predictions of \PWGPYT, \PWGHRW, \MGHPYT and \MINPYT. In the right panels, results are presented with the \PWGPYT predictions subtracted. The \PWGPYT prediction is displayed with ME scale and PDF uncertainties. The values of \DE are displayed for each measurement.}
\label{results_fig_6_2}
\end{figure*}

\begin{figure}[htp!]
\centering
\includegraphics[width=0.48\textwidth]{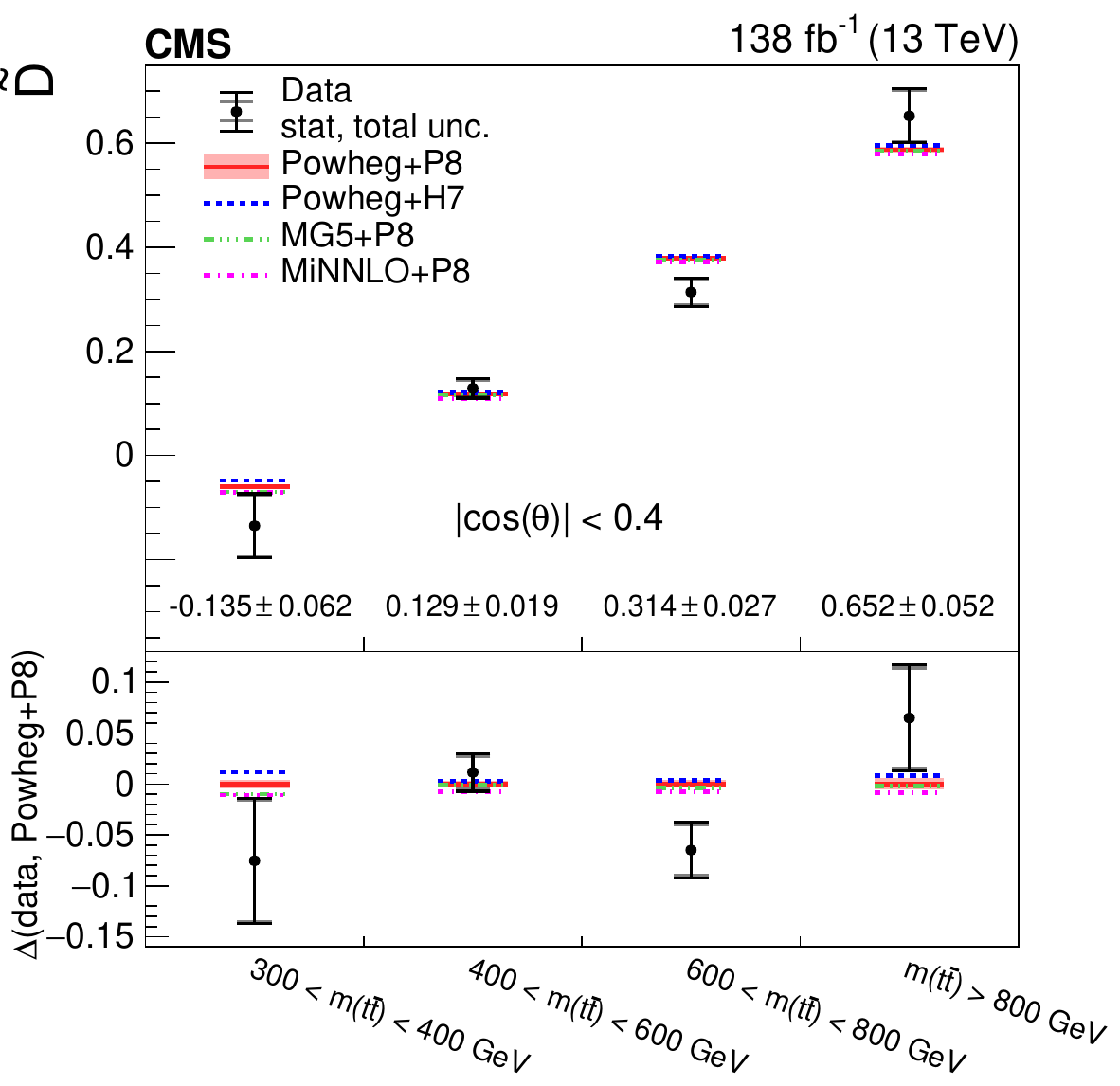}
\caption{Results of the \Dtilde measurements in bins of \ttm for $\absct<0.4$. The measurements (markers) are shown with the statistical uncertainty (inner error bars) and total uncertainty (outer error bars) and compared to the predictions of \PWGPYT, \PWGHRW, \MGHPYT and \MINPYT. In the lower panel results are presented with the \PWGPYT predictions subtracted. The \PWGPYT prediction is displayed with ME scale and PDF uncertainties.}
\label{results_fig_6_1}
\end{figure}

We further study the results in the region $\absct<0.4$. These are shown in bins of \ttm in Figs.~\ref{results_fig_6}--\ref{results_fig_6_2} for the full matrix and in Fig.~\ref{results_fig_6_1} for the \Dtilde measurement. The most significant observation (expectation) of entanglement is obtained for $\ttm>800\GeV$ and $\absct<0.4$, with 6.7 (5.6) standard deviations for the full matrix measurement and 6.1 (5.5) standard deviations for the \Dtilde measurement. The \Qone measurement does not have any sensitivity for entanglement in this region, since the diagonal elements of the $C$ matrix do not all have the same sign.

\begin{figure}[!tp]
\centering
\includegraphics[width=0.48\textwidth]{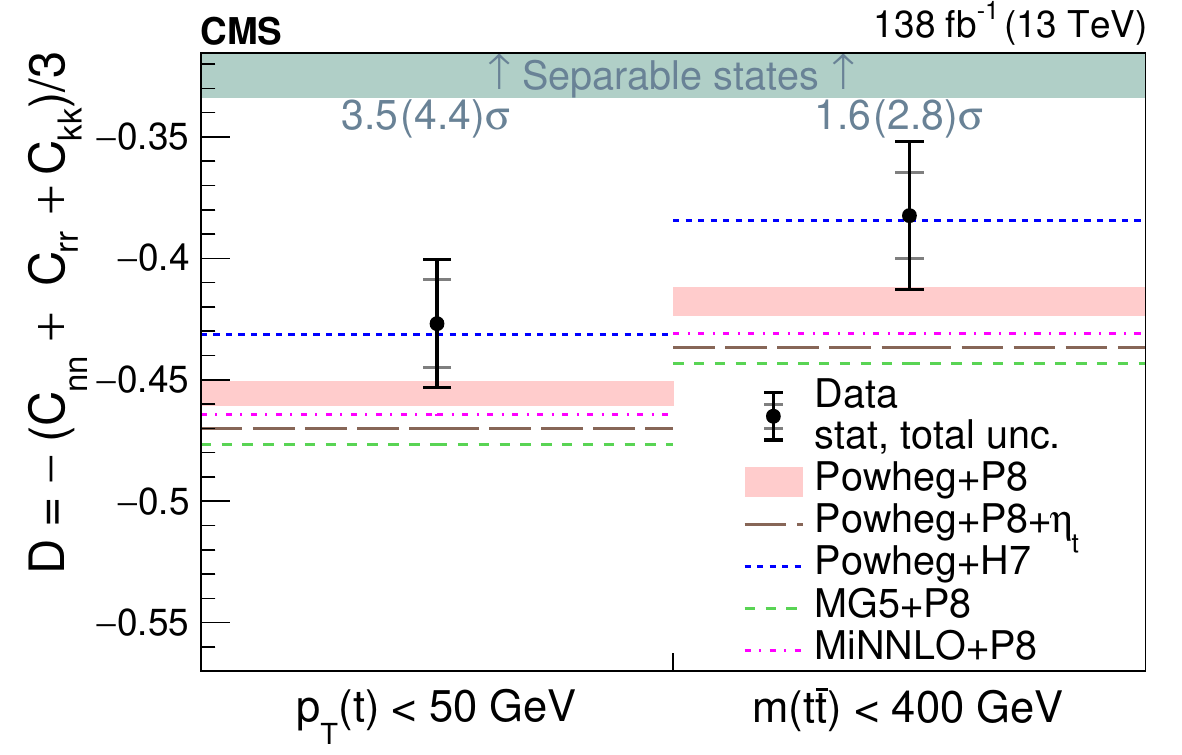}%
\hfill%
\includegraphics[width=0.48\textwidth]{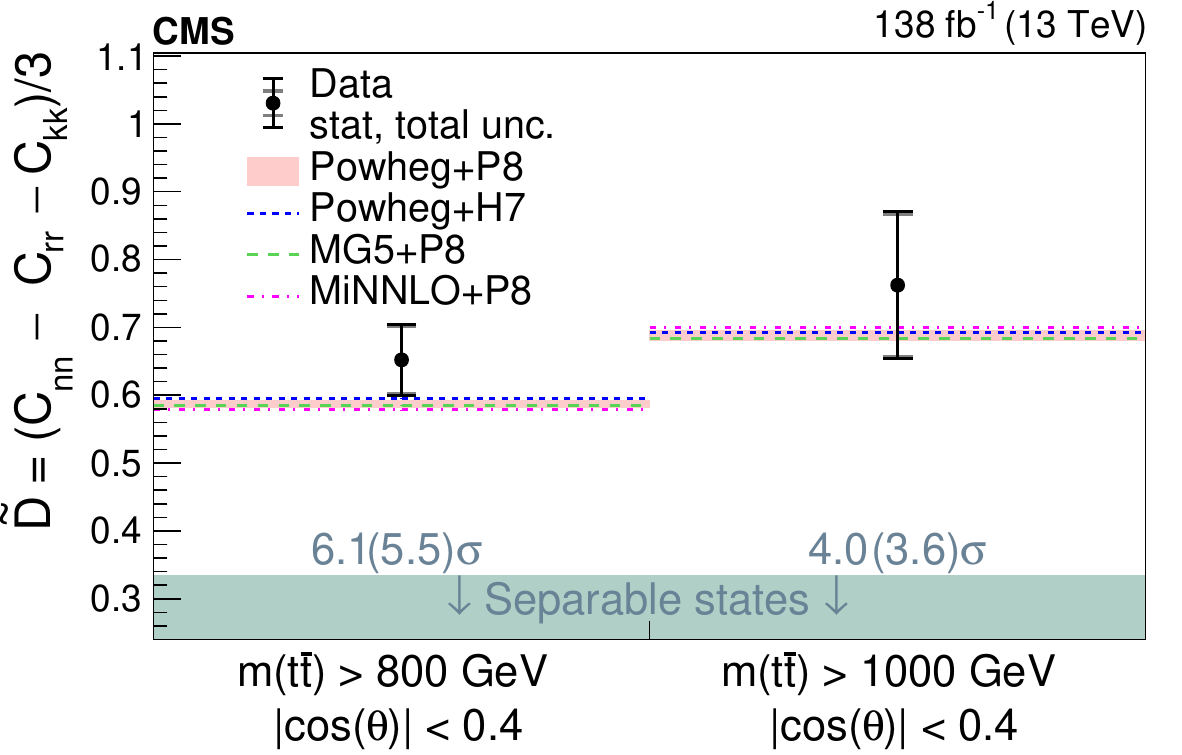} \\
\includegraphics[width=0.48\textwidth]{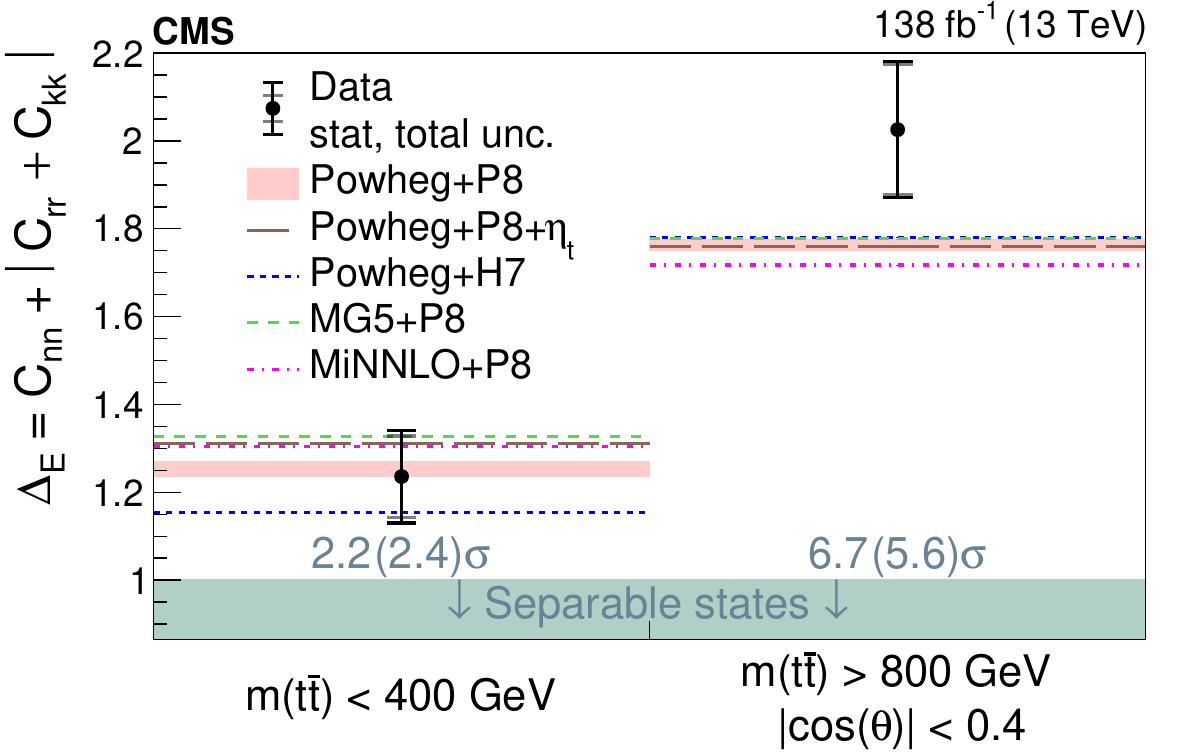}
\caption{Entanglement results for the \Qone measurement in the threshold region (\cmsUpperLeft), \Dtilde measurement in the high-\ttm region (\cmsUpperRight), and the full matrix measurement in different \ttm regions (lower). The measurements (points) are shown with the statistical uncertainty (inner error bars) and total uncertainty (outer error bars) and compared to the predictions of \PWGPYT, \PWGPYTETA, \PWGHRW, \MGHPYT, and \MINPYT. The \PWGPYT prediction is displayed with the ME scale and PDF uncertainties, while for all other predictions only the central values are indicated. The observed (expected) significance of the deviation from the boundary of separable states (green region) is quoted in standard deviations ($\sigma$).
}
\label{resultsdata_fig_18}
\end{figure}

Figure~\ref{resultsdata_fig_18} summarizes the observed (expected) significance of the measured entanglement variables. In the upper panels of Fig.~\ref{resultsdata_fig_18}, the measured results for \Qone (left) near the \ttbar production threshold and for \Dtilde (right) at high \ttm and low \absct are shown. For the full matrix measurement (lower), the \DE results with the highest expected significance are shown for the threshold and high-\ttm kinematic regions.

\begin{figure}[htp!]
\centering
\includegraphics[width=0.48\textwidth]{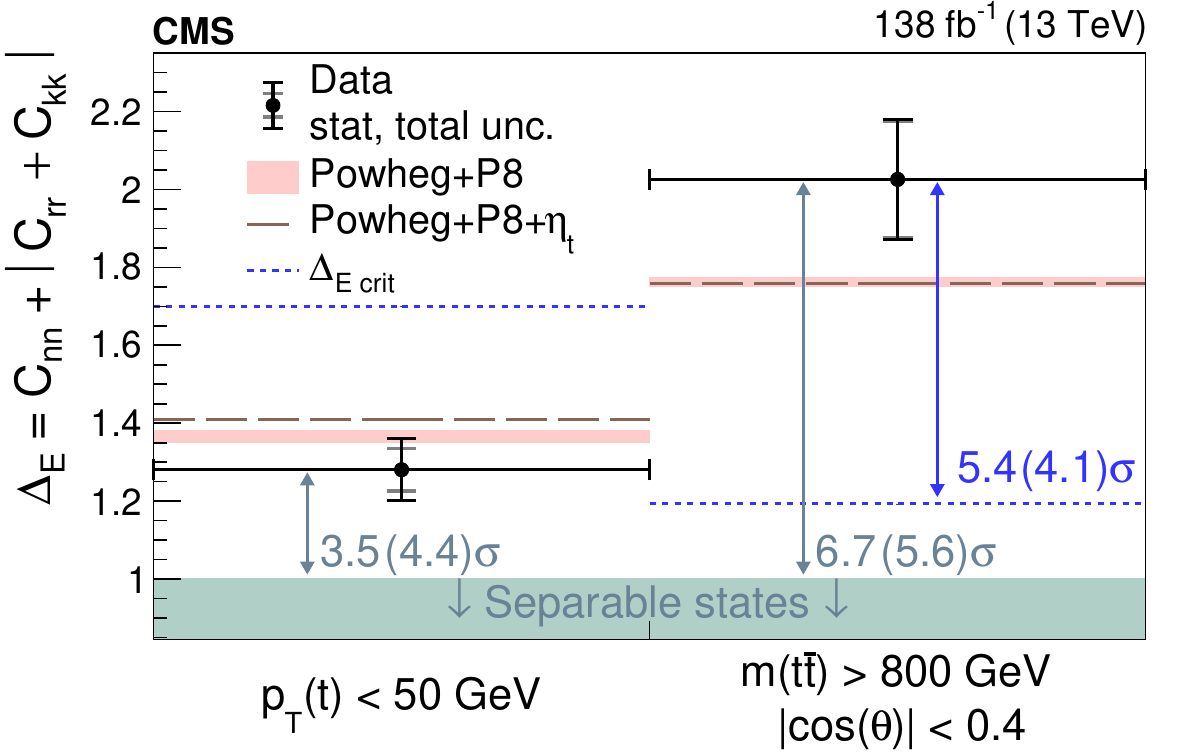}
\caption{The observed levels of entanglement characterized by \DE are shown in the threshold region using the \Qone measurement (first bin), and in the high-\ttm region using the full matrix measurement (second bin). The measurements (points) are shown with the statistical uncertainty (inner error bars) and total uncertainty (outer error bars) and compared to the predictions of \PWGPYT, \PWGPYTETA. The \PWGPYT prediction is displayed with the ME scale and PDF uncertainties. The horizontal blue lines correspond to the maximum level of entanglement \DEcrit that can be explained by the exchange of information between \PQt and \PAQt at the speed of light. The significance in standard deviations ($\sigma$) by which the measurement exceeds \DEcrit and unity is quoted in blue and light green, respectively, and indicated by the corresponding arrows.}
\label{resultsdata_fig_20}
\end{figure}

Previous measurements of the \ttbar entanglement by ATLAS~\cite{ATLAS:2023fsd} and CMS~\cite{CMS-PAS-TOP-23-001} were performed only at the threshold of \ttbar production, where the relative velocity of the top quarks is low. In this analysis, we additionally measure the entanglement at high \ttm, where, in most of the selected events, the top quark and antiquark decays are space-like separated because of their high relative velocity. Using the decay products, the spin correlation is measured when the top quarks decay. From simulations we know that for $\ttm>800\GeV$ the fraction of space-like separated decays is about 90\%~\cite{Severi:2021cnj}. An observation of entanglement could be explained by an unobserved exchange of classical information between the decaying top quarks. Therefore, we introduce a more stringent criterion for entanglement that cannot be explained by such an exchange of information at $v \leq c$ alone (``critical entanglement''). For this, the time-like separated decays are assumed to contribute with the maximum possible value for entanglement of $\DEmax=3$, while the space-like separated decays should at least fulfill the condition $\DEsep=1$. Therefore, the lower boundary of critical entanglement \DEcrit can be defined for a given fraction $f$ of space-like separated decays as follows:
\begin{equation}
    \DEcrit=f\DEsep+(1-f)\DEmax.
\end{equation}
As was shown in Ref.~\cite{regina_locality}, using top quark decays in the definition of $f$ results in the most stringent criterion \DEcrit. The most sensitive measurements in the threshold and high-\ttm kinematic regions are summarized in Fig.~\ref{resultsdata_fig_20} together with the value for \DEcrit for each case, and the significance with respect to critical entanglement and separable states. The first bin $\tpt<50\GeV$ was obtained using the \Qone measurement, and the second bin for $\ttm>800\GeV$ and $\absct<0.4$ was obtained using the full matrix measurement. In the second bin, the space-like fraction is $f=90\%$, which corresponds to a $\DEcrit=1.2$. The measured (expected) value exceeds this limit by 5.4 (4.1) standard deviations as shown by the blue vertical arrow.

\begin{figure}[tp!]
\centering
\includegraphics[width=0.48\textwidth]{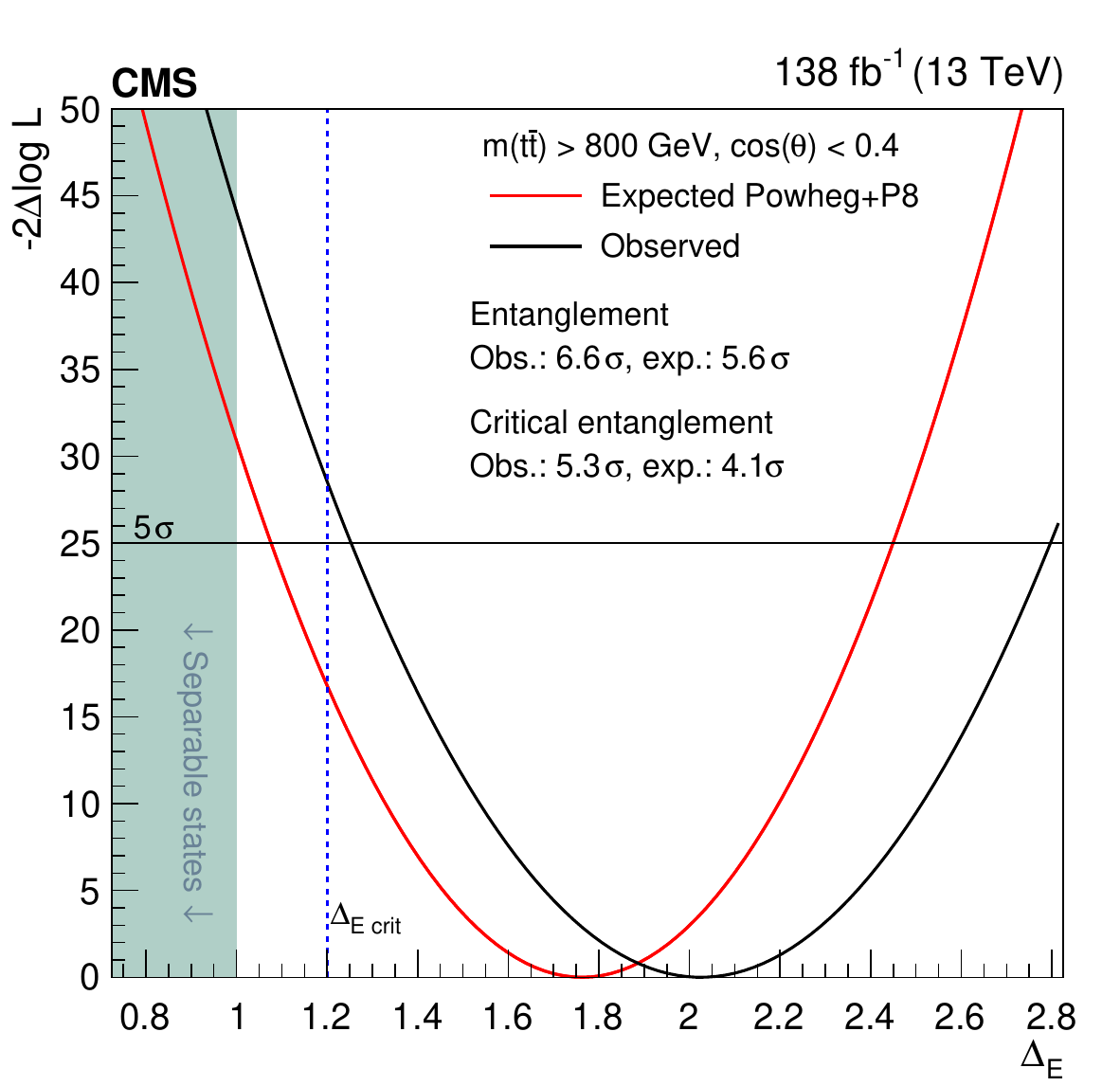}
\caption{Results of the profile likelihood scans. The quantity $-2\Delta\log(L)$ is shown as function of \DE in the bin with $\ttm>800$\GeV and $\absct<0.4$ for data (black line) and the \PWGPYT simulation (red line). The observed and expected significances in standard deviations ($\sigma$) for \DE exceeding unity and \DEcrit are quoted.}
\label{results_fig_descan}
\end{figure}

To validate the Gaussian approximation used in calculating these significances, we perform profile likelihood fits for fixed values of \DE in the bin with $\ttm>800\GeV$ and $\absct<0.4$. For this test, the parameter transformation $\Cnn=\DE-\abs{\Crr+\Ckk}$ is applied to make \DE a parameter of $L$. The significances are calculated as $\sqrt{\smash[b]{-2\Delta\log(L)}}$, where $\Delta\log(L)$ is the difference between the profiled likelihood values for a fixed value of \DE and the global maximum. In Fig.~\ref{results_fig_descan}, the scan of $-2\Delta\log(L)$ is shown as a function of \DE. The extracted significances for $\DE=1$ and 1.2 are in close agreement with those obtained with the approximate method.

\clearpage
\section{Summary}
\label{summary}

The polarization and spin correlation in top quark pair (\ttbar) production are measured in events with an electron or a muon plus jets in the final state. The entanglement between the spins of the top quark and antiquark is determined from the measured spin correlation by applying the Peres--Horodecki criterion. The measurements are based on proton-proton collision data at $\sqrt{s}=13\TeV$ collected by the CMS experiment at the LHC, corresponding to an integrated luminosity of 138\fbinv. The decay products of the top quarks are identified using an artificial neural network. The coefficients of the polarization vectors and the spin correlation matrix are extracted simultaneously from the angular distributions of \ttbar decay products using a binned likelihood fit. This is done both inclusively and in various regions of the phase space. The observed polarization and spin correlation are in agreement with the standard model expectations. The standard model predicts entangled \ttbar states at the production threshold and at high masses of the \ttbar system. Entanglement is observed in events with high \ttbar mass, with an observed (expected) significance of 6.7 (5.6) standard deviations, while in events with low transverse momentum of the top quark a significance of 3.5 (4.4) standard deviations is observed (expected). This is the first observation of entanglement at high \ttbar mass where in about 90\% of the observed \ttbar events the decays of the top quark and antiquark are space-like separated.

\begin{acknowledgments}
    We congratulate our colleagues in the CERN accelerator departments for the excellent performance of the LHC and thank the technical and administrative staffs at CERN and at other CMS institutes for their contributions to the success of the CMS effort. In addition, we gratefully acknowledge the computing centers and personnel of the Worldwide LHC Computing Grid and other centers for delivering so effectively the computing infrastructure essential to our analyses. Finally, we acknowledge the enduring support for the construction and operation of the LHC, the CMS detector, and the supporting computing infrastructure provided by the following funding agencies: SC (Armenia), BMBWF and FWF (Austria); FNRS and FWO (Belgium); CNPq, CAPES, FAPERJ, FAPERGS, and FAPESP (Brazil); MES and BNSF (Bulgaria); CERN; CAS, MoST, and NSFC (China); MINCIENCIAS (Colombia); MSES and CSF (Croatia); RIF (Cyprus); SENESCYT (Ecuador); ERC PRG, RVTT3 and MoER TK202 (Estonia); Academy of Finland, MEC, and HIP (Finland); CEA and CNRS/IN2P3 (France); SRNSF (Georgia); BMBF, DFG, and HGF (Germany); GSRI (Greece); NKFIH (Hungary); DAE and DST (India); IPM (Iran); SFI (Ireland); INFN (Italy); MSIP and NRF (Republic of Korea); MES (Latvia); LMTLT (Lithuania); MOE and UM (Malaysia); BUAP, CINVESTAV, CONACYT, LNS, SEP, and UASLP-FAI (Mexico); MOS (Montenegro); MBIE (New Zealand); PAEC (Pakistan); MES and NSC (Poland); FCT (Portugal); MESTD (Serbia); MCIN/AEI and PCTI (Spain); MOSTR (Sri Lanka); Swiss Funding Agencies (Switzerland); MST (Taipei); MHESI and NSTDA (Thailand); TUBITAK and TENMAK (Turkey); NASU (Ukraine); STFC (United Kingdom); DOE and NSF (USA).
    
    \hyphenation{Rachada-pisek} Individuals have received support from the Marie-Curie program and the European Research Council and Horizon 2020 Grant, contract Nos.\ 675440, 724704, 752730, 758316, 765710, 824093, 101115353, 101002207, and COST Action CA16108 (European Union); the Leventis Foundation; the Alfred P.\ Sloan Foundation; the Alexander von Humboldt Foundation; the Science Committee, project no. 22rl-037 (Armenia); the Belgian Federal Science Policy Office; the Fonds pour la Formation \`a la Recherche dans l'Industrie et dans l'Agriculture (FRIA-Belgium); the F.R.S.-FNRS and FWO (Belgium) under the ``Excellence of Science -- EOS" -- be.h project n.\ 30820817; the Beijing Municipal Science \& Technology Commission, No. Z191100007219010 and Fundamental Research Funds for the Central Universities (China); the Ministry of Education, Youth and Sports (MEYS) of the Czech Republic; the Shota Rustaveli National Science Foundation, grant FR-22-985 (Georgia); the Deutsche Forschungsgemeinschaft (DFG), among others, under Germany's Excellence Strategy -- EXC 2121 ``Quantum Universe" -- 390833306, and under project number 400140256 - GRK2497; the Hellenic Foundation for Research and Innovation (HFRI), Project Number 2288 (Greece); the Hungarian Academy of Sciences, the New National Excellence Program - \'UNKP, the NKFIH research grants K 131991, K 133046, K 138136, K 143460, K 143477, K 146913, K 146914, K 147048, 2020-2.2.1-ED-2021-00181, and TKP2021-NKTA-64 (Hungary); the Council of Science and Industrial Research, India; ICSC -- National Research Center for High Performance Computing, Big Data and Quantum Computing and FAIR -- Future Artificial Intelligence Research, funded by the NextGenerationEU program (Italy); the Latvian Council of Science; the Ministry of Education and Science, project no. 2022/WK/14, and the National Science Center, contracts Opus 2021/41/B/ST2/01369 and 2021/43/B/ST2/01552 (Poland); the Funda\c{c}\~ao para a Ci\^encia e a Tecnologia, grant CEECIND/01334/2018 (Portugal); the National Priorities Research Program by Qatar National Research Fund; MCIN/AEI/10.13039/501100011033, ERDF ``a way of making Europe", and the Programa Estatal de Fomento de la Investigaci{\'o}n Cient{\'i}fica y T{\'e}cnica de Excelencia Mar\'{\i}a de Maeztu, grant MDM-2017-0765 and Programa Severo Ochoa del Principado de Asturias (Spain); the Chulalongkorn Academic into Its 2nd Century Project Advancement Project, and the National Science, Research and Innovation Fund via the Program Management Unit for Human Resources \& Institutional Development, Research and Innovation, grant B39G670016 (Thailand); the Kavli Foundation; the Nvidia Corporation; the SuperMicro Corporation; the Welch Foundation, contract C-1845; and the Weston Havens Foundation (USA).
\end{acknowledgments}

\bibliography{auto_generated}
\cleardoublepage \appendix\section{The CMS Collaboration \label{app:collab}}\begin{sloppypar}\hyphenpenalty=5000\widowpenalty=500\clubpenalty=5000
\cmsinstitute{Yerevan Physics Institute, Yerevan, Armenia}
{\tolerance=6000
A.~Hayrapetyan, A.~Tumasyan\cmsAuthorMark{1}\cmsorcid{0009-0000-0684-6742}
\par}
\cmsinstitute{Institut f\"{u}r Hochenergiephysik, Vienna, Austria}
{\tolerance=6000
W.~Adam\cmsorcid{0000-0001-9099-4341}, J.W.~Andrejkovic, L.~Benato\cmsorcid{0000-0001-5135-7489}, T.~Bergauer\cmsorcid{0000-0002-5786-0293}, S.~Chatterjee\cmsorcid{0000-0003-2660-0349}, K.~Damanakis\cmsorcid{0000-0001-5389-2872}, M.~Dragicevic\cmsorcid{0000-0003-1967-6783}, P.S.~Hussain\cmsorcid{0000-0002-4825-5278}, M.~Jeitler\cmsAuthorMark{2}\cmsorcid{0000-0002-5141-9560}, N.~Krammer\cmsorcid{0000-0002-0548-0985}, A.~Li\cmsorcid{0000-0002-4547-116X}, D.~Liko\cmsorcid{0000-0002-3380-473X}, I.~Mikulec\cmsorcid{0000-0003-0385-2746}, J.~Schieck\cmsAuthorMark{2}\cmsorcid{0000-0002-1058-8093}, R.~Sch\"{o}fbeck\cmsAuthorMark{2}\cmsorcid{0000-0002-2332-8784}, D.~Schwarz\cmsorcid{0000-0002-3821-7331}, M.~Sonawane\cmsorcid{0000-0003-0510-7010}, W.~Waltenberger\cmsorcid{0000-0002-6215-7228}, C.-E.~Wulz\cmsAuthorMark{2}\cmsorcid{0000-0001-9226-5812}
% ...
\par}
\vskip\cmsinstskip
\dag:~Deceased\\
$^{1}$Also at Yerevan State University, Yerevan, Armenia\\
$^{2}$Also at TU Wien, Vienna, Austria\\
% ...
\end{sloppypar}
\end{document}